\documentclass[11pt]{article}
\usepackage{amssymb,amsmath}
\usepackage{graphicx}
\usepackage{cite}
\usepackage{xcolor}
\definecolor{darkblue}{rgb}{0,0,0.8}
\usepackage[dvipdfmx,breaklinks,colorlinks=true,linktoc=page,urlcolor=darkblue,linkcolor=darkblue,citecolor=darkblue]{hyperref}
\usepackage{overpic}
\usepackage{url}

\numberwithin{equation}{section}
\newcommand{\SUSY}{{\bf Q}}
\newcommand{\BRST}{{\bf Q}_\text{B}}
\newcommand{\QTOT}{\widehat{\bf Q}}
\newcommand{\SLASH}[1]{\hskip.5mm/\hskip-2.4mm#1}
\newcommand{\mini}[1]{{\scalebox{0.6}{$#1$}}}
\newcommand{\dotp}{{\hskip.2mm{\raisebox{-0.5mm}{\scalebox{1.3}{$\cdot$}}}\hskip.2mm}}
\begin{document}

\thispagestyle{empty}

\vskip1cm
\begin{center}
{\Large \bf Supersymmetric vortex loops in 3D gauge theories}
\vskip1.5cm
\renewcommand{\thefootnote}{\fnsymbol{footnote}}
Kazuo Hosomichi\footnote{hosomiti@nda.ac.jp} and
Kohei Suzuki\footnote{em57042@nda.ac.jp}

\bigskip
{\it Department of Applied Physics, National Defense Academy,\\
1-10-20 Hashirimizu, Yokosuka-city, Kanagawa 239-8686 Japan}

\end{center}

\vskip1cm
\begin{abstract}
\noindent
We give a precise definition of BPS vortex loops in 3D non-abelian gauge theories with ${\cal N}=2$ SUSY by the path integral over fields with a prescribed singular behavior. We compute the expectation value of a BPS vortex loop on an ellipsoid. Using the result we revisit the known equivalence between Wilson and vortex loops in pure Chern-Simons theory. Naive computations of expectation values in ${\cal N}=2$ theory leads to an unwanted shift of parameters in the rule of correspondence. We resolve the problem by relating the shift to the global anomaly of ${\cal N}=2$ SUSY quantum mechanics. For theories with $U(N)$ gauge group we also develop an alternative description of vortex loops in terms of 1D ${\cal N}=2$ SUSY quantum mechanics on their worldline. For vortex loops in ${\cal N}=4$ theories, our construction reproduces some of the quiver GLSMs of Assel and Gomis.
\end{abstract}
\newpage
\setcounter{tocdepth}{2}
\tableofcontents
%%%%%%%%%%%%%%%%%%%%%%%%%%
\renewcommand{\thefootnote}{\arabic{footnote}}
\setcounter{footnote}{0}

\section{Introduction}\label{sec:intro}

Vortex loops and Wilson loops \cite{Wilson:1974sk} play an equally important role in the study of 3D gauge theories. Expectation values of these operators and their relations in supersymmetric gauge theories have been studied extensively in recent years. The first exact formula was obtained for BPS Wilson loops in ${\cal N}=2$ Chern-Simons(CS)-matter theories on $S^3$ in \cite{Kapustin:2009kz}. On the other hand, there remain a number of problems to be resolved for vortex loops.

There are two common definitions of vortex loops. The first is that a gauge field has a prescribed singularity along a loop in space. The other is that the loop supports a quantum mechanics interacting with the 3D fields. The second definition is expected to coincide with the first upon integrating out the quantum mechanical variables. Both have supersymmetric extensions so as to describe BPS vortex loops. All these are similar to how surface operators are defined for 4D gauge theories \cite{Gukov:2006jk}.

Based on the first definition, BPS vortex loops were first studied in \cite{Drukker:2008jm} for ABJM model \cite{Aharony:2008ug}. Exact computation of their expectation values was initiated in \cite{Kapustin:2012iw,Drukker:2012sr}, but so far it has been mostly restricted to abelian gauge theories. Moreover, the results seem to indicate that BPS vortex loops in abelian ${\cal N}=2$ gauge theories are trivial; namely, as far as supersymmetric observables are concerned, they are equivalent to the identity operator. Regarding the second definition, a systematic identification of the quantum mechanics on various vortex loops in 3D ${\cal N}=4$ theories was made in \cite{Assel:2015oxa} using mirror symmetry \cite{Intriligator:1996ex} and type IIB brane construction \cite{Hanany:1996ie,deBoer:1996mp,deBoer:1996ck}. However, generalization of their result to theories with less SUSY does not seem straightforward. Also, the correspondence between the two definitions is not fully clear yet.

This paper has two main purposes. The first is to give a fully precise definition of the path integral in the presence of vortex loops. We present our definition in Section \ref{sec:Localization} and apply it to the expectation value of a BPS vortex loop on an ellipsoid. Then in Section \ref{sec:pureCS} we test our result against the known equivalence of Wilson and vortex loops in pure CS theory \cite{Moore:1989yh}. The original proof of the equivalence \cite{Moore:1989yh} used the coadjoint orbit quantization for representing Wilson loops. It can actually be thought of as a prototypical example of a quantum mechanics on a loop interacting with the field theory in 3D space. By understanding the equivalence of the Wilson and vortex loops we make the first precise correspondence between the two definitions of vortex loops explained above.

In fact, by a naive comparison in ${\cal N}=2$ CS theory we find there is an unwanted shift of parameters in the equivalence relation. This was already pointed out in \cite{Fan:2018wya}. At the end of Section \ref{sec:pureCS} we propose a resolution which relates the shift to the global anomaly in ${\cal N}=2$ SUSY quantum mechanics \cite{Hori:2014tda}.

The second purpose is to make correspondence of the two definitions of vortex loops in a wider class of ${\cal N}=2$ theories. For this purpose we will focus on vortex loops in $U(N)$ gauge theories. We begin in Section \ref{sec:GLSM-Vortex} by developing the description of coadjoint orbit quantum mechanics as quiver gauged linear sigma models (GLSMs) of the kind studied in \cite{Donagi:2007hi,Gadde:2013dda}. It turns out that there is a number of ${\cal N}=2$ GLSMs for a single coadjoint orbit. We suspect that they are dual to one another. Then in Section \ref{sec:add-matters} we identify the extensions of these GLSMs that account for the addition of various matter chiral multiplets on the vortex background. This will be done for the matters in the adjoint, fundamental and anti-fundamental representations of $U(N)$. Vortex loops in ${\cal N}=4$ theories are studied in Section \ref{sec:N=4} where our construction reproduces some of the GLSMs for vortex loops that are identified in \cite{Assel:2015oxa}. We conclude in Section \ref{sec:conclusion} with a discussion of possible future directions. In Appendix \ref{sec:U(1)} we discuss the vortex loops in abelian theories in detail and compare our result with those of \cite{Kapustin:2012iw, Drukker:2012sr}.

\paragraph{Conventions.}

We denote by $\gamma^a$ the 3D gamma matrices satisfying $\{\gamma^a,\gamma^b\}=2\delta^{ab}$ and $\gamma^{ab}\equiv\frac12(\gamma^a\gamma^b-\gamma^b\gamma^a)=i\varepsilon^{abc}\gamma^c$, where $\varepsilon^{abc}$ is the totally antisymmetric symbol with $\varepsilon^{123}=1$. We set $\gamma^a$ equal to Pauli's matrices when its explicit form is needed. We also use Dirac's slash notation such as $\SLASH{K}\equiv\gamma^aK^a$. The indices for spinors are always suppressed. For their bilinear, we use the notation 
\[
 \xi\psi \equiv \xi_\alpha C^{\alpha\beta}\psi_\beta,\qquad
 \xi\gamma^a\psi \equiv \xi_\alpha C^{\alpha\beta}(\gamma^a)_\beta^{~\gamma}\psi_\gamma,\qquad\text{etc.}
\]
where $C$ is an anti-symmetric matrix with $C^{12}=-C^{21}=1$.

\section{Exact computation on an ellipsoid}\label{sec:Localization}

In this section we give our definition of BPS vortex loops in ${\cal N}=2$ SUSY gauge theories. This involves specifying the singular behavior of the gauge field as well as the fluctuation of all the fields near the loop. As a concrete problem to work with, we consider the expectation value of a BPS vortex loop on an ellipsoid.

\subsection{Definition of the theory and vortex loops}\label{sec:def3dtheory}

3D ${\cal N}=2$ supersymmetric field theories can be defined on a curved Riemannian manifold if the background fields $H, K_m, V_m$ are turned on so that the following Killing spinor equations have solutions \cite{Closset:2012ru}:
\begin{align}
 \nabla_m\xi &~\equiv~ \bigg(\partial_m+\frac14\omega_m^{ab}\gamma^{ab}-iV_m\bigg)\xi ~=~ \frac i2\gamma_m(H-i\SLASH{K})\xi,\nonumber \\
 \nabla_m\bar\xi &~\equiv~ \bigg(\partial_m+\frac14\omega_m^{ab}\gamma^{ab}+iV_m\bigg)\bar\xi ~=~ \frac i2\gamma_m(H+i\SLASH{K})\bar\xi,
\label{KSsugra}
\end{align}
where $\SLASH{K}\equiv \gamma^mK_m$. Note that $V_m$ is the gauge field for the $U(1)$ R-symmetry and $\xi,\bar\xi$ have R-charges $+1,-1$. This paper focuses on an ellipsoid \cite{Hama:2011ea}
\begin{equation}
 \frac{x_1^2+x_2^2}{\tilde\ell^2}+\frac{x_3^2+x_4^2}{\ell^2}=1\quad\text{in}\quad\mathbb R^4.
\end{equation}
Using a suitable set of coordinates $(\theta,\varphi,\tau)$, the dreibein and the background fields are expressed as follows:
\begin{align}
&e^1=f(\theta)\mathrm d\theta,\quad e^2=\tilde\ell\sin\theta\mathrm d\varphi,\quad e^3=\ell\cos\theta\mathrm d\tau,\qquad f(\theta)=\sqrt{\tilde\ell^2\cos^2\theta+\ell^2\sin^2\theta},\nonumber \\[1mm]
&H=\frac1f,\quad K=0,\quad V=\frac12\Big(1-\frac{\tilde\ell}f\Big)\mathrm d\varphi+\frac12\Big(1-\frac\ell{f}\Big)\mathrm d\tau\,.
\label{ellipsoid}
\end{align}
The coordinates $\varphi,\tau$ correspond to rotations within $(x_1,x_2)$ and $(x_3,x_4)$-planes, whereas $\theta$ takes values between $0$ and $\pi/2$. The above ellipsoid background has the Killing spinors
\begin{equation}
 \xi = e^{\frac i2(\varphi+\tau)}\left(\begin{array}{r}\cos\frac\theta2\\i\sin\frac\theta2\end{array}\right),\qquad
 \bar\xi = e^{-\frac i2(\varphi+\tau)}\left(\begin{array}{r}i\sin\frac\theta2\\\cos\frac\theta2\end{array}\right).
\label{KS-ell}
\end{equation}
They are normalized to satisfy $\xi\bar\xi=-\bar\xi\xi=1$.

For ${\cal N}=2$ supersymmetric theories on this ellipsoid, the square of the SUSY involves translation along a Killing vector
\begin{equation}
v\equiv\bar\xi\gamma^m\xi\partial_m = -\frac1{\tilde\ell}\partial_\varphi -\frac1\ell\partial_\tau.
\end{equation}
The worldline of one-dimensional BPS objects has to be invariant under the translation along $v$. If $\ell,\tilde\ell$ are incommensurable, there are only two circles on which a closed BPS loops of finite length can be wrapped. One is $S^1_{(\tau)}$ (the circle parameterized by $\tau$) at $\theta=0$, and the other is $S^1_{(\varphi)}$ at $\theta=\pi/2$. In the following we will focus on a single loop operator wrapped on $S^1_{(\tau)}$ at $\theta=0$.

\paragraph{3D ${\cal N}=2$ multiplets.}

The 3D ${\cal N}=2$ theories of our interest are made of two kinds of multiplets. A vectormultiplet for a gauge group $G$ consists of a gauge field $A_m$, a scalar $\sigma$, spinors $\lambda,\bar\lambda$ and an auxiliary field $D$, all valued in the Lie algebra $\mathfrak{g}=\text{Lie}(G)$. They transform under SUSY $\SUSY$ as
\begin{align}
\SUSY A_m &= -\frac i2(\bar\xi\gamma_m\lambda+\xi\gamma_m\bar\lambda),\nonumber\\
\SUSY \sigma &= \frac12(\xi\bar\lambda-\bar\xi\lambda),\nonumber\\
\SUSY\lambda &= \frac12\gamma^{mn}\xi F_{mn}-\xi D-i\SLASH\nabla\sigma\cdot\xi+\sigma\kappa,\nonumber\\
\SUSY\bar\lambda &= \frac12\gamma^{mn}\bar\xi F_{mn}+\bar\xi D+i\SLASH\nabla\sigma\cdot\bar\xi-\sigma\bar\kappa,\nonumber\\
\SUSY D &= \frac i2(\xi\SLASH\nabla\bar\lambda-\bar\xi\SLASH\nabla\lambda)+\frac i2(\xi[\sigma,\bar\lambda]+\bar\xi[\sigma,\lambda])+\frac14(\kappa\bar\lambda-\bar\kappa\lambda),
\label{Q3d-v}
\end{align}
where
\begin{alignat}{2}
 \kappa&\equiv(H-i\SLASH K)\xi,\qquad&
 F_{mn}&\equiv\partial_mA_n-\partial_nA_m-i[A_m,A_n],\nonumber\\
 \bar\kappa&\equiv(H+i\SLASH K)\bar\xi,\qquad&
 \nabla_m\sigma&\equiv\partial_m\sigma-i[A_m,\sigma].
\end{alignat}
A chiral multiplet in a representation $\Lambda$ of $G$ consists of a scalar $\phi$, spinor $\psi$ and an auxiliary field $F$. Their conjugate $(\bar\phi,\bar\psi,\bar F)$ form an anti-chiral multiplet in the representation $\bar\Lambda$. They transform under supersymmetry as
\begin{align}
\SUSY\phi &= \xi\psi,\nonumber\\
\SUSY\bar\phi &=\bar\xi\bar\psi,\nonumber\\
\SUSY\psi &= i(\SLASH\nabla\phi+\sigma\phi)\bar\xi-r\phi\bar\kappa+F\xi,\nonumber\\
\SUSY\bar\psi &= i(\SLASH\nabla\bar\phi+\bar\phi\sigma)\xi-r\bar\phi\kappa+\bar F\bar\xi,\nonumber\\
\SUSY F &= i\bar\xi(\SLASH\nabla\psi-\sigma\psi)-i\bar\xi\bar\lambda\phi+(r-\tfrac12)\bar\kappa\psi,\nonumber\\
\SUSY\bar F &= i\xi(\SLASH\nabla\bar\psi-\bar\psi\sigma)+i\xi\bar\phi\lambda+(r-\tfrac12)\kappa\bar\psi,
\label{Q3d-c}
\end{align}
where $r$ is the R-charge of $\phi$. Note that here we regarded the fields in $\Lambda$ ($\bar\Lambda$) as column vectors (resp. row vectors) on which the elements of $\mathfrak g$ act from the left (right).

Note that $\xi,\bar\xi$ in our convention are Grassmann-even spinor fields with values (\ref{KS-ell}), so $\SUSY$ flips the statistics of the fields. It is also important in later discussions that the square of $\SUSY$ acts as a sum of bosonic symmetries
\begin{align}
 \SUSY^2 &\;=\; i{\cal L}_v+i(\sigma-iv^mA_m)-\frac12\Big(\frac1{\,\tilde\ell\,}+\frac1{\,\ell\,}\Big)\text{R}_{U(1)} \nonumber \\&\;=\; -\frac i{\,\tilde\ell\,}{\cal L}_{\partial_\varphi}-\frac i{\,\ell\,}{\cal L}_{\partial_\tau}+i\Big(\sigma+\frac i{\,\tilde\ell\,}A_\varphi+\frac i{\,\ell\,} A_\tau\Big)-\frac12\Big(\frac1{\,\tilde\ell\,}+\frac1{\,\ell\,}\Big)\text{R}_{U(1)}
\label{Q^2}
\end{align}
on all the fields. Here the Lie derivative ${\cal L}_X$ along a Killing vector $X$ is defined to act on fields with local Lorentz (vector or spinor) indices as follows.
\begin{align}
 {\cal L}_Xe^a &= X^n\Big(\partial_ne^a_m+\omega^{ab}_ne^b_m+\Theta^{ab}e^b_m\Big)\mathrm dx^m+e_m^a\partial_nX^m\mathrm dx^n,
\nonumber\\
 {\cal L}_X\xi &= X^n\Big(\partial_n+\frac14\omega^{ab}_n\gamma^{ab}+\frac14\Theta^{ab}\gamma^{ab}\Big)\xi\,.
\qquad\big(\Theta^{ab}\equiv e^{am}e^{bn}\nabla_{[m}X_{n]}\big)
\end{align}
Using these relations one can check that $e^a_m,\xi,\bar\xi$ are all invariant
%\footnote{They are of course invariant under $\SUSY^2$ because $\SUSY$ is defined to act only on dynamical fields. But they are also invariant if one lets each term in the RHS of (\ref{Q^2}) act on them.}
under $\SUSY^2$.

\paragraph{BPS vortex loops.}

Vortex loops are one-dimensional defects in 3D gauge theories characterized by a singular behavior of the gauge field. Suppose there is a vortex line along the $x^3$-axis of $\mathbb R^3$. Let $r,\varphi$ be the polar coordinates for the transverse $(x^1,x^2)$-plane. Then the gauge field behaves near it as
\begin{equation}
A \sim \beta\,\mathrm d\varphi
\label{Def-Vortex}
\end{equation}
and $F_{12} = 2\pi\beta\delta^2(x^1,x^2)$. The parameter $\beta$ of the vortex line, called {\it vorticity}, can be gauge-rotated to be in a Cartan subalgebra $\mathfrak h\subset\mathfrak g$. Along the vortex line the gauge group $G$ is broken to a subgroup $K$ which is the centralizer of $\beta$. $K$ is $U(1)^r$ $(r=\mathrm{rk}G)$ for generic choice of $\beta$, but it can be non-abelian for special $\beta$.

The vortex configuration can be made supersymmetric by turning on a suitable imaginary value for $D$. For example, one can show using (\ref{Q3d-v}) that by setting $D=iF_{12}$ the SUSY corresponding to
\begin{equation}
 \gamma^3\xi=+\xi,\qquad \gamma^3\bar\xi=-\bar\xi
\label{xi-BPS}
\end{equation}
is unbroken. Note that our Killing spinor (\ref{KS-ell}) satisfies the same condition along the circle $S^1_{(\tau)}$ at $\theta=0$. So the BPS vortex loops of our interest preserve {\it all} the SUSY on the ellipsoid.

Note that the vorticity $\beta$ in (\ref{Def-Vortex}) is periodic because, if $u\in\mathfrak h$ is such that $\exp(2\pi iu)=1$, $\beta$ can be shifted by $u$ by the large gauge transformation $\exp(iu\varphi)$. But the periodicity will be lost by setting $D=2\pi i\beta\delta^2(x^1,x^2)$.

\paragraph{Lagrangians and boundary terms.}

As supersymmetric Lagrangian we take a sum of the Chern-Simons, Fayet-Iliopoulos, Yang-Mills and the matter kinetic terms.
\begin{align}
{\cal L}_\text{CS} \;=\;& \frac{ik}{4\pi}\mathrm{Tr}\left[\varepsilon^{mnp}(A_m\partial_nA_p-\frac{2i}3A_mA_nA_p)-\bar\lambda\lambda-2\sigma D\right],
\nonumber\\
{\cal L}_\text{FI} \;=\;& \frac{i\zeta}{2\pi}\big(D+H\sigma\big),
\nonumber\\
{\cal L}_\text{YM} \;=\;& \frac1{g^2}\mathrm{Tr}\bigg[
 \frac12F_{mn}^2+(\nabla_m\sigma)^2+\big(D-H\sigma\big)^2
 +\frac i2\bar\lambda\gamma^m\nabla_m\lambda-\frac i2\nabla_m\bar\lambda\gamma^m\lambda-i\bar\lambda[\sigma,\lambda]-\frac H2\bar\lambda\lambda\bigg],
\nonumber\\
{\cal L}_\text{mat} \;=\;& \nabla_m\bar\phi\nabla^m\phi+\bar\phi\sigma^2\phi+i(2r-1)H\bar\phi\sigma\phi-\frac{r(2r-1)}2H^2\bar\phi\phi+\frac r4R\bar\phi\phi-i\bar\phi D\phi+\bar FF
\nonumber\\ &
-\frac i2\bar\psi\gamma^m\nabla_m\psi+\frac i2\nabla_m\bar\psi\gamma^m\psi+i\bar\psi\sigma\psi-\frac{2r-1}2H\bar\psi\psi+i\bar\psi\bar\lambda\phi-i\bar\phi\lambda\psi.
\label{Lagrangians}
\end{align}
Naive volume integral of these terms may be divergent in the presence of vortex loops. As in \cite{Drukker:2012sr}, we regularize it by removing a tubular neighborhood of the loop $\theta\le\epsilon$ from the integration domain and adding appropriate boundary terms at $\theta=\epsilon$, so that the sum of bulk and boundary terms
\[
 S+S_\text{B} = \int_{\theta\ge\epsilon}\mathrm dV {\cal L} + \int_{\theta=\epsilon}\mathrm dS {\cal L}_\text{B}\qquad
\left(\mathrm dV\equiv e^1e^2e^3,\quad\mathrm dS\equiv e^2e^3\right)
\]
is SUSY invariant.

For some of the Lagrangians listed above, the boundary terms can be found by using the following fact: the F-component of a gauge-invariant chiral multiplet $(\Phi,\Psi,F_\Phi)$ with $r=2$ is $\SUSY$-exact up to a total derivative. More explicitly, the following holds:
\begin{equation}
\SUSY(-\bar\xi\Psi) = F_\Phi + \nabla_m(-i\bar\xi\gamma^m\bar\xi\Phi),\qquad
\SUSY(\xi\bar\Psi) = \bar F_\Phi+\nabla_m(i\xi\gamma^m\xi\bar\Phi).
\end{equation}
As an example, ${\cal L}_\text{YM}$ can be expressed as $F_\Phi+\bar F_\Phi$ for a gauge invariant chiral field $\Phi=\frac1{4g^2}\mathrm{Tr}\lambda\lambda$ and its conjugate $\bar\Phi=\frac1{4g^2}\mathrm{Tr}\bar\lambda\bar\lambda$. This implies the exactness of
\[
 {\cal L}_\text{YM}-\frac{i}{4g^2}\nabla_m\Big(\bar\xi\gamma^m\bar\xi\,\mathrm{Tr}\lambda\lambda-\xi\gamma^m\xi\,\mathrm{Tr}\bar\lambda\bar\lambda\Big). 
\]
The boundary term for ${\cal L}_\text{YM}$ is thus given by
\begin{equation}
 {\cal L}_\text{YM,B} = -\frac{i}{4g^2}\Big(\bar\xi\gamma^1\bar\xi\,\mathrm{Tr}\lambda\lambda-\xi\gamma^1\xi\,\mathrm{Tr}\bar\lambda\bar\lambda\Big)\,.
\end{equation}
Similarly, by setting $\Phi=\frac12\bar F\phi, \bar\Phi=\frac12\bar\phi F$ one obtains ${\cal L}_\text{mat}$ as their F-components up to total derivatives. This allows us to determine the boundary term for ${\cal L}_\text{mat}$ as follows:
\begin{equation}
 {\cal L}_\text{mat,B}= \frac12\Big(i\xi\gamma^1\xi\cdot\bar\phi F-i\bar\xi\gamma^1\bar\xi\cdot\bar F\phi-\nabla^1(\bar\phi\phi)\Big)\,.
\end{equation}

The boundary terms for ${\cal L}_\text{FI}$ and ${\cal L}_\text{CS}$ can be constructed using the following argument. Generally, supersymmetric bulk Lagrangian ${\cal L}$ satisfies $\SUSY{\cal L}=\nabla_m{\cal V}^m$ for some ${\cal V}^m$. If ${\cal V}^1$ is $\SUSY$-exact, the boundary term can be determined from ${\cal V}^1=-\SUSY{\cal L}_\text{B}$. By applying this to ${\cal L}_\text{FI}$ we find
\begin{align*}
 {\cal V}^1_\text{FI} &= \frac{\zeta}{4\pi}(\bar\xi\gamma^1\lambda-\xi\gamma^1\bar\lambda) = -\frac\zeta{4\pi}v_m(\bar\xi\gamma^m\gamma^1\lambda+\xi\gamma^m\gamma^1\bar\lambda) = \frac{i\zeta}{4\pi}\varepsilon^{1mn}v_m(\bar\xi\gamma_n\lambda+\xi\gamma_n\bar\lambda)\nonumber\\
&=-\frac{\zeta}{2\pi}\SUSY(w^nA_n),\qquad w^n\equiv\varepsilon^{1mn}v_m.
\end{align*}
Here we used $\SLASH v\xi=-\xi$, $\SLASH v\bar\xi=\bar\xi$ at the second equality and $v^1=0$ at the third equality. Similar analysis can be performed also for ${\cal L}_\text{CS}$. We thus find the following boundary terms:
\begin{equation}
{\cal L}_\text{FI,B}=\frac{\zeta}{2\pi}w^mA_m,\qquad
{\cal L}_\text{CS,B}=\frac{ik}{4\pi}\mathrm{Tr}\Big[w^mA_m(2i\sigma+v^nA_n)\Big].
\label{LB-FI-CS}
\end{equation}
The derivation of both requires $v^1=0$, which means that the Killing vector $v$ has to lie along the boundary in order for SUSY-preserving boundary terms to exist.

\subsection{Path integration with localization technique}\label{sec:exactPI}

Supersymmetric path integrals localize to $\SUSY$-invariant field configurations or {\it saddle points}, so that the sum of Gaussian path-integrals (\!{\it one-loop determinants}) on each saddle point gives an exact answer. See \cite{Pestun:2016zxk} for a review of localization techniques in SUSY gauge theories. Saddle point configurations are the solution of $\SUSY\Psi=0$ for all the fermions $\Psi$ of the theory. For the gauge theories on an ellipsoid with a vortex loop, the saddle points are given by
\begin{equation}
 A_\varphi=\beta~(\text{constant})\in\mathfrak h,\quad
 \sigma\;(\text{constant})\in\mathfrak h,\quad
 D=\frac\sigma f,\quad
 \phi=F=0.
\label{saddlept}
\end{equation}
The FI and CS actions take the following classical value on these saddle points\footnote{Note that the boundary at $\theta=\epsilon$ is oriented in such a way that $\int_{\theta=\epsilon}\mathrm d\varphi\mathrm d\tau=-4\pi^2$.}.
\begin{equation}
S_\text{FI}+S_\text{FI,B} = 2\pi i\zeta\ell\tilde\ell\Big(\sigma+\frac{i\beta}{\tilde\ell}\Big),\qquad
S_\text{CS}+S_\text{CS,B} = -i\pi k\ell\tilde\ell\,\mathrm{Tr}\Big(\sigma+\frac{i\beta}{\tilde\ell}\Big)^2\,.
\label{Scl}
\end{equation}
The YM and matter kinetic actions vanish on the saddle points since they are $\SUSY$-exact.

\paragraph{One-loop determinants: chiral multiplet.}

One-loop determinants can be computed most easily by a suitable change of path-integration variables. Let us first explain this procedure for the theory of a chiral multiplet of unit $U(1)$ charge, with the $U(1)$ vectormultiplet fields fixed at a saddle point (\ref{saddlept}). The problem is already Gaussian, but it can be simplified further by rewriting in terms of the so-called {\it cohomological variables}
\begin{equation}
 \Psi\equiv \SUSY\phi=\xi\psi,\quad
 \Psi'\equiv-\bar\xi\psi,\quad
 F'\equiv \SUSY\Psi' = F+{\cal J}\phi\,;\quad{\cal J}\equiv-i\bar\xi\gamma^m\bar\xi\nabla_m\,.
\label{defcoh}
\end{equation}
The change of path integration variables from $(\phi,\psi,F)$ to $(\phi,\Psi,\Psi',F')$ is invertible and the Jacobian is trivial. $\phi$ is Grassmann-even and its superpartner $\Psi$ is odd, and they are both scalars of R-charge $r$. Likewise, $\Psi'$ (odd) and its superpartner $F'$ (even) are both scalars of R-charge $r-2$. We denote the Hilbert spaces of their wavefunctions as
\[
 \phi,\Psi\in{\cal H}, \qquad
 \Psi',F'\in{\cal H}'.
\]
Physically this means that the fields $\phi$ and $\Psi$ are to be mode-expanded using the same set of basis wavefunctions of ${\cal H}$, and similarly for $\Psi'$ and $F'$ in ${\cal H}'$.

The one-loop determinant $\Delta_\text{1-loop}$ can be computed by path integrating over the fields $(\phi,\Psi,\Psi',F')$ and their conjugates with a suitable choice of localizing Lagrangian ${\cal L}$. Any ${\cal L}$ will do as long as it is $\SUSY$-exact and its bosonic part is bounded from below. Let us take\footnote{One can check that $\SUSY^2$ and $\overline{\SUSY^2}$ commute. Note also that there is no issue of boundary terms for this ${\cal L}$ since $\SUSY^2$ and $\overline{\SUSY^2}$ contain no $\theta$-derivatives.}
\[
 {\cal L} = \SUSY\Big(\bar\phi\cdot\overline{\SUSY^2}\Psi+\bar\Psi' F'\Big)
 = \bar\Psi\cdot\overline{\SUSY^2}\Psi+\bar\phi\cdot\overline{\SUSY^2}\SUSY^2\phi+\bar F'F'-\bar\Psi'\SUSY^2\Psi',
\]
Then the Gaussian integration gives the ratio of determinants
\begin{equation}
 \Delta_\text{1-loop} \;=\; \frac{\mathrm{Det}(\overline{\SUSY^2})_{\cal H}\cdot\mathrm{Det}(-\SUSY^2)_{{\cal H}'}}{\mathrm{Det}(\overline{\SUSY^2}\SUSY^2)_{\cal H}}\;=\; \frac{\mathrm{Det}(\SUSY^2)_{{\cal H}'}}{\mathrm{Det}(\SUSY^2)_{\cal H}\,}.
\end{equation}
The last equality holds up to a sign factor $\mathrm{Det}(-1)_{{\cal H}'}$ which we have just dropped. So $\Delta_\text{1-loop}$ can be computed from the spectrum of $\SUSY^2$ on ${\cal H}$ and ${\cal H}'$. Furthermore, one can check that the map ${\cal J}:{\cal H}\to{\cal H}'$ commutes with $\SUSY^2$, which is as expected because it is made only of $\SUSY^2$-invariant background fields. The $\SUSY^2$-eigenmodes in ${\cal H}$ and ${\cal H}'$ paired by ${\cal J}$ make no net contribution to $\Delta_\text{1-loop}$. Hence we only need the spectrum of $\SUSY^2$ on the kernel and cokernel of $\cal J$. In other words,
\begin{equation}
 \Delta_\text{1-loop} \;=\;\frac{\mathrm{Det}(\SUSY^2)_{\mathrm{coker}({\cal J})}}{\mathrm{Det}(\SUSY^2)_{\mathrm{ker}({\cal J})}~~}\,.
\end{equation}

To work out the basis wavefunctions of $\mathrm{ker}({\cal J})$ and $\mathrm{coker}({\cal J})=\mathrm{ker}(\bar{\cal J})$, we need the explicit form of ${\cal J}$ and its conjugate $\bar{\cal J}$.
\begin{align}
 {\cal J} &= -ie^{-i(\varphi+\tau)}\left[-\frac1f\partial_\theta+\frac{i\cos\theta}{\tilde\ell\sin\theta}(\partial_\varphi-i\beta-irV_\varphi)-\frac{i\sin\theta}{\ell\cos\theta}(\partial_\tau-irV_\tau)\right],
\nonumber \\
 \bar{\cal J} &= +ie^{+i(\varphi+\tau)}\left[-\frac1f\partial_\theta-\frac{i\cos\theta}{\tilde\ell\sin\theta}(\partial_\varphi-i\beta-i(r-2)V_\varphi)+\frac{i\sin\theta}{\ell\cos\theta}(\partial_\tau-i(r-2)V_\tau)\right].
\end{align}
In fact $\bar{\cal J}$ can be expressed as $\bar{\cal J}=-i\xi\gamma^m\xi\nabla_m$. The zeromode equations ${\cal J}\Phi=0, \bar{\cal J}\Phi'=0$ can be reduced to ODEs for functions of $\theta$ by assuming that $\Phi,\Phi'$ have definite $\varphi$ and $\tau$-momenta. The resulting ODEs actually need not be solved explicitly, but the behavior of the solutions at $\theta=0$ and $\pi/2$ are important. They are summarized as follows.
\begin{alignat}{3}
\Phi=\hat\Phi(\theta)e^{im\varphi+in\tau}~~&\in \text{ker}({\cal J}) \quad&\Longrightarrow\quad&&
\hat\Phi(\theta)&\sim (\sin\theta)^{\beta-m}(\cos\theta)^{-n},
\nonumber\\[1mm]
\Phi'=\hat\Phi'(\theta)e^{im'\varphi+in'\tau}&\in \text{ker}(\bar{\cal J}) \quad&\Longrightarrow\quad&&
\hat\Phi'(\theta)&\sim (\sin\theta)^{m'-\beta}(\cos\theta)^{n'}.
\label{zeromodes}
\end{alignat}

In the absence of the vortex loop one should require $m,n\le0$ and $m',n'\ge0$ so that the zeromodes are regular. The zeromodes all have definite eigenvalues of $\SUSY^2$ (\ref{Q^2}). By multiplying all the eigenvalues we obtain
\begin{equation}
\Delta_\text{1-loop} = \frac{\prod_{m',n'\ge0}\frac{m'}{\tilde\ell}+\frac{n'}{\ell}+i\sigma-\frac{r-2}2\Big(\frac1{\tilde\ell}+\frac1{\ell}\Big)}{\prod_{m,n\le0}\frac{m}{\tilde\ell}+\frac{n}{\ell}+i\sigma-\frac{r}2\Big(\frac1{\tilde\ell}+\frac1{\ell}\Big)}.
\end{equation}
By introducing $b\equiv(\ell/\tilde\ell)^\frac12,\;Q\equiv b+b^{-1},\;\hat\sigma\equiv\sqrt{\ell\tilde\ell}\sigma$ and neglecting sign factors, this can be expressed in terms of the double sine function $s_b(x)$.
\begin{align}
\Delta_\text{1-loop}&~=~ \prod_{m,n\ge0}\frac{mb+nb^{-1}+i\hat\sigma-\frac Q2(r-2)}{mb+nb^{-1}-i\hat\sigma+\frac Q2r}~=~ s_b\Big(\frac{i(1-r)Q}2-\hat\sigma\Big),\nonumber \\
s_b(x)&~=~ \prod_{m,n\ge0}\frac{mb+nb^{-1}-ix+\frac Q2}{mb+nb^{-1}+ix+\frac Q2}.
\end{align}
See \cite{Teschner:2003en,Kharchev:2001rs} for more detail on the function $s_b(x)$.

In the presence of a vortex loop with non-integer $\beta$, not only the zeromodes (\ref{zeromodes}) but all the eigenfunctions of $\bar{\cal J}{\cal J}$ or ${\cal J}\bar{\cal J}$, which are the natural basis wavefunctions of ${\cal H}$ or ${\cal H}'$, behave as fractional power of $\theta$ near $\theta=0$ \cite{Hosomichi:2017dbc}. In this case, the simplest boundary condition requiring the wavefunctions of both ${\cal H}$ and ${\cal H}'$ to vanish at $\theta=0$ is inconsistent for the following reason. In order for the $\SUSY$-transformation to be well-defined, the Hilbert spaces ${\cal H}$, ${\cal H}'$ need to satisfy
\begin{equation}
 {\cal J}{\cal H}\subset{\cal H}',\qquad
 \bar{\cal J}{\cal H}'\subset{\cal H}\,.
\end{equation}
Also, the operators ${\cal J},\bar{\cal J}$ contain $\theta$-derivatives which generically lower the power of $\theta$ by 1. Suppose a wavefunction $\Phi\in{\cal H}$ vanishes as $\theta^\gamma~(0<\gamma<1)$ near $\theta=0$. Then ${\cal J}\Phi$, if nonzero, would have to be in ${\cal H}'$ and diverge as $\theta^{-(1-\gamma)}$ at $\theta=0$. Similar argument holds with the role of ${\cal H}$ and ${\cal H}'$ exchanged.

As was proposed in \cite{Hosomichi:2017dbc} for a similar problem in two dimensions, there are two consistent boundary conditions for chiral multiplet fields at $\theta=0$.\\[2mm]
%%%
~~{\bf BC1.}~~ $\Phi\in{\cal H}$ is finite. $\Phi'\in{\cal H}'$ may diverge mildly but $\bar{\cal J}\Phi'$ is finite.\\[2mm]
~~{\bf BC2.}~~ $\Phi'\in{\cal H}'$ is finite. $\Phi\in{\cal H}$ may diverge mildly but ${\cal J}\Phi$ is finite.\\[2mm]
%%%
The mild divergence here means the behavior $\theta^{-\gamma}~(0<\gamma<1)$, which is not forbidden by the normalizability of wavefunctions. Note that {\it ``is finite''} can be replaced by {\it ``vanishes''} for non-integer $\beta$.

Let us compute $\Delta_\text{1-loop}$ for the chiral multiplet in the presence of a vortex loop. First, under the boundary condition {\bf BC1}, the physical zeromodes of ${\cal J},\bar{\cal J}$ are those in (\ref{zeromodes}) with
\[
 \beta-m\ge0\,,\quad -n\ge0\,;\quad m'-\beta>-1\,,\quad n'\ge0\,.
\]
The first and the third inequalities are equivalent to $m\le\lfloor\beta\rfloor$ and $m'\ge\lfloor\beta\rfloor$. These zeromodes all have definite $\SUSY^2$-eigenvalues which are now $\beta$-dependent. By multiplying all of them one obtains the one-loop determinant of a chiral multiplet on a vortex background:
\begin{equation}
\textbf{BC1}~~\Longrightarrow~~
 \Delta_\text{1-loop} = s_b\Big(\frac{i(1-r)Q}2-\hat\sigma-ib\beta+ib\lfloor\beta\rfloor\Big)\,.
\end{equation}
The computation is similar for the boundary condition {\bf BC2}. In this case, the integers $m,m'$ in (\ref{zeromodes}) are bounded as $\beta-m>-1$ and $m'-\beta\ge0$, or equivalently $m\le\lceil\beta\rceil$ and $m'\ge\lceil\beta\rceil$.
\begin{equation}
\textbf{BC2}~~\Longrightarrow~~
 \Delta_\text{1-loop} = s_b\Big(\frac{i(1-r)Q}2-\hat\sigma-ib\beta+ib\lceil\beta\rceil\Big)\,.
\end{equation}
Note that $\Delta_\text{1-loop}$ is a periodic function of $\beta$ for both boundary conditions. This is a consequence of large gauge invariance.

The above result can be easily generalized to the theory of chiral multiplet in a representation $\Lambda$ of the gauge group $G$. The one-loop determinant is then given by a product over weights $\mu$ of $\Lambda$.
\begin{equation}
 \Delta_\text{1-loop}^\text{c}(\beta) = \prod_\mu s_b\Big(\frac{i(1-r)Q}2-\mu\dotp(\hat\sigma+ib\beta)+ib[\mu\dotp\beta]\Big).
\label{Det-c}
\end{equation}
Here $[\cdots]$ is the floor or ceiling functions depending on the choice of boundary condition.

\paragraph{One-loop determinants: vectormultiplet.}

Let us next study the integration over fluctuations of vectormultiplet fields around a saddle point (\ref{saddlept}). In what follows we denote the saddle-point value of a field $\Phi$ by $\langle\Phi\rangle$ and its fluctuation by $\delta\Phi$. As in \cite{Hosomichi:2017dbc}, we first introduce the Faddeev-Popov ghost $c$, antighost $\bar c$ and an auxiliary field $B$ and then move to cohomological variables.

The system of physical fields and ghosts has a nilpotent BRST symmetry $\BRST$. It acts on all the physical fields as gauge transformation with parameter $c$, whereas the ghost fields transform as
\begin{equation}
\BRST c=ic^2,\quad
\BRST\bar c=B,\quad
\BRST B=0.
\end{equation}
It is also known from \cite{Pestun:2007rz} that if we set
\begin{equation}
\SUSY c= -\delta\big(\sigma-iv^mA_m\big),\quad
\SUSY\bar c=0,\quad
\SUSY B = iv^m\partial_m\bar c+i\big[\langle\sigma-iv^mA_m\rangle\,,\,\bar c\,\big],
\end{equation}
then the combined supercharge $\QTOT\equiv\SUSY+\BRST$ acts on all the fields as
\begin{equation}
 \QTOT^2 \;=\; i{\cal L}_v+i\langle\sigma-iv^mA_m\rangle -\frac12\Big(\frac1{\tilde\ell}+\frac1\ell\Big)\text{R}_{U(1)}\,.
\end{equation}
One may use $\QTOT$ as the localizing supercharge and study its action on the fluctuation of fields under linear approximation. The problem thus becomes essentially the same as that of path-integral over matter fields coupled to a fixed vectormultiplet field.

We next move from $(\delta A_m,\delta\sigma,\lambda,\bar\lambda,\delta D\,;\,c,\bar c,B)$ to cohomological variables. They are given by 3 Grassmann-even plus 3 Grassmann-odd adjoint scalars
\begin{alignat}{2}
A_\mini{+}&\equiv\xi\gamma^m\xi\cdot\delta A_m,\qquad\qquad& \bar c,\nonumber \\
A_\mini{-}&\equiv\bar\xi\gamma^m\bar\xi\cdot\delta A_m,\qquad& c,\nonumber \\
A_v&\equiv\bar\xi\gamma^m\xi\cdot\delta A_m,\qquad&
\Lambda&\equiv\bar\xi\lambda+\xi\bar\lambda,
\label{defCV}
\end{alignat}
and their $\QTOT$-superpartners. It is straightforward to check that the change of variables is invertible and the Jacobian is trivial. The one-loop determinant for a vectormultiplet is thus given by
\begin{equation}
 \Delta_\text{1-loop}^\text{v} ~=~ \left(\frac{\mathrm{Det}(\QTOT^2)_{{\cal H}(\bar c)\oplus{\cal H}(c)\oplus{\cal H}(\Lambda)}}{\mathrm{Det}(\QTOT^2)_{{\cal H}(A_+)\oplus{\cal H}(A_-)\oplus{\cal H}(A_v)}}\right)^{\frac12}.
\end{equation}
Since $A_\mini{\pm}$ have R-charge $\pm2$ and $A_v,\bar c, c, \Lambda$ have R-charge $0$, this actually equals the one-loop determinant for an adjoint chiral multiplet with $r=2$.

In the presence of a vortex loop at $\theta=0$, the Cartan part of $\xi\gamma^m\xi A_m$ and $\bar\xi\gamma^m\bar\xi A_m$ diverge as $\theta^{-1}$ but $\bar\xi\gamma^m\xi A_m$ is finite. It is therefore natural to allow mild divergence for $A_\mini{\pm}$ but require $A_v$ to be finite at $\theta=0$. Also, the cohomological variables transform under $\QTOT$ as
\begin{alignat}{2}
\QTOT A_\mini{+} &\;\simeq\; i\xi\lambda+i\bar{\cal J}c,\qquad&
\QTOT\bar c&\;=\; B,
\nonumber \\[1mm]
\QTOT A_\mini{-} &\;\simeq\; -i\bar\xi\bar\lambda +i{\cal J}c,\qquad&
\QTOT c&\;\simeq\;-\delta\sigma+iA_v,
\nonumber \\
\QTOT A_v &\;\simeq\; \frac i2(\bar\xi\lambda-\xi\bar\lambda)+{\cal L}_vc-i\big[v^m\langle A_m\rangle,\,c\,\big],\qquad&
\QTOT\Lambda &\;\simeq\; 2\delta\Big(D-\frac\sigma f\Big)+i{\cal J}A_\mini{+}-i\bar{\cal J}A_\mini{-} +\frac{4i}fA_v,
\nonumber
\end{alignat}
where $\simeq$ stands for the equality up to linear order in the fluctuation. This implies the relations among Hilbert spaces
\[
 {\cal H}(A_\mini{+})~\begin{array}{c}\xrightarrow{~~{\cal J}~~}\\[-4mm] \xleftarrow[~~\bar{\cal J}~~]{}\end{array}~{\cal H}(A_v,c,\Lambda)~\begin{array}{c}\xrightarrow{~~{\cal J}~~}\\[-4mm] \xleftarrow[~~\bar{\cal J}~~]{}\end{array}~{\cal H}(A_\mini{-})
\]
So we require $c,\Lambda$ and $\bar c$ be finite at $\theta=0$ as well. We thus conclude that, as far as the computation of one-loop determinant is concerned, a vectormultiplet is equivalent to an adjoint chiral multiplet with $r=2$ obeying \textbf{BC2}.

\paragraph{Partition function and vortex loop VEVs.}

Now we are ready to present exact formulae for supersymmetric observables on an ellipsoid. First, the partition function can be expressed as \cite{Hama:2011ea}
\begin{equation}
 Z ~=~ \frac1{|{\cal W}|}\int\mathrm d^r\hat\sigma\,e^{-S}\cdot\Delta_\text{1-loop}^\text{v}\cdot\Delta_\text{1-loop}^\text{c},
\label{Zell}
\end{equation}
where $r=\mathrm{rk}(G)$ and ${\cal W}$ is the Weyl group of $G$. $S$ is the sum of the classical FI and CS actions evaluated at saddle points,
\begin{equation}
 S_\text{FI} = 2\pi i\hat\zeta\hat\sigma,\qquad
 S_\text{CS} = -i\pi k\mathrm{Tr}\big(\hat\sigma^2\big),
\label{Scl2}
\end{equation}
where $\hat\zeta\equiv\sqrt{\ell\tilde\ell}\,\zeta$ is the dimensionless FI coupling. The one-loop determinants $\Delta_\text{1-loop}^\text{c,v}$ in the absence of vortex loop take the following form
\begin{align}
 \Delta_\text{1-loop}^\text{c}(\beta=0) &~= \prod_{\mu}s_b\Big(\frac{i(1-r)Q}2-\mu\dotp\hat\sigma\Big),
\nonumber \\
 \Delta_\text{1-loop}^\text{v}(\beta=0) &~= \prod_{\alpha\in\Delta}s_b\Big({-}\frac{iQ}2-\alpha\dotp\hat\sigma\Big)~=~\prod_{\alpha\in\Delta^+}2\sinh(\pi b\alpha\dotp\hat\sigma)\cdot 2\sinh(\pi b^{-1}\alpha\dotp\hat\sigma),
\label{DetV0}
\end{align}
where $\Delta$ is the set of roots and $\Delta^+$ the set of positive roots.

The expectation value of a vortex loop can be expressed in a similar way,
\begin{equation}
 \langle V_\beta\rangle = \frac1{|{\cal W}_K|}\int\mathrm d^r\hat\sigma e^{-S-S_\text{B}}\cdot\Delta^\text{v}_\text{1-loop}(\beta)\cdot\Delta^\text{c{\bf 1}}_\text{1-loop}(\beta)\cdot\Delta^\text{c{\bf 2}}_\text{1-loop}(\beta),
\label{VbetaDEF}
\end{equation}
where ${\cal W}_K$ is the Weyl group of $K$ (the centralizer of $\beta$) or equivalently the subgroup of ${\cal W}$ which leaves $\beta$ invariant. We also separate the one-loop determinant of chiral multiplets according to the type of boundary conditions. We notice that the classical actions (\ref{Scl}) remain the same as (\ref{Scl2}) if one redefines $\hat\sigma+ib\beta$ as $\hat\sigma$. Under the same redefinition of $\hat\sigma$, the one-loop determinants for vector and chiral multiplets become
\begin{align}
 \Delta_\text{1-loop}^\text{v}(\beta) &\;=\; \prod_{\alpha\in\Delta}s_b\Big({-}\frac{iQ}2-\alpha\dotp\hat\sigma+ib\lceil\alpha\dotp\beta\rceil\Big),
\nonumber \\
 \Delta_\text{1-loop}^\text{c{\bf 1}}(\beta) &\;=\; \prod_{\mu}s_b\Big(\frac{i(1-r)Q}2-\mu\dotp\hat\sigma+ib\lfloor\mu\dotp\beta\rfloor\Big),
\nonumber \\
 \Delta_\text{1-loop}^\text{c{\bf 2}}(\beta) &\;=\; \prod_{\mu}s_b\Big(\frac{i(1-r)Q}2-\mu\dotp\hat\sigma+ib\lceil\mu\dotp\beta\rceil\Big).
\label{DetwV}
\end{align}
Since the double sine function satisfies
\begin{align}
%s_b(x)= \frac1{s_b(-x)},\qquad
%s_b\Big(\frac{ib}2-x\Big)s_b\Big(\frac{ib}2+x\Big)=\frac1{2\cosh\pi bx},
%\quad s_b\Big(\frac{i}{2b}-x\Big)s_b\Big(\frac{i}{2b}+x\Big)=\frac1{2\cosh\frac{\pi x}b}.
\frac{s_b(x\pm ib)}{s_b(x)} ~=~ i\cdot\left(2\sinh\pi b\Big(x\pm\frac{iQ}2\Big)\right)^{\mp1},
\end{align}
the one-loop determinants (\ref{DetwV}) essentially differ from those at $\beta=0$ only by a product of $\sinh$ functions. Thus the expectation value of a vortex loop can be expressed as
\begin{equation}
\langle V_\beta\rangle ~=~ \frac1{|{\cal W}|}\int\mathrm d^r\hat\sigma\,e^{-S}\cdot\Delta_\text{1-loop}^\text{v}\Delta_\text{1-loop}^\text{c}\cdot V_\beta(\hat\sigma),
\label{VbetaDEF2}
\end{equation}
where $S$ and $\Delta_\text{1-loop}^\text{v,c}$ are the same as those in the partition function (\ref{Zell}), and $V_\beta(\hat\sigma)$ is the function which encodes the effects of insertion of a vortex loop.
\begin{align}
 V_\beta(\hat\sigma) &~=~ \frac{|{\cal W}|}{|{\cal W}_K|}\cdot V_\beta^\text{v}(\hat\sigma)\cdot V_\beta^\text{c{\bf 1}}(\hat\sigma)\cdot V_\beta^\text{c{\bf 2}}(\hat\sigma) \nonumber \\ &~=~ \frac{|{\cal W}|}{|{\cal W}_K|}\cdot\frac{\Delta_\text{1-loop}^\text{v}(\beta)}{\Delta_\text{1-loop}^\text{v}(0)}\cdot\frac{\Delta_\text{1-loop}^\text{c{\bf 1}}(\beta)}{\Delta_\text{1-loop}^\text{c{\bf 1}}(0)}\cdot\frac{\Delta_\text{1-loop}^\text{c{\bf 2}}(\beta)}{\Delta_\text{1-loop}^\text{c{\bf 2}}(0)}\,.
\label{Vbeta0}
\end{align}

Note that, since we have redefined $\hat\sigma$, the contour of integration is now $\hat\sigma\in\mathfrak h+ib\beta$. In the following we will assume that it can be brought back to $\mathfrak h$ without problem. This is the case for pure YM-CS theories since $\Delta_\text{1-loop}^\text{v}(\beta)$ has no poles. For theories with chiral multiplets this would lead to constraints on their R-charges $r$, representation $\Lambda$ as well as $\beta$ which we will not go into details.

Using Weyl group, generic $\beta$ can be brought into a Weyl chamber so that $\alpha\dotp\beta>0$ for all positive roots $\alpha$. For non-generic $\beta$ one has $\alpha\dotp\beta\ge0$ for all positive roots $\alpha$ but $\alpha\dotp\beta=0$ for some $\alpha$, corresponding to the enhanced unbroken symmetry $K$. In what follows we make further simplifying assumption that $\beta$ {\it is small}:
\begin{equation}
 -1<\alpha\dotp\beta<1 \quad\text{for all roots}~~\alpha,\qquad
 -1<\mu\dotp\beta<1\quad\text{for all weights}~~\mu.
\label{betasmall}
\end{equation}
Then $V_\beta(\hat\sigma)$ is the product of the following functions.
\begin{align}
 V_\beta^\text{v}(\hat\sigma)&~=~\prod_{\alpha\dotp\beta>0}\Big(2\sinh\pi b\alpha\dotp\hat\sigma\Big)^{-1},
\nonumber \\
 V_\beta^\text{c{\bf 1}}(\hat\sigma)&~=~\prod_{\mu\dotp\beta<0}2\sinh\pi b\Big(\mu\dotp\hat\sigma+\frac{irQ}2\Big),
\nonumber \\
 V_\beta^\text{c{\bf 2}}(\hat\sigma)&~=~
\prod_{\mu\dotp\beta>0}\Big(2\sinh\pi b\Big(\mu\dotp\hat\sigma-\frac{i(2-r)Q}2\Big)\Big)^{-1}\,.
\label{Vbeta1}
\end{align}
Here we neglected all the signs and powers of $i$'s which can be absorbed into redefinition of the loop operator.

In the following sections we test the above formulae against several consistency checks. In fact, we will find that all these formulae need to be corrected.

\section{Pure Chern-Simons theories}\label{sec:pureCS}

The (bosonic) CS theory is a topological field theory which provides a physical description of a wide class of topological invariants associated to knots or links in 3-manifolds or the manifolds themselves. The theory was exactly solved in \cite{Witten:1988hf} by using non-perturbative methods and its relation to 2D conformal field theory with $G$ symmetry.

${\cal N}=2$ pure CS theories are essentially the same as the bosonic CS theories, because all the vectormultiplet fields except for the gauge field $A_m$ are auxiliary fields. Some of the known formulae for observables in the bosonic CS theory can be reproduced using the results of the previous section. For example, the ellipsoid partition function of ${\cal N}=2$ CS theories is given by the following integral
\begin{align}
 Z &~=~ \frac1{|{\cal W}|}\int \mathrm d^r\hat\sigma e^{-S_\text{CS}}\cdot\Delta_\text{1-loop}^\text{v}
\nonumber \\[1mm] &
~=~ \frac1{|{\cal W}|}\int \mathrm d^r\hat\sigma e^{i\pi k\mathrm{Tr}(\hat\sigma^2)}\prod_{\alpha\in\Delta^+}2\sinh(\pi b\alpha\dotp\hat\sigma)\cdot 2\sinh(\pi b^{-1}\alpha\dotp\hat\sigma).
\label{ZS3CS}
\end{align}
The result of \cite{Witten:1988hf} for the sphere partition function can be reproduced up to overall coefficients by setting $b=1$ and performing explicit $\hat\sigma$-integration with the help of Weyl's denominator formula
\begin{equation}
 \prod_{\alpha\in\Delta^+}2\sinh(\pi\alpha\dotp\hat\sigma) ~=~ \sum_{w\in{\cal W}}\epsilon(w)e^{2\pi w(\rho)\dotp\hat\sigma},
\end{equation}
where $\rho\equiv\frac12\sum_{\alpha\in\Delta^+}\alpha$ is the Weyl vector and $\epsilon(w)=\pm1$ is the parity of $w\in{\cal W}$. Likewise, the expectation value of an unknot can be reproduced as that of a BPS Wilson loop in ${\cal N}=2$ theory,
\begin{equation}
 W_\Lambda(C) \equiv \mathrm{Tr}_\Lambda\mathrm P\exp i\oint_C (A_mv^m+i\sigma)a\mathrm dt\,,
\end{equation}
where $a$ is an arbitrary real constant and $C$ is an integral curve of $\frac{\mathrm d}{\mathrm dt}x^m=av^m(x)$. As an example, take $C=S^1_{(\tau)}$ oriented in the increasing direction of $\tau$ (which is opposite to the direction of $v^m$). The Wilson loop expectation value is then given by an integral of the form (\ref{ZS3CS}) with an additional insertion of
\begin{equation}
 W_\lambda(\widehat\sigma) = \mathrm{Tr}_\Lambda e^{2\pi\widehat\sigma} = \frac{\sum_{w\in{\cal W}}\epsilon(w)e^{2\pi w(\rho+\lambda)\dotp\widehat\sigma}}{\prod_{\alpha\in\Delta^+}2\sinh\pi\alpha\dotp\widehat\sigma}.
\label{WLambda}
\end{equation}
Here $\lambda$ is the highest weight of the representation $\Lambda$. Also, hereafter we will use a new dimensionless field $\widehat\sigma$ 
\begin{equation}
 \widehat\sigma = b\hat\sigma = \ell\sigma,
\end{equation}
which is more suitable than $\hat\sigma$ for the discussion of circular vortex loops of radius $\ell$.

An important remark is in order. Many exact formulae for observables in bosonic CS theory depends on the CS coupling through the combination $k+h^\vee$, where $h^\vee$ is the dual Coxeter number of $G$. This can be understood as a perturbative correction at one-loop. But such shift of $k$ does not occur in ${\cal N}=2$ CS theories due to the presence of auxiliary fields \cite{Kao:1995gf}. Later we will encounter a similar difference between bosonic and ${\cal N}=2$ theories concerning the shift of the label $\lambda$ of Wilson loops \cite{Fan:2018wya}.

\subsection{Equivalence of Wilson and vortex loops}\label{sec:WVequiv}

An interesting fact known in bosonic CS theories is that vortex loops are equivalent to Wilson loops in the representation with the highest weight $\lambda=k\beta/2$. We will first review how the equivalence works in bosonic CS theories, and then attempt to reproduce it in ${\cal N}=2$ supersymmetric setting.

\paragraph{Quantization of (co)adjoint orbits.}

It is known that, for every irreducible representation $\Lambda$ of a compact group $G$, there is a symplectic manifold $(M,\omega)$ which gives $\Lambda$ as the Hilbert space of its geometric quantization. Using this, one can express a Wilson loop for arbitrary $G$ and $\Lambda$ by a suitable quantum mechanics on the loop interacting with the bulk gauge field. We summarize the basic idea here by going through one simple example. For more details of geometric quantization, see \cite{Woodhouse:1980pa, BlauGQ}.

Let us take $G=SU(2)$ and $\Lambda=$ spin-$s$ representation. The symplectic manifold for this case is $M=S^2$ and the symplectic form $\omega=\hbar s\sin\theta\mathrm d\theta\mathrm d\varphi$, where $\theta,\varphi$ are the usual polar coordinates. We will keep the $\hbar$-dependence of various formulae for the next few paragraphs. The Hamiltonian functions (moment maps) and corresponding vector fields generating $SU(2)$ symmetry are given by
\begin{alignat}{2}
 P^1 &= -\hbar s\,\sin\theta\cos\varphi,\qquad&
 X(P^1) &= -\sin\varphi\frac\partial{\partial\theta}-\cot\theta\cos\varphi\frac\partial{\partial\varphi},\nonumber \\
 P^2 &= -\hbar s\,\sin\theta\sin\varphi,\qquad&
 X(P^2) &= +\cos\varphi\frac\partial{\partial\theta}-\cot\theta\sin\varphi\frac\partial{\partial\varphi},\nonumber \\
 P^3 &= -\hbar s\,\cos\theta,\qquad&
 X(P^3) &= \frac\partial{\partial\varphi}.
\end{alignat}
They are related to each other by $\mathrm dP^a+\imath_{X(P^a)}\omega=0$. The Poisson bracket on this $M$ is such that
\[
 \{\varphi,\theta\}=\frac1{\hbar s\sin\theta},\qquad
 \{P^a,P^b\} = \varepsilon^{abc}P^c.
\]

In geometric quantization, Hilbert space is constructed in two steps. The first step, called {\it prequantization}, defines a map from functions $f,g,\cdots$ on $M$ to operators $\hat f,\hat g,\cdots$ acting on certain Hilbert space ${\cal H}$ of wave functions by the formula
\begin{equation}
 \hat f\;\equiv\; -i\hbar X(f)-\imath_{X(f)}\vartheta+f.
\end{equation}
Here $\vartheta$ is a one-form satisfying $\mathrm d\vartheta=\omega$, which is necessary in order that $\{f_1,f_2\}=f_3$ lead to $[\hat f_1,\hat f_2]=i\hbar\hat f_3$. But such a $\vartheta$ exists in general only locally. This makes the wave functions not ordinary functions on $M$ but sections of a line bundle $B$, called {\it prequantum bundle}, with connection $\nabla=\mathrm d-i\hbar^{-1}\vartheta$. The symplectic form $\omega$ is then subject to the quantization condition
\[
 c_1(B)= \Big[\frac{\omega}{2\pi\hbar}\Big]\;\in\; H^2(M,\mathbb Z)\,.
\]
In the present case it gives $\int_{S^2}\frac{\omega}{2\pi\hbar} = 2s\in\mathbb Z$.

The second step is to choose an integrable Lagrangian subbundle $P$ of $TM^{\mathbb C}$ called {\it polarization} and require the {\it quantum} wave functions to be covariantly constant along $\bar P$. This is the generalization of the familiar fact that wave functions depend only on half of the phase space coordinates, and the complexification is to accommodate generalizations of coherent state quantization of harmonic oscillator. Various choice of $P$ is possible for a given $(M,\omega)$, but for a K\"ahler manifold $M$ a particularly convenient one is in which the quantum wave functions depend only on holomorphic coordinates. For the present example, $M=S^2$ can be covered by two coordinate patches $z=\tan\frac\theta2e^{i\varphi}$ and $w=\cot\frac\theta2e^{-i\varphi}=z^{-1}$. In the gauge
\[
 \vartheta_{[z]}= -2i\hbar s\frac{\bar z\,\mathrm dz}{1+z\bar z},\qquad
 \vartheta_{[w]}= -2i\hbar s\frac{\bar w\,\mathrm dw}{1+w\bar w},
\]
quantum wave functions $\Psi$ are holomorphic functions in the respective coordinate patches. Moreover, $\Psi_{[z]}$ and $\Psi_{[w]}$ are related by $\Psi_{[w]}=z^{-2s}\Psi_{[z]}$, so they are both polynomials of degree $\le 2s$. Quantum Hilbert space thus becomes $(2s+1)$-dimensional as required for the spin-$s$ representation.

The above simple problem can also be studied using path integral formalism \cite{Alekseev:1988vx}. The appropriate Lagrangian for the quantum mechanics of $\theta$ and $\varphi$ is (hereafter we are back in $\hbar=1$ units)
\begin{equation}
L=-s\cos\theta\dot\varphi + \gamma\dot\varphi\,, 
\label{Lgq1}
\end{equation}
where $\gamma$ is a constant satisfying the quantization condition $s\pm\gamma\in\mathbb Z$. Note that the first term in (\ref{Lgq1}) gives the correct Poisson bracket of $\theta$ and $\varphi$ in the same way that $\{q,p\}=1$ follows from $L=p\dot q$. The second term and quantization condition are necessary for $\exp(i\int \mathrm dt L)$ to be a continuous functional of the path $\{\theta(t),\varphi(t)\}$. It can be understood by thinking of continuous deformations of a path such that its winding number around the points $\theta=0$ or $\pi$ jumps.

For $\gamma=-s$ the above $L$ and $P^a$ can be expressed as
\begin{align}
& L = 2i\,\mathrm{Tr}(\lambda g^{-1}\dot g),\quad
 P^a = \mathrm{Tr}(\lambda g^{-1}\sigma^ag),
\end{align}
where $\sigma^a$ are Pauli's matrices and $\lambda,g$ are the following $2\times2$ matrices.
\begin{equation}
\lambda=\frac s2\sigma^3,\quad g=\left(\begin{array}{rr}\sin\frac\theta2 & e^{-i\varphi}\cos\frac\theta2\\-e^{i\varphi}\cos\frac\theta2 & \sin\frac\theta2\end{array}\right).
\label{MtoG}
\end{equation}
Using these quantities, one can express the Wilson loop as a path integral of a quantum mechanical system coupled to the 3D gauge field.
\begin{align}
 W_\lambda(C) &\;=\; \mathrm{Tr}_\Lambda \mathrm P\exp\bigg( i\oint_C \mathrm dx^m A^a_m\,T^a\bigg)\nonumber \\&\;=\;\int{\cal D}g\exp\int\mathrm dt\,\mathrm{Tr}\Big({-}2\lambda g^{-1}(\dot g-i\dot x^mA_mg)\Big).
\label{WilsonQM}
\end{align}

The $S^2$ in the above discussion is the simplest example of {\it adjoint orbit}\footnote{Throughout this paper we work with the natural identification of adjoint and coadjoint orbits.}. The adjoint orbit of a Lie algebra element $\lambda\in\mathfrak g=\text{Lie}(G)$ is defined by
\begin{equation}
 \mathrm{Ad}_G(\lambda)\;\equiv\; \{g\lambda g^{-1}\,|\,g\in G\}.
\label{AdGl}
\end{equation}
The irreducible representation of a Lie group with highest weight $\lambda$ can be obtained from geometric quantization of the adjoint orbit $\mathrm{Ad}_G(\lambda)$, where the weight $\lambda\in\mathfrak{h}^\ast$ and the Lie algebra element $\lambda\in\mathfrak h$ are identified via
\begin{equation}
\lambda\dotp\sigma = 2\mathrm{Tr}(\lambda\sigma)\,.\qquad\big(\forall \sigma\in\mathfrak h\big)
\label{hhast}
\end{equation}
The formula (\ref{WilsonQM}) works for arbitrary gauge groups and representations. General properties of adjoint orbits will be discussed in more detail later.

\paragraph{Boundary terms in CS theories revisited.} In the previous section we determined the boundary term for the CS action (\ref{LB-FI-CS}) from SUSY invariance. We are now in a position to argue this was not enough, and explain what needs to be added. Our argument is based on \cite{Moore:1989yh,Elitzur:1989nr} which carefully studied the canonical quantization of CS theories.

For simplicity, let us first consider the theory on $\mathbb R^3$ with a BPS vortex line satisfying (\ref{Def-Vortex}), (\ref{xi-BPS}) lying along the $x^3$-axis. So $M$ is an $\mathbb R^3$ with the tubular neighborhood of the vortex line removed. We use $t$ for the coordinate along the vortex line and the polar coordinate $r,\varphi$ for the transverse two dimensions, so that $\partial M$ is the cylinder at $r=\epsilon$ parameterized by $\varphi,t$. Our formula (\ref{LB-FI-CS}) for the boundary term for ${\cal N}=2$ CS theory becomes in this case
\begin{equation}
 S_\text{CS,B} = -\frac{ik}{4\pi}\int_{\partial M}\mathrm d\varphi\mathrm dt\,\mathrm{Tr}\Big[A_\varphi(A_t-2i\sigma)\Big]\,.
\label{SCSB-R3}
\end{equation}
Let us examine if the variational problem is well-defined under this choice of boundary term.

Recall that the variation of the bosonic CS action gives
\begin{equation}
 \delta S_\text{CS}\;=\;\delta\bigg\{\frac{ik}{4\pi}\int_M\mathrm{Tr}\Big(A\mathrm dA-\frac{2i}3A^3\Big)\bigg\} \;=\; \frac{ik}{2\pi}\int_M\mathrm{Tr}\big(\delta A\wedge F\big)+\frac{ik}{4\pi}\int_{\partial M}\mathrm{Tr}\big(\delta A\wedge A\big).
\end{equation}
The first term in the RHS vanishes due to the equation of motion $F=0$. The second term can be rewritten as
\[
 \frac{ik}{4\pi}\int_{\partial M}\mathrm d\varphi\mathrm dt\,\mathrm{Tr}\Big(\delta A_\varphi A_t-A_\varphi\delta A_t\Big).
\]
The variational problem becomes well-defined by requiring that one of the two gauge field components $A_\varphi,A_t$ vanish on $\partial M$. Alternatively, one can specify nonzero boundary value for $A_t$ by adding a boundary term
\begin{equation}
 S_\text{CS,B}~=~-\frac{ik}{4\pi}\int_{\partial M}\mathrm d\varphi\mathrm dt\,\mathrm{Tr}\big(A_\varphi A_t\big),
\label{SCSB-Moore}
\end{equation}
which is in fact a part of (\ref{SCSB-R3}). Somewhat confusingly, the boundary term for specifying $A_\varphi$ is different from this $S_\text{CS,B}$ by minus sign. One can indeed check $\delta(S_\text{CS}+S_\text{CS,B})$ vanishes if $F=0$ holds in the bulk and $\delta A_t=0$ on the boundary. Also, the addition of (\ref{SCSB-Moore}) has an effect of changing the bulk Lagrangian
\[
 {\cal L}_\text{CS}= -\frac{ik}{4\pi}\mathrm{Tr}\Big(A_\varphi \dot A_t-A_t\dot A_\varphi+\cdots\Big)~~\longrightarrow~~
 {\cal L}'_\text{CS}= -\frac{ik}{4\pi}\mathrm{Tr}\Big(2A_\varphi \dot A_t+\cdots\Big),
\]
where the dots above $A_t,A_\varphi$ stand for $r$-derivatives. Therefore, if the theory is radially quantized with the Lagrangian ${\cal L}'_\text{CS}$, $A_t$ plays the role of canonical coordinate and $A_\varphi$ the momentum. The wave functions describing states on equal-$r$ surfaces are functionals of $A_t$. This is in accord with the fact that one can set the value of $A_t$ on the boundary at will.

Suppose that, instead of vortex singularity, a quantum mechanics with $G$ symmetry is introduced along the $x^3$-axis. Let $S_\text{QM}$ be the action describing the quantum mechanics interacting with the $G$-gauge field $A_t$ in the bulk $\mathbb R^3$. Then one can define a 1D-3D coupled system by the path integral of $e^{-S_\text{CS}-S_\text{CS,B}-S_\text{QM}}$ with respect to the quantum mechanical variables and the 3D gauge field. The boundary term which is appropriate for this construction is again (\ref{SCSB-Moore}).

Now that we have already chosen (\ref{SCSB-Moore}) as the boundary term, what can we do to impose the boundary condition on $A_\varphi$? The answer is simply to set
\begin{equation}
S_\text{QM}=-ik\int\mathrm dt\,\mathrm{Tr}\big(\beta A_t\big).
\end{equation}
Then the variation of the whole action
\[
 \delta\big(S_\text{CS}+S_\text{CS,B}+S_\text{QM}\big) \;=\;\frac{ik}{2\pi}\int_M\mathrm{Tr}\big(\delta A\wedge F\big)-\frac{ik}{2\pi}\int_{\partial M}\mathrm d\varphi\mathrm dt\,\mathrm{Tr}\big(A_\varphi\delta A_t\big)-ik\int\mathrm dt\,\mathrm{Tr}\big(\beta\delta A_t\big)
\]
gives $A_\varphi\big|_\mini{\partial M}=\beta$ as an equation of motion. Furthermore, according to \cite{Moore:1989yh} one should average the boundary condition over the orbit of $\beta$, namely to modify the boundary condition as $A_\varphi\big|_\mini{\partial M}=g\beta g^{-1}$ for a $t$-dependent element $g\in G$ and integrate over $g(t)$. This can be done by modifying $S_\text{QM}$ as follows:
\begin{equation}
 S_\text{QM}[g] \;=\;k\int\mathrm dt\,\mathrm{Tr}\Big(\beta g^{-1}\Big(\frac{\mathrm d}{\mathrm dt} g-iA_tg\Big)\Big).
\end{equation}
Here the kinetic term for $g(t)$ has been added to make $S_\text{QM}$ gauge-invariant. We thus arrived at a description of vortex loops in terms of a quantum mechanics of $g(t)$ coupled to 3D gauge field. Moreover, the quantum mechanics is the same as the one for the Wilson loops (\ref{WilsonQM}) if their parameters $\lambda, \beta$ are related as
\begin{equation}
 \lambda = \frac{k\beta}2.
\label{lambda-beta}
\end{equation}
So, in bosonic CS theory with coupling $k$, a vortex loop with vorticity $\beta$ is equivalent to a Wilson loop for the representation with the highest weight $\lambda=k\beta/2$. Note that this leads to a quantization of $\beta$ in CS theories.

\vskip3mm

Let us come back to the ${\cal N}=2$ CS theories on an ellipsoid with a BPS vortex loop along $S^1_{(\tau)}$ at $\theta=0$. The supersymmetric boundary term is (\ref{LB-FI-CS}) instead of (\ref{SCSB-R3}). The role of $A_t, A_\varphi$ in the previous discussion is now played by
\[
 -v^mA_m = \frac1{\tilde\ell}A_\varphi +\frac1\ell A_\tau\,,\qquad
 \ell\tilde\ell\sin\theta\cos\theta\cdot w^mA_m =\ell\cos^2\theta A_\varphi-\tilde\ell\sin^2\theta A_\tau.
\]
To describe a vortex loop with vorticity $\beta$, one needs to introduce
\begin{equation}
 S_\text{QM} = k\int\mathrm d\tau\mathrm{Tr}\,\Big[\beta\ell(iv^mA_m-\sigma)\Big],
\label{N2-SCSQM}
\end{equation}
or the averaged version
\begin{equation}
 S_\text{QM}[g] = k\int\mathrm d\tau\mathrm{Tr}\,\Big[\beta g^{-1}\frac{\mathrm d}{\mathrm d\tau}g+\beta g^{-1}\ell(iv^mA_m-\sigma)g\Big].
\label{N2-SCSQMg}
\end{equation}
Note that we included $\sigma$ in these formulae to make $S_\text{QM}$ supersymmetric.

We believe that both of the above boundary terms lead to consistent descriptions of vortex loops. The boundary term $S_\text{QM}$ (\ref{N2-SCSQM}) sets the boundary condition $A_\varphi=\beta$ and leads to the definition of a vortex loop by a singular behavior of the gauge field. On the other hand, the averaged version $S_\text{QM}[g]$ describes a vortex loop in terms of a quantum mechanics coupled to the bulk gauge field. In the latter description of vortex loops, one usually does not assume singular behavior for the gauge field before integrating out the quantum mechanical degrees of freedom. These may sound somewhat empirical, but we would like to show in the following that the above two definitions indeed lead to the same result for the expectation value of a vortex loop.

\subsection{Path integral over fields with singularity}\label{sec:PIsing}

Here we compute the expectation value of a BPS vortex loop on an ellipsoid using the boundary term without averaging, i.e. $S_\text{QM}$ (\ref{N2-SCSQM}).
Its value on the saddle point (\ref{saddlept}) is
\[
S_\text{QM} = -k\int\mathrm d\tau\,\mathrm{Tr}\Big[\beta\ell\Big(\sigma+\frac{i\beta}{\ell}\Big)\Big] = -2\pi k\mathrm{Tr}(\beta\widehat\sigma)= -\pi k\beta\dotp\widehat\sigma.
\]
Note that we shifted $\sigma$ as explained after (\ref{VbetaDEF}) and then used (\ref{hhast}).
This corrects our previous formula for $V_\beta(\hat\sigma)$ (\ref{Vbeta0}) and $V_\beta^\text{v}(\hat\sigma)$ (\ref{Vbeta1}) as follows:
\begin{align}
 V_\beta(\widehat\sigma) = \frac{|{\cal W}|}{|{\cal W}_K|}V_\beta^\text{v}(\widehat\sigma),\qquad
 V_\beta^\text{v}(\widehat\sigma)\;=\; \frac{e^{\pi k\beta\dotp\widehat\sigma}}{\prod_{\alpha\dotp\beta>0}2\sinh\pi\alpha\dotp\widehat\sigma}\,.
\label{Vbeta2}
\end{align}
Recall that $\beta$ was gauge-rotated so that $\alpha\dotp\beta\ge0$ for all the positive roots. Those which are orthogonal to $\beta$, if any, are the positive roots of the subgroup $K\subset G$ left unbroken by the vortex loop.

We would like to compare this with the function $W_\lambda(\widehat\sigma)$ (\ref{WLambda}) for a Wilson loop in the representation $\Lambda$. We decompose the Weyl vector as $\rho= \rho_\mini{K}+\tilde\rho$, where
\begin{equation}
\rho_\mini{K}=\frac12\sum_{\alpha\in\Delta^+_\mini{K}}\alpha,\qquad
\tilde\rho=\frac12\sum_{\alpha\in\Pi^+}\alpha\,.\qquad
\left(
\begin{array}{r}
\Delta_\mini{K}^+ \equiv \{\alpha\in\Delta^+\,|\,\alpha\dotp\lambda=0\} \\[1mm]
\Pi^+ \equiv \{\alpha\in\Delta^+\,|\,\alpha\dotp\lambda>0\}
\end{array}
\right)
\label{rhotilde}
\end{equation}
Then
\begin{align}
 W_\lambda(\widehat\sigma) &= \frac{\sum_{w\in{\cal W}/{\cal W}_K}\sum_{w'\in{\cal W}_K}\epsilon(w)\epsilon(w')e^{2\pi w'(\rho+\lambda)\dotp w(\widehat\sigma)}}{\prod_{\alpha\in\Delta^+}2\sinh\pi\alpha\dotp\widehat\sigma}
\nonumber \\
 &= \sum_{w\in{\cal W}/{\cal W}_K}\frac{\sum_{w'\in{\cal W}_K}\epsilon(w')e^{2\pi (w'(\rho_K)+\tilde\rho+\lambda)\dotp w(\widehat\sigma)}}{\prod_{\alpha\in\Delta^+}2\sinh\pi\alpha\dotp w(\widehat\sigma)}
\nonumber \\
 &= \sum_{w\in{\cal W}/{\cal W}_K}e^{2\pi(\tilde\rho+\lambda)\dotp w(\widehat\sigma)}\,\frac{\prod_{\alpha\in\Delta^+_K}2\sinh\pi\alpha\dotp w(\widehat\sigma)}{\prod_{\alpha\in\Delta^+}2\sinh\pi\alpha\dotp w(\widehat\sigma)}
\nonumber \\
 &= \sum_{w\in{\cal W}/{\cal W}_K}\frac{e^{2\pi(\tilde\rho+\lambda)\dotp w(\widehat\sigma)}}{\prod_{\alpha\in\Pi^+}2\sinh\pi\alpha\dotp w(\widehat\sigma)}\,,
\label{Wlambda}
\end{align}
where ${\cal W}_K$ was defined at (\ref{VbetaDEF}).

The expectation values of a Wilson loop in a representation with highest weight $\lambda$ and a vortex loop with vorticity $\beta$ are given respectively by integrals of $W_\lambda(\widehat\sigma)$ and $V_\beta(\widehat\sigma)$ over $\mathfrak h$ with a measure (\ref{VbetaDEF2}). Inside such an integral, the summation over the images of ${\cal W}/{\cal W}_K$ is the same as the multiplication by $|{\cal W}|/|{\cal W}_K|$. So the above result implies an equivalence between Wilson and vortex loops
\begin{equation}
V_\beta(\widehat\sigma)\simeq W_\lambda(\widehat\sigma)\qquad\text{for}\qquad
\lambda+\tilde\rho=\frac{k\beta}2\,.
\label{corr01}
\end{equation}
Note that there is a correction to the rule of correspondence compared to that for bosonic theory (\ref{lambda-beta}). This looks problematic because the trivial Wilson loop ($\lambda=0$) does not correspond to the trivial vortex loop $(\beta=0)$.

\subsection{1D-3D coupled system}\label{sec:1D3D}

Next we study the description of a vortex loop using the averaged version (\ref{N2-SCSQMg}) of the boundary term. The quantization of (\ref{N2-SCSQMg}) itself would give the representation with the highest weight $\lambda=k\beta/2$, because it is identical to the action (\ref{WilsonQM}) for the adjoint orbit quantization. We would like to do something slightly different here. As the bulk CS theory was promoted to a 3D ${\cal N}=2$ theory, one can also promote the quantum mechanics on the vortex worldline to a 1D ${\cal N}=2$ SUSY theory. The interaction between 1D and 3D fields can be chosen in such a way that the whole system is invariant under a SUSY that acts on both 1D and 3D fields at the same time. The path integral of the combined system can be performed exactly.

\subsubsection{Adjoint orbits} \label{sec:Adj-Orb}

We begin by summarizing basic properties of general adjoint orbits\footnote{For more detailed reviews of the mathematical properties of adjoint orbits, see \cite{Alekseevsky:1997zb,Arvanitoyeorgos2003AnIT}.}. The adjoint orbit $M=\mathrm{Ad}_G(\lambda)$ for $\lambda\in\mathfrak g$ can be viewed as the coset space $M=G/K$, where $K$ is the centralizer of $\lambda$:
\begin{equation}
 \mathrm{Ad}_G(\lambda)=\{g\lambda g^{-1}\,|\,g\in G\}=G/K,\quad
 K=\{h\in G\,|\,h\lambda h^{-1}=\lambda\}.
\end{equation}
To describe mathematical properties of $M$, it is convenient to think of a map $g(x)~(x\in M,\,g\in G)$ such as the $SU(2)$-valued function $g(\theta,\varphi)$ (\ref{MtoG}). The action of a Lie group element $g_0\in G$ on $M$ translates into a coordinate transformation $x^m\to x'^m$ according to the relation
\begin{equation}
 g_0\cdot g(x) = g(x')\cdot h(x,g_0)\qquad\big(h(x,g_0)\in K\big)\,.
\label{Gaction}
\end{equation}
As an infinitesimal version of this, multiplication of Lie algebra generators $T^a\in\mathfrak g$ translates into the action of vector fields $X^a=X^{am}(x)\partial_m$,
\begin{equation}
 X^ag(x) = -iT^a g(x)+ig(x)H^a(x)\,,\qquad\big(H^a(x)\in\mathfrak k\big)
\label{vecG}
\end{equation}
where $\mathfrak k\subset\mathfrak g$ is the Lie algebra of $K$. The corresponding moment map function $P^a$ is given by
\begin{equation}
 P^a = 2\mathrm{Tr}\big[\lambda g^{-1}T^ag\big].
\label{MMonM}
\end{equation}
It satisfies $\mathrm dP^a+\imath_{X^a}\omega=0$, where $\omega$ is the $G$-invariant symplectic form on $M$ called the {\it Kirillov-Kostant-Souriau 2-form}.
\begin{equation}
 \omega = -2i\,\mathrm{Tr}\big[\lambda(g^{-1}\mathrm dg)^2\big].
\label{KSSform}
\end{equation}

Next we turn to complex structures on $M$. Let $\mathfrak n$ be the orthogonal complement of $\mathfrak k$ with respect to the Killing form. We are interested in the cases where $G/K$ is reductive, that is when the decomposition $\mathfrak g=\mathfrak k\oplus\mathfrak n$ is such that $[\mathfrak k,\mathfrak n]\subset\mathfrak n$. Note that $\mathfrak n$ is identified with the tangent space at $\lambda\in M$. To define a complex structure $J$ on $M$, one first needs a decomposition of $\mathfrak n^\mathbb{C}$ into two subspaces $\mathfrak n_\pm$ of definite eigenvalues of $J$. By transporting this decomposition of $T_\lambda M^\mathbb{C}$ to all other points on $M$ by the action of $G$ (\ref{Gaction}) one obtains an almost complex structure on $M$.  It is integrable if the set of holomorphic (or antiholomorphic) vector fields on $M$ is closed under Lie bracket, which simply amounts to $[\mathfrak n_\pm,\mathfrak n_\pm]\subset \mathfrak n_\pm$.

Recall that $\lambda$ was chosen to be in a Cartan subalgebra $\mathfrak h$, and $\alpha\dotp\lambda\ge0$ for all the positive roots $\alpha\in\Delta^+$. This leads to a decomposition $\mathfrak g^\mathbb{C}=\mathfrak k^\mathbb{C}\oplus\mathfrak n_+\oplus \mathfrak n_-$, where
\begin{equation}
\mathfrak k^\mathbb{C}=\mathfrak h^\mathbb{C}\oplus\sum_{\alpha\in\Delta^+_K}(\mathbb{C}E_\alpha+\mathbb{C}E_{-\alpha})\,,\qquad
\mathfrak n_+= \sum_{\alpha\in\Pi^+}\mathbb{C}E_\alpha\,,\qquad
\mathfrak n_-= \sum_{\alpha\in\Pi^+}\mathbb{C}E_{-\alpha}\,
\end{equation}
and $\Pi^+$ was defined in (\ref{rhotilde}). The symplectic form (\ref{KSSform}) is of type $(1,1)$ under the complex structure thus defined, so $M$ is a K\"ahler manifold. Note that there are in general multiple complex structures for a single coset space $G/K$. For example,
\[
\lambda_1=\mathrm{diag}(1,1,0,0,0,-1,-1),\qquad
\lambda_2=\mathrm{diag}(3,3,0,0,-2,-2,-2)
\]
both break $G=SU(7)$ to $K=SU(3)\times SU(2)^2\times U(1)^2$ but lead to different complex structures on $G/K$.

A useful fact is that $G/K$ can be thought of as the {\it flag manifold} $G^\mathbb{C}/P$, where $P$ is a {\it parabolic subgroup} of $G$ corresponding to the Lie algebra $\mathfrak k^\mathbb{C}\oplus\mathfrak n_-$. (When $K$ equals a maximal torus of $G$, $P$ is called {\it Borel subgroup}.) This implies that any complex coordinate on $N_+$, the Lie group corresponding to $\mathfrak n_+$, can be used as a complex coordinate on $M$. Moreover, under such a choice of coordinate on $M$, the vector fields $X^a$ (\ref{vecG}) become holomorphic Killing vector fields which preserve the K\"ahler metric on $M$.

\subsubsection{${\cal N}=2$ SUSY quantum mechanics on $M$}\label{sec:QM-Adj-Orb}

Let us now turn to the 1D ${\cal N}=2$ supersymmetric quantum mechanics with the target space $M=\mathrm{Ad}_G(\lambda)$ and its quantization. As $M$ is K\"ahler and we are gauging its isometry, we need chiral and vectormultiplets.

Take a complex coordinate $z^\mini{I}$ on $M$ such that its metric and K\"ahler form are given in terms of the K\"ahler potential $K(z,\bar z)$ as follows.
\begin{equation}
\mathrm ds^2 = g_\mini{I\bar J}(z,\bar z)\mathrm dz^\mini{I}\mathrm d\bar z^\mini{\bar J},\quad
\omega = ig_\mini{I\bar J}(z,\bar z)\mathrm dz^\mini{I}\mathrm\wedge d\bar z^\mini{\bar J}\,;\qquad
g_\mini{I\bar J}(z,\bar z) = \frac{\partial^2 K(z,\bar z)}{\partial z^\mini{I}\partial \bar z^\mini{\bar J}}\,.
\end{equation}
The isometry of $M$ is generated by holomorphic Killing vectors
\[
 X^a = X^{a\mini{I}}(z)\frac{\partial}{\partial z^\mini{I}} + \bar X^{a\mini{\bar J}}(\bar z)\frac\partial{\partial\bar z^\mini{\bar J}}
\]
satisfying $[X^a,X^b]=-f^{abc}X^c$. To each $X^a$ there is a corresponding moment map $P^a$ satisfying $\mathrm dP^a+\imath_{X^a}\omega=0$, or in components
\begin{equation}
\partial_\mini{I}P^a = ig_\mini{I\bar J}\bar X^{a\mini{\bar J}},\qquad
\bar\partial_\mini{\bar J}P^a = -ig_\mini{I\bar J}X^{a\mini{I}}.
\end{equation}
Using $g_\mini{I\bar J}=\partial_\mini{I}\bar\partial_\mini{\bar J}K$ and the holomorphicity of Killing vector one can integrate these equalities to determine $P^a$ up to constant shifts, which in turn can be fixed by requiring $P^a$ to transform in the adjoint representation. For a suitable $K$, $P^a$ can be written as
\begin{equation}
 P^a = -iX^{a\mini{I}}\partial_\mini{I}K = i\bar X^{a\mini{\bar J}}\bar\partial_\mini{\bar J}K\,.
\end{equation}

A (1D) vectormultiplet consists of a gauge field $A_t$, bosons $\sigma,D$ and fermions $\lambda,\bar\lambda$ transforming as
\begin{alignat}{3}
\SUSY A_t &= \frac i2(\epsilon\bar\lambda+\bar\epsilon\lambda), \qquad &
\SUSY\lambda &= \epsilon(-i\nabla_t\sigma-D),\qquad&
\SUSY D &= \frac i2\nabla_t(\bar\epsilon\lambda-\epsilon\bar\lambda)-\frac i2[\sigma,\bar\epsilon\lambda-\epsilon\bar\lambda],\nonumber\\
\SUSY \sigma &= \frac12(\epsilon\bar\lambda+\bar\epsilon\lambda),\qquad &
\SUSY\bar\lambda &= \bar\epsilon(-i\nabla_t\sigma+D),
\label{1Dvec}
\end{alignat}
where $\epsilon,\bar\epsilon$ are Grassmann-even constant SUSY parameters. All the fields are Lie algebra valued, so one can express them using the set of generators $T^a$ as follows.
\[
 A_t=A_t^aT^a,\quad \sigma=\sigma^aT^a,\quad\text{etc.}\qquad\Big([T^a,T^b]=if^{abc}T^c\Big)
\]
The complex coordinates $z^\mini{I}$ on $M$ are promoted to chiral multiplets. Each chiral multiplet consists of a boson $z^\mini{I}$ and its superpartner $\chi^\mini{I}$. They transform as
\begin{alignat}{3}
\SUSY z^\mini{I} &=\epsilon\chi^\mini{I},\qquad
\SUSY\chi^\mini{I} &= -i\bar\epsilon\big(\nabla_tz^\mini{I}-i\sigma^aX^{a\mini{I}}\big),\qquad
\nabla_tz^\mini{I} &\equiv \dot z^\mini{I}+A_t^aX^{a\mini{I}},\nonumber\\
\SUSY\bar z^\mini{\bar J} &=\bar\epsilon\bar\chi^\mini{\bar J},\qquad
\SUSY\bar\chi^\mini{\bar J} &= -i\epsilon\big(\nabla_t\bar z^\mini{\bar J}-i\sigma^a\bar X^{a\mini{\bar J}}\big),\qquad
\nabla_t\bar z^\mini{\bar J} &\equiv\dot{\bar z}^\mini{\bar J}+A_t^a\bar X^{a\mini{\bar J}}.
\label{1dchiral}
\end{alignat}
The SUSY-invariant kinetic Lagrangian for the chiral multiplets is given by
\begin{align}
 L_\text{kin} ~=~& g_\mini{I\bar J}\nabla_t\bar z^\mini{\bar J}\nabla_t z^\mini{I}+g_\mini{I\bar J}\bar X^{a\mini{\bar J}}X^{b\mini{I}}\sigma^a\sigma^b+iD^aP^a-g_\mini{I\bar J}\bar X^{a\mini{\bar J}}\lambda^a\chi^\mini{I}+g_\mini{I\bar J}\bar\chi^\mini{\bar J}\bar\lambda^aX^{a\mini{I}}\nonumber\\
&-ig_\mini{I\bar J}\bar\chi^\mini{\bar J}\nabla_t\chi^\mini{I}+g_\mini{I\bar J}\bar\chi^\mini{\bar J}\partial_\mini{K}X^{a\mini{I}}\sigma^a\chi^\mini{K}+g_\mini{I\bar J,K}\bar\chi^\mini{\bar J}X^{a\mini{K}}\sigma^a\chi^\mini{I},
\nonumber \\[1mm] &~\nabla_t\chi^\mini{I}\equiv\dot\chi^\mini{I}+A_t^a\partial_\mini{K}X^{a\mini{I}}\chi^\mini{K}+\Gamma^\mini{I}_\mini{KL}\nabla_tz^\mini{K}\chi^\mini{L}.
\label{Lkin1d}
\end{align}
Another invariant can be constructed using the one-form $\vartheta=\vartheta_\mini{I}\mathrm dz^\mini{I}+\vartheta_\mini{\bar J}\mathrm d\bar z^\mini{\bar J}$ satisfying $\mathrm d\vartheta=\omega$.
\begin{align}
 L_\text{top} &= ig_\mini{I\bar J}\chi^\mini{I}\bar\chi^\mini{\bar J}-i\vartheta_\mini{I}\big(\nabla_tz^\mini{I}-i\sigma^aX^{a\mini{I}}\big)-i\vartheta_\mini{\bar J}\big(\nabla_t\bar z^\mini{\bar J}-i\sigma^a\bar X^{a\mini{\bar J}}\big) \nonumber \\ &= ig_\mini{I\bar J}\chi^\mini{I}\bar\chi^\mini{\bar J}-i(\vartheta_\mini{I}\dot z^\mini{I}+\vartheta_\mini{\bar J}\dot{\bar z}^\mini{\bar J})-(\sigma^a+iA_t^a)P^a.
\label{Ltop1d}
\end{align}

What we actually need to do is to gauge the isometry of the adjoint orbit $M$ by the 3D gauge field and not by an independent 1D vector field. To do this in a supersymmetric manner, we recall the transformation rule of cohomological variables constructed from the 3D vectormultiplet fields.
\begin{align}
\SUSY (-v^mA_m) &= \frac i2(\xi\bar\lambda-\bar\xi\lambda),\quad&
\SUSY (-\bar\xi\lambda) &= iv^m\nabla_m\sigma-\tilde D,\quad&
\SUSY\tilde D &= \frac i2v^m\nabla_m(\xi\bar\lambda+\bar\xi\lambda)
\nonumber \\
\SUSY\sigma &= \frac12(\xi\bar\lambda-\bar\xi\lambda),\quad&
\SUSY(\xi\bar\lambda) &= iv^m\nabla_m\sigma+\tilde D.\quad&
&~+\frac i2[\sigma,\xi\bar\lambda+\bar\xi\lambda].
\label{Q3DVch}
\end{align}
Here $\tilde D\equiv D-\frac1f\sigma+\frac12\bar\xi\gamma^{mn}\xi F_{mn}$.
By comparing this with (\ref{1Dvec}) one finds that the 3D fields
\[
-v^mA_m\;,\;\sigma\;,\;-\bar\xi\lambda\;,\;\xi\bar\lambda\;,\;\tilde D 
\]
transform under the 3D SUSY in the same way that the 1D vectormultiplet transforms under 1D SUSY with $\epsilon=\bar\epsilon=1$. The 1D-3D coupling is thus obtained by identifying $t$ with $\ell\tau$, replacing the vectormultiplet fields in (\ref{Lkin1d}), (\ref{Ltop1d}) by the above 3D fields and regarding $\SUSY^\text{(3D)}+\SUSY^\text{(1D)}_{(\epsilon=\bar\epsilon=1)}$ as the SUSY of the total system. Recalling (\ref{MMonM}) and (\ref{KSSform}) one finds that the bosonic part of $L_\text{top}$ (\ref{Ltop1d}) agrees precisely with the action $S_\text{QM}[g]$ (\ref{N2-SCSQMg}) for the quantum mechanics on vortex loops, and the fermions appear in $L_\text{top}$ as auxiliary fields.

The Lagrangians $L_\text{top}$ and $L_\text{kin}$ play a role similar to that of
$S_\text{CS}$ and $S_\text{YM}$ for the 3D gauge field. First, the fermions $\chi^\mini{I},\bar\chi^\mini{\bar J}$ are auxiliary variables in the theory without $L_\text{kin}$. Second, $L_\text{top}=\SUSY \Psi_\text{top}$ but $\Psi_\text{top}$ depends on the components of $\vartheta$
\[
 \Psi_\text{top} = \vartheta_\mini{I}\chi^\mini{I}+\vartheta_\mini{\bar J}\bar\chi^\mini{\bar J},
\] 
which are defined only up to (K\"ahler) gauge transformations. As a consequence, $L_\text{top}$ takes different nonzero values on different saddle points, whereas $L_\text{kin}$ vanishes at every saddle point.

\paragraph{Witten index.}

Let us compute the Witten index, i.e. the $S^1$ partition function of the quantum mechanics on a vortex loop. It is a SUSY quantum mechanics with the target space $M=\mathrm{Ad}_G(\lambda)$ coupled to 3D vectormultiplet field. The 3D fields are fixed at a saddle point (\ref{saddlept}) with $\beta=0$. So we only need to study the 1D theory defined by (\ref{1dchiral}), (\ref{Lkin1d}) and (\ref{Ltop1d}) with all the vectormultiplet fields turned off except for constant $\sigma$, which we may assume to be in $\mathfrak h$.

According to (\ref{1dchiral}), the saddle point condition for our quantum mechanics is
\[
 \dot z^\mini{I}-i\sigma^aX^{a\mini{I}}=0,\qquad\dot{\bar z}^\mini{\bar J}-i\sigma^a\bar X^{a\mini{\bar J}}=0.
\]
In terms of the original coordinate $g$ on $M$, these become
\[
 \frac{\mathrm d}{\mathrm dt}(g\lambda g^{-1})-[\sigma, g\lambda g^{-1}]=0.
\] 
If $\sigma$ and the periodicity of $t$ take generic values, this can only be solved by requiring the two terms on the LHS vanish independently. So, $g\lambda g^{-1}$ is a constant element of $\mathfrak h$ at saddle points. Since $\lambda$ is also an element of $\mathfrak h$, $g\lambda g^{-1}$ has to be an image of $\lambda$ under Weyl group.

Let us study the saddle point $g=\mathrm{id}~(g\lambda g^{-1}=\lambda)$ in detail. The neighborhood of this point can be covered by a local complex coordinate system $z^\alpha$ such that
\begin{equation}
 g = \exp i\sum_{\alpha\in\Pi^+}(z^\alpha E_\alpha+\bar z^\alpha E_{-\alpha}).
\end{equation}
Then the K\"ahler form and metric around this point are approximately given by
\begin{equation}
 \omega \simeq i\sum_{\alpha\in\Pi^+}g_{\alpha\bar\alpha}\mathrm dz^\alpha\wedge \mathrm d\bar z^\alpha,\qquad
 g_{\alpha\bar\alpha} \equiv 2\lambda\dotp\alpha\,\mathrm{Tr}\big(E_\alpha E_{-\alpha}\big).
\end{equation}
Note the positive definiteness of the metric. The moment map and the Killing vector corresponding to $\sigma=\sigma_iH_i\in\mathfrak h$ read
\begin{equation}
 \sigma_iP_i \simeq \lambda\dotp\sigma-\sum_{\alpha\in\Pi^+}(\alpha\dotp\sigma) z^\alpha\bar z^\alpha g_{\alpha\bar\alpha},\qquad
 \sigma_iX_i = -i\sum_{\alpha\in\Pi^+}\alpha\dotp\sigma\Big(z^\alpha\frac{\partial}{\partial z^\alpha}-\bar z^\alpha\frac{\partial}{\partial\bar z^\alpha}\Big),
\label{PXonM}
\end{equation}
where we used $[H_i,E_\alpha]=\alpha_iE_\alpha$. Note that the expression for the Killing vector is exact. The value of the action (the integral of $L_\text{top}$) on this saddle point is
\begin{equation}
 e^{-S_\text{QM}} = e^{2\pi\ell\lambda\dotp\sigma} = e^{2\pi\lambda\dotp\widehat\sigma}\,.
\end{equation}
The one-loop determinant $\Delta_\text{1-loop}$ at this saddle point can be computed using the SUSY-exact localizing Lagrangian $L_\text{kin}$, which takes the approximate form
\begin{equation}
 L_\text{kin}\simeq \sum_{\alpha\in\Pi^+}g_{\alpha\bar\alpha}\Big\{\dot{\bar z}^\alpha\dot z^\alpha+(\alpha\dotp\sigma)^2\bar z^\alpha z^\alpha-i\bar\chi^\alpha\dot\chi^\alpha-i(\alpha\dotp\sigma)\bar\chi^\alpha\chi^\alpha\Big\}.
\end{equation}
The Gaussian integration over $z^\alpha$ and $\chi^\alpha$ can be easily performed using $\mathrm{det}\big(\frac{\mathrm d}{\mathrm dt}+\omega\big)=2\sinh\pi\ell\omega$ (if $t\sim t+2\pi\ell$). The contribution of this saddle point finally becomes
\begin{equation}
 e^{-S_\text{QM}}\cdot\Delta_\text{1-loop}\,\Big|_{g\lambda g^{-1}=\lambda}~=~\frac{e^{2\pi\lambda\dotp\widehat\sigma}}{\prod_{\alpha\in\Pi^+}2\sinh\pi\alpha\dotp\widehat\sigma}.
\label{QMsaddle}
\end{equation}

Other saddle points are all characterized by the equation $g\lambda g^{-1}=w(\lambda)$ for some element $w$ of the Weyl group. Their contribution can be computed by repeating the above steps with the replacement $\lambda\to w(\lambda)$. But once this replacement is made, the set of positive roots also needs to be redefined so that $\alpha\dotp w(\lambda)\ge0$ for all $\alpha\in\Delta^+_\text{(new)}$. So the contribution from other saddle points are obtained from (\ref{QMsaddle}) by replacing $\lambda\to w(\lambda)$ and $\alpha\to w(\alpha)$, or more simply by the replacement $\widehat\sigma\to w^{-1}(\widehat\sigma)$. The full partition function is thus obtained by summing over different saddle points labeled by $w\in {\cal W}/{\cal W}_K$. The index finally becomes
\begin{align}
 I_\lambda(\widehat\sigma) &~=~ \int {\cal D}[z,\chi]\exp\left(-S_\text{QM}\right) \nonumber \\ &~=~ \sum_{w\in{\cal W}/{\cal W}_K}\frac{e^{2\pi\lambda\dotp w(\widehat\sigma)}}{\prod_{\alpha\in\Pi^+}2\sinh\pi\alpha\dotp w(\widehat\sigma)}.
\label{VbetaQM}
\end{align}

\vskip4mm

\paragraph{Comparison of the results.}

The functions $V_\beta(\widehat\sigma)$ (\ref{Vbeta2}) and $I_\lambda(\widehat\sigma)$ are to be integrated over $\mathfrak h$ with a ${\cal W}$-invariant measure (\ref{VbetaDEF2}) to give the expectation value of a vortex loop defined in two different ways. Taking account of the fact that the sum over the Weyl images is redundant inside the integral, one finds
\begin{equation}
 V_\beta(\widehat\sigma) \simeq I_\lambda(\widehat\sigma)\qquad\text{for}\qquad\lambda=\frac{k\beta}2.
\end{equation}
This gives a precise correspondence between the two definitions of a BPS vortex loop, namely the boundary condition $A_\varphi=\beta$ versus an ${\cal N}=2$ SUSY quantum mechanics with the target $M=\mathrm{Ad}_G(\lambda)$.

On the other hand, the insertion of a BPS Wilson loop in the representation with the highest weight $\lambda$ is described by $\mathrm{Tr}_\lambda e^{2\pi\widehat\sigma}=W_\lambda(\widehat\sigma)$ (\ref{Wlambda}). This function can be reproduced from a {\it non-supersymmetric} quantum mechanics with the target $M=\mathrm{Ad}_G(\lambda)$ and the action (\ref{WilsonQM}). Our computation shows that the partition functions of the bosonic and supersymmetric quantum mechanics with the same target $M=\mathrm{Ad}_G(\lambda)$ are slightly different:
\begin{equation}
I_{\lambda}(\widehat\sigma)=W_{\lambda-\tilde\rho}(\widehat\sigma) \,.
\end{equation}
Similar shift of parameter was noticed and studied in some earlier works \cite{Szabo:1996md,Fan:2018wya}. This result may look strange since the bosonic model was supersymmetrized by adding fermions as auxiliary fields. However, when computing $I_\lambda$ we perturbed the theory further by $L_\text{kin}$, and as a consequence the fermions became dynamical. In fact, the problem is similar to the evaluation of perturbative correction to the CS coupling of SUSY YM-CS theory \cite{Kao:1995gf}. For the simplest case $G=SU(2)$ it was shown by an explicit one loop analysis that the added fermions give rise to a shift of the spin $s$ by $-1/2$ \cite{Fan:2018wya}.

\subsection{Resolution of the unwanted parameter shift}\label{sec:match}

As we have seen, there is a subtle difference between the bosonic and ${\cal N}=2$ theories which appears as the shift $\lambda\to\lambda-\tilde\rho$ in the formulae for observables. Here we would like to argue that one can (and should) nevertheless relate the Wilson and vortex loops in ${\cal N}=2$ theory by the same formula $\lambda=k\beta/2$ as in bosonic theory. For this purpose, we need to explain the effect of the added fermions in more detail.

It is worth noting that the partition function $I_\lambda$ of the ${\cal N}=2$ SUSY quantum mechanics agrees precisely with that of geometric quantization with the so-called {\it metaplectic correction} taken into account. The importance of metaplectic correction is often skipped over, but when applied to the system of harmonic oscillator, it gives the correct account of its zero-point energy from the requirement of internal consistency alone. The origin of the metaplectic correction can be understood by studying how the quantum Hilbert spaces corresponding to different polarizations are related to each other, and in particular how the group of canonical transformations (the symplectic group) is represented. See for example \cite{Woodhouse:1980pa} for more detail. The upshot is that, if the quantum Hilbert spaces are constructed from the space of sections of the prequantum bundle $B$, the symplectic group will be represented only projectively. But it can be improved by replacing $B$ by $B\otimes K^{1/2}$, where $K$ is the canonical bundle of the target space $M$. Note that $K$ does not always have a well-defined square root, and $K^{1/2}$ may not be unique even if it exists.

Let us calculate the metaplectic correction for the case $M=\mathrm{Ad}_G(\lambda)$. Since the correction should preserve the property of $M$ as a homogeneous manifold with $G$-symmetry, it should at most modify the parameter $\lambda$. Take a function $\sigma_iP_i$ and the vector field $\sigma_iX_i$ in (\ref{PXonM}), and consider the action of the corresponding operator $\sigma_i\hat P_i$ on quantum wave functions in the holomorphic polarization. Before the metaplectic correction, $\sigma_i\hat P_i$ is the following differential operator near $z^\alpha=0$.
\begin{align}
 \sigma_i\hat P_i &~=~ -i\sum_{\alpha\in\Pi^+}\sigma_iX_i^\alpha\Big(\frac{\partial}{\partial z^\alpha}-i\vartheta_\alpha\Big)+\sigma_iP_i
\nonumber \\ &~=~ \lambda\dotp\sigma-\sum_{\alpha\in\Pi^+}(\alpha\dotp\sigma)z^\alpha\frac\partial{\partial z^\alpha}.
\end{align}
After the metaplectic correction, wave functions transform differently under infinitesimal coordinate transformations. So the definition of the operator is also modified accordingly.
\begin{align}
 \sigma_i\hat P_i &~=~ -i\sum_{\alpha\in\Pi^+}\left[\sigma_iX_i^\alpha\Big(\frac{\partial}{\partial z^\alpha}-i\vartheta_\alpha\Big)+\frac{1}2\frac{\partial(\sigma_iX^\alpha_i)}{\partial z^\alpha}\right]+\sigma_iP_i
\nonumber \\ &~=~ \lambda\dotp\sigma-\frac12\sum_{\alpha\in\Pi^+}(\alpha\dotp\sigma)-\sum_{\alpha\in\Pi^+}(\alpha\dotp\sigma)z^\alpha\frac\partial{\partial z^\alpha}.
\end{align}
This shows that the shift $\lambda\to\lambda-\tilde\rho$ can indeed be explained by metaplectic correction.

Another important effect of the fermions in ${\cal N}=2$ SUSY quantum mechanics is the global anomaly \cite{Hori:2014tda}. The fact that the highest weight $\lambda$ receives quantum correction implies that the $G$-symmetry of the quantum mechanics may be anomalous, because $\lambda-\tilde\rho$ is not always a weight of $G$. The anomaly arises from quantization of the fermions. Consider a theory with fermions $\chi,\bar\chi$ valued in linear spaces $V_\text{F}, V_\text{F}^\ast$ and a Lagrangian of the form
\begin{equation}
 L=i\bar\chi\nabla_t\chi+\cdots.
\end{equation}
Quantization of the fermions leads to the Hilbert space of fermionic states
\begin{equation}
 {\cal H}_\text{F} = \mathrm{det}^{-\frac12}V_\text{F}\otimes\wedge V_\text{F}.
\end{equation}
If $V_\text{F}$ represents a symmetry, then the symmetry has an anomaly unless $\mathrm{det}^\frac12V_\text{F}$ gives a well-defined one-dimensional representation. For ${\cal N}=2$ SUSY non-linear sigma model (NLSM) with the target space $M$ discussed in Section \ref{sec:QM-Adj-Orb}, the fermions $\chi$ take values on the pull back of the holomorphic tangent bundle $T_M$ by the boson $z$. The Hilbert space of this model is thus identified with the space of sections of the bundle
\begin{equation}
 K^{1/2}\otimes \wedge T_M \otimes B.
\label{hilbN2}
\end{equation}
The model has an anomaly unless this is a well-defined vector bundle. Note the similarity of (\ref{hilbN2}) with the metaplectic correction. As an example, for the case $M=S^2$ with $\omega=s\sin\theta\mathrm d\theta\mathrm d\varphi$ one can show by canonical quantization that the Hilbert spaces of the bosonic and ${\cal N}=2$ supersymmetric NLSMs are spanned by monopole harmonics \cite{Fan:2018wya}. They can therefore be decomposed into irreducible representations of $SU(2)$:
\begin{align}
{\cal H}_{{\cal N}=0} &= \bigoplus_{n\in\mathbb Z_{\ge0}}(\text{spin}~s+n),
 \nonumber \\
{\cal H}_{{\cal N}=2} &= \bigg[\bigoplus_{n\in\mathbb Z_{\ge0}}(\text{spin}~s-\tfrac12+n)\bigg]_\text{boson}\;\oplus\;\bigg[\bigoplus_{n\in\mathbb Z_{\ge0}}(\text{spin}~s+\tfrac12+n)\bigg]_\text{fermion}\,.
\end{align}
Note that these Hilbert spaces are for NLSMs which have a mixture of the first and second order kinetic terms for bosons. As the second order kinetic term is turned off, only the representation with the lowest spin remains and others are all lifted up to extremely high energy. This is another way to see the shift $s\to s-1/2$.

The global anomaly in ${\cal N}=2$ SUSY quantum mechanics can be canceled by turning on a suitable Wilson line \cite{Hori:2014tda}. This is because the introduction of a Wilson line with charge $q$,
\[
\exp\left(-\int \mathrm dt L_\text{WL}\right)= \exp\left(iq\int\mathrm dt A_t\right),
\]
has an effect to shift the charge of all the states uniformly by $q$. In fact, $L_\text{top}$ (\ref{Ltop1d}) can be regarded as a Wilson line in which the pull back of $\vartheta$ plays the role of $A_t$. This can be used to cancel the unwanted shift of $\lambda$ while maintaining the relation $\lambda=k\beta/2$. We define the BPS vortex loop with vorticity $\beta$ by a 1D ${\cal N}=2$ SUSY quantum mechanics with the target $M=\mathrm{Ad}_G(\lambda),~\lambda=k\beta/2$ and the Wilson line which precisely cancels the shift $\lambda\to\lambda-\tilde\rho$. As we will see in the next sections, this definition turns out to be more convenient when describing the quantum mechanics on vortex loops in terms of gauged linear sigma models.

\paragraph{An example: $\mathbb C\mathbb P^{N-1}$.}

We close this section with one concrete example. Take $G=SU(N)$ and
\begin{equation}
\lambda=m\left(\frac{N-1}N,-\frac1N,\cdots,-\frac1N\right)\in\mathfrak h^\ast,\qquad
k\beta = m\cdot\mathrm{diag}\Big(\frac{N-1}N,-\frac1N,\cdots,-\frac1N\Big)\in\mathfrak h.
\label{lambdaSUN}
\end{equation}
The corresponding adjoint orbit is $\mathbb C\mathbb P^{N-1}$ with the prequantum bundle $B={\cal O}(m)$. The quantum mechanical partition function is supposed to reproduce the character for the $m$-th symmetric tensor representation of $SU(N)$.

We start from the Euclidean action (\ref{N2-SCSQMg}) for the vortex loop along $S^1_{(\tau)}$ at $\theta=0$:
\begin{equation}
 S ~=~ k\int\mathrm d\tau\mathrm{Tr}\Big[\beta g^{-1}\Big(\frac{\mathrm d}{\mathrm d\tau}-iA_\tau-\widehat\sigma\Big)g\Big]\,.
\label{SSUN}
\end{equation}
We assume that the values of the 3D vectormultiplet fields $A_\tau$ and $\widehat\sigma=\ell\sigma$ are constant, and they take the following diagonal form.
\begin{equation}
 A_\tau = \mathrm{diag}(A^0_\tau,\cdots,A^{N-1}_\tau),\qquad
\widehat\sigma=\mathrm{diag}(\widehat\sigma^0,\cdots,\widehat\sigma^{N-1}).
\end{equation}
Let $\bar{\bf Z}\equiv(\bar Z_0,\cdots,\bar Z_{N-1})^\text{T}$ be the first column of $g$ and ${\bf Z}\equiv(Z_0,\cdots,Z_{N-1})$ the first row of $g^{-1}$. The above action can be rewritten as
\begin{equation}
 S = m\int\mathrm d\tau\,{\bf Z}\Big(\frac{\mathrm d}{\mathrm d\tau}-iA_\tau-\widehat\sigma\Big)\bar{\bf Z},\qquad |{\bf Z}|^2=1.
\end{equation}
The field $\bf Z$ transforms as anti-fundamental of the $SU(N)$. One can regard it as the homogeneous coordinate on $\mathbb C\mathbb P^{N-1}$. In terms of $z^\mini{I}\equiv Z_\mini{I}/Z_0$ the above action can be further rewritten as
\begin{equation}
 S = \int\mathrm d\tau\left\{-i(\vartheta_\mini{I}\nabla_\tau z^\mini{I}+\vartheta_{\bar{\mini{J}}}\nabla_\tau\bar z^{\bar{\mini{J}}})-\widehat\sigma^aP^a\right\},
\end{equation}
from which one can read off the 1-form $\vartheta$, Killing vector $X^a$ and the moment map $P^a$.
\begin{align}
\vartheta &= \frac{im}2\,\frac{z^\mini{I}\mathrm d\bar z^{\bar{\mini{I}}}-\mathrm dz^\mini{I}\bar z^{\bar{\mini{I}}}}{1+z^\mini{I}\bar z^{\bar{\mini{I}}}},
\nonumber \\
X^{a\mini{I}}\partial_\mini{I} &= i\big(T^a_\mini{IJ}z^\mini{I}+T^a_{0\mini{J}}-T^a_{\mini{I}0}z^\mini{I}z^\mini{J}-T^a_{00}z^\mini{J}\big)\partial_\mini{J},
\nonumber \\[.8mm]
P^a &= \frac{m}{1+z^\mini{I}\bar z^{\bar{\mini{I}}}}\Big\{(T^a)_\mini{I\bar J}z^\mini{I}\bar z^\mini{\bar J}+(T^a)_{\mini{I}0}z^\mini{I}+(T^a)_{0\mini{\bar J}}\bar z^\mini{\bar J}+(T^a)_{00}\Big\}\,.
\end{align}
Here $T^a$ are $N\times N$ matrices representing the generators of $SU(N)$, and $I,\bar{J}=1,\cdots,{N-1}$.

The supersymmetrized theory has $N$ saddle points. One of them corresponds to ${\bf Z}=(1,0,\cdots,0)$, and the others are all related to it by permutations of the $N$ components. The classical value of the action on this saddle point is $S=-2\pi m(\widehat\sigma^0+iA_\tau^0)$. The localizing Lagrangian near $z^\mini{I}=\bar z^\mini{\bar I}=0$ looks like
\begin{align}
 L_\text{kin}&\simeq m\sum_{I=1}^{N-1}\Big[\big\{\dot{\bar z}^\mini{\bar I}-i(A_\tau^\mini{I}-A_\tau^0)\bar z^{\bar{\mini{I}}}\big\}\big\{\dot z^\mini{I}+i(A_\tau^\mini{I}-A_\tau^0)z\big\}+(\widehat\sigma^\mini{I}-\widehat\sigma^0)^2\bar z^{\bar{\mini I}}z^\mini{I}
\nonumber \\ & \hskip18mm -i\bar\chi^{\bar{\mini I}}\big\{\dot\chi^\mini{I}+i(A_\tau^\mini{I}-A_\tau^0)-(\widehat\sigma^\mini{I}-\widehat\sigma^0)\chi^\mini{I}\}\Big]\,.
\end{align}
So the contribution to partition function from this saddle point is
\[
 e^{-S}\prod_{I=1}^{N-1}\frac{\mathrm{Det}\Big[\frac{\mathrm d}{\mathrm d\tau}+i(A_\tau^\mini{I}-A_\tau^0)-(\widehat\sigma^\mini{I}-\widehat\sigma^0)\Big]}{\mathrm{Det}\Big[\big(\frac{\mathrm d}{\mathrm d\tau}+i(A_\tau^\mini{I}-A_\tau^0)\big)^2-(\widehat\sigma^\mini{I}-\widehat\sigma^0)^2\Big]}~=~
 \frac{e^{2\pi m\widehat u^0}}{\prod_{I=1}^{N-1}2\sinh\pi(\widehat u^0-\widehat u^\mini{I})},
\]
where $\widehat u\equiv\widehat\sigma+iA_\tau$. It depends holomorphically on $\widehat u$, which is as expected because we started from the action (\ref{SSUN}). Summing up the contributions from all saddle points one obtains the full partition function
\begin{equation}
 \int{\cal D}g e^{-S} ~=~ \sum_{w\in{\cal W}/{\cal W}_K}\frac{e^{2\pi\lambda\dotp w(\widehat u)}}{\prod_{\alpha\in\Pi^+}2\sinh\pi\alpha\dotp w(\widehat u)}~=~ \sum_{I=0}^{N-1}\frac{e^{2\pi m\widehat u^\mini{I}}}{\prod_{J\ne I}2\sinh(\widehat u^\mini{I}-\widehat u^\mini{J})}
\,.
\end{equation}
This is {\it not} the character for the $m$-th symmetric tensor representation of $SU(N)$. One way to fix the mismatch would be to start with the orbit of $\lambda+\tilde\rho$ instead of $\lambda$, where
\begin{equation}
 \tilde\rho\;=\;\frac12\sum_{\alpha\in\Pi^+}\alpha \;=\; \frac12\sum_{I=1}^{N-1}({\bf e}_0-{\bf e}_\mini{I})\;=\;\left(\frac{N-1}2,-\frac12,\cdots,-\frac12\right).
\end{equation}
In other words, replace $m$ by $m+N/2$ at the beginning. Our resolution is not to shift $m$, but to cancel the anomaly by turning on the Wilson line with ``charge'' $N/2$.

\section{GLSM on vortex loops}\label{sec:GLSM-Vortex}

In the remaining sections we develop further the description of vortex loops as 1D-3D coupled systems using gauged linear sigma models (GLSMs). These models generally have an independent 1D gauge symmetry in addition to the (global) $G$ symmetry that is gauged by the 3D vectormultiplet. We will see that the Wilson line that cancels the global anomaly for this 1D gauge symmetry naturally resolves the problem of the unwanted shift $\lambda\to\lambda-\tilde\rho$.

We begin by reviewing 1D ${\cal N}=2$ supersymmetric GLSMs and an exact formula for the Witten indices.

\subsection{1D ${\cal N}=2$ SUSY GLSMs}\label{sec:1DGLSM}

A 1D ${\cal N}=2$ supersymmetric GLSM consists of a vectormultiplet $(A_t,\sigma,\lambda,\bar\lambda,D)$ (\ref{1Dvec}) for some gauge group $G$ and matter chiral multiplets $(\phi,\psi)$ and Fermi multiplets $(\eta,F)$ in some representations of $G$. The fields in chiral and Fermi multiplets transform under SUSY as
\begin{alignat}{2}
\SUSY\phi &= \epsilon\psi,\qquad&
\SUSY\psi &= \bar\epsilon(-i\nabla_t\phi+i\sigma\phi), \nonumber\\
\SUSY\bar\phi &= \bar\epsilon\bar\psi,\qquad&
\SUSY\bar\psi &= \epsilon(-i\nabla_t\bar\phi-i\bar\phi\sigma),\nonumber \\
\SUSY\eta &= \epsilon F+\bar\epsilon E,\qquad&
\SUSY F &= \bar\epsilon(-i\nabla_t\eta+i\sigma\eta-\Psi),\nonumber \\
\SUSY\bar\eta &= \bar\epsilon\bar F+\epsilon\bar E,\qquad&
\SUSY\bar F &= \epsilon(-i\nabla_t\bar\eta-i\bar\eta\sigma-\bar\Psi).
\label{SUSYGLQM}
\end{alignat}
Here $E$ is a composite field made only of chiral fields of the theory and $\Psi$ is its superpartner.

There are various $\SUSY$-invariants which can be used for Lagrangian. First, there are kinetic terms for the three multiplets,
\begin{align}
 L_\text{v} &= \mathrm{Tr}\big[(\nabla_t\sigma)^2-i\bar\lambda\nabla_t\lambda+i\bar\lambda[\sigma,\lambda]+D^2\big],\nonumber\\
 L_\text{c} &= \nabla_t\bar\phi\nabla_t\phi-i\bar\psi\nabla_t\psi+\bar\phi\sigma^2\phi-i\bar\phi D\phi-i\bar\psi\sigma\psi-i\bar\phi\lambda\psi-i\bar\psi\bar\lambda\phi,\nonumber \\
 L_\text{f} &= -i\bar\eta\nabla_t\eta+i\bar\eta\sigma\eta-\bar FF+\bar EE-\bar\eta\Psi+\bar\Psi\eta.
\label{LkinGLQM}
\end{align}
Also, supersymmetric interaction terms of chiral multiplets $(\phi_i,\psi_i)$ and Fermi multiplets $(\eta_i,F_i;E_i)$ can be constructed according to the formula:
\begin{equation}
 L_\text{int} = \sum_i\bigg(J_iF_i+\bar J_i\bar F_i\bigg)+\sum_{i,j}\bigg(\psi_j\frac{\partial J_i}{\partial\phi_j}\eta_i+\bar\psi_j\frac{\partial{\bar J}_i}{\partial\phi_j}\bar\eta_i\bigg),
\end{equation}
where $J_i$ is a composite of chiral fields such that $\sum_iJ_iE_i=0$. This can be regarded as the F-term of the Fermi multiplet with the lowest component (superpotential) $W=\sum_iJ_i\eta_i$. In addition, for $U(1)$ vectormultiplets, the Fayet-Iliopoulos term (with coupling $\zeta$) and the Wilson line (with charge $q$) are also invariant.
\begin{equation}
 L_\text{FI} = i\zeta D,\qquad
 L_\text{WL}= -q(iA_t+\sigma).
\end{equation}

An important role of Wilson lines in 1D GLSMs is to cancel global anomaly. Sometimes Wilson lines with fractional charges become necessary. For example, for a $U(N)$ gauge theory with $N_\text{f}$ fundamental chirals, $N_\text{a}$ anti-fundamental chirals, $\widetilde N_\text{f}$ fundamental Fermis and $\widetilde N_\text{a}$ anti-fundamental Fermis, the diagonal $U(1)$ subgroup is anomaly free if the Wilson line with the following $U(1)$ charge $q$ is added.
\begin{equation}
q\in -\frac12(N_\text{f}-N_\text{a}+\widetilde N_\text{f}-\widetilde N_\text{a})+\mathbb Z.
\end{equation}

\paragraph{Witten index.}

A powerful formula for the Witten index of 1D ${\cal N}=2$ GLSMs was obtained in \cite{Hori:2014tda}. The derivation uses the localization of path integral that follows from the $\SUSY$-exactness of the Lagrangians (\ref{LkinGLQM}).

At saddle points, $\sigma$ and $A_t$ are mutually commuting constants and all other fields must vanish. One can gauge-rotate $\sigma$ into a Cartan subalgebra $\mathfrak h\subset \mathrm{Lie}(G)$, and $A_t$ then takes values in the corresponding maximal torus. The pair $(\sigma,A_t)$ is further subject to the identification by the action of Weyl group ${\cal W}$. The space of saddle points thus becomes a real $2r$-dimensional orbifold, where $r=\mathrm{rk}(G)$. It is useful to define a complex coordinate $u\equiv\sigma+iA_t$ on this space. At this stage, one may also deform the theory by gauging its global symmetry $G_\mathrm{F}$ by a background vectormultiplet $\widehat u\equiv\widehat\sigma+i\widehat A_t$ satisfying the saddle point condition. Also, for convenience we rescale all the fields and the coordinate $t$ so that the time circle has unit radius.

The index can be obtained by evaluating the one-loop determinant $\Delta(u,\widehat u)$, multiplying by the Wilson line $e^{-S_\text{WL}}$ and then integrating over $u$. Due to the fact that $u$ is $\SUSY$-closed but $\bar u$ is not, the index $I(\widehat u)$ is expressed (up to an overall $\pm$ sign) as a multiple contour integral of a holomorphic function
\begin{align}
I(\widehat u) &~=~ \frac1{i^r|{\cal W}|}\int\mathrm d^ru\, e^{-S_\text{WL}(u)}\Delta(u,\widehat u),
\nonumber\\[1mm]
\Delta(u,\widehat u) &~=~ \frac{\displaystyle\prod_{\alpha}2\sinh\pi(\alpha\dotp u)\prod_{i}2\sinh\pi\big(\nu_i\dotp u+\widehat\nu_i\dotp\widehat u\big)}{\displaystyle\prod_j2\sinh\pi\big(\mu_j\dotp u+\widehat\mu_j\dotp\widehat u\big)}.
\label{Delta1D}
\end{align}
Here $(\mu_j,\widehat\mu_j)$ runs over the weights of the representation of $G\times G_\mathrm{F}$ furnished by chiral multiplets, and similarly $(\nu_i,\widehat\nu_i)$ is for the Fermi multiplets.

The contour integral can be performed using the operation called the {\it Jeffrey-Kirwan (JK) residue}, which means that one only has to collect residue of the poles meeting certain requirement \cite{Benini:2013nda,Benini:2013xpa}. To simplify the discussion, let us assume that all the poles of $\Delta$ are transverse intersection of $r$ singular hyperplanes. Each singular hyperplane is of the form
\[
 \mu_j\dotp u+\widehat\mu_j\dotp\widehat u=ik\,\quad \big(k\in\mathbb Z\big),
\]
and is labeled by a charge vector $\mu_j\in\mathfrak h^\ast$. Now, the evaluation of JK-residue integral begins by choosing an arbitrary reference charge vector $\delta\in\mathfrak h^\ast$. Then a pole contributes to the integral if $\delta$ is contained in the cone spanned by the $r$ charge vectors labeling the pole. Note that the set of poles contributing to the integral depends on the choice of $\delta$, but the final result of the integral is independent of $\delta$.

The function $\Delta(u,\widehat u)$ has poles in the interior of the space of saddle points as well as at infinity. As was studied in detail in \cite{Hori:2014tda}, the residue of the pole at infinity may or may not contribute depending on the choice of $\delta$ as well as the value of the FI coupling $\zeta$. In particular, they do not contribute if $\delta$ is set equal to $\zeta$, so it is customary to set $\delta$ as such when studying Witten indices of 1D GLSMs. Note that this implies that the Witten indices do depend on $\zeta$ although the FI Lagrangian is $\SUSY$-exact. The GLSMs in general are known to exhibit different behavior depending on the values of $\zeta$, and accordingly the space of FI couplings is divided into several regions or ``phases''. The index may jump as $\zeta$ is varied across phase boundaries. See \cite{Hori:2014tda} for more detail.

\paragraph{An example: $\mathbb C\mathbb P^{N-1}$.}

The GLSM is given by a $U(1)$ gauge theory with $N$ chiral multiplets of charge $+1$ and a positive FI coupling. We turn on the Wilson line with charge $q$ and gauge the flavor $SU(N)$ symmetry by a constant background vectormultiplet $\widehat u=\mathrm{diag}(\widehat{u}^{0},\cdots,\widehat{u}^{N-1})$. The Witten index is then given by a contour integral
\begin{equation}
 I(\widehat u) ~=~ \int\frac{\mathrm du}{i}\frac{e^{2\pi qu}}{\prod_{J=0}^{N-1}2\sinh\pi(u-\widehat u^\mini{J})}~=~ \sum_{I=0}^{N-1}\frac{e^{2\pi q\widehat u^\mini{I}}}{\prod_{J\ne I}2\sinh\pi(\widehat u^\mini{I}-\widehat u^\mini{J})}\,.
\end{equation}
The JK-residue integral picks up the contribution of all the $N$ poles $u=\widehat u^\mini{I}$. Without the Wilson line, the integrand is {\it not} invariant under a large gauge transformation $u\to u+i$ for odd $N$. This is an example of global anomaly. To obtain the character for the $m$-th symmetric tensor representation of $SU(N)$ one has to set $q=m+N/2$. We would like to view it as the model with $q=m$ whose anomaly is canceled by the additional Wilson line with $q=N/2$.

\subsection{GLSM for vortex worldline quantum mechanics}\label{sec:GLSM-defect}

Let us now turn to the SUSY quantum mechanics on the worldline of vortex loops. We first consider the case where the 3D gauge theory is made of vectormultiplet only. So we take the 3D ${\cal N}=2$ CS theory with $G=SU(N)$ at level $k$, and put a vortex loop with
\begin{align}
 \beta ~=~ \mathrm{diag}(\beta_1,\cdots,\beta_N)~=~&\mathrm{diag}\big(\,\underbrace{\beta_{(1)},\cdots,\beta_{(1)}}_{n_1}\,,\,\underbrace{\beta_{(2)},\cdots,\beta_{(2)}}_{n_2}\,,\,\cdots\,,\,\underbrace{\beta_{(p)},\cdots,\beta_{(p)}}_{n_p}\,\big)\,,
\nonumber \\ &
\beta_{(1)}>\beta_{(2)}>\cdots>\beta_{(p)}
\label{betaSUN}
\end{align}
which breaks $G$ to $K=S[U(n_1)\times\cdots\times U(n_p)]$. For later use let us introduce
\[
 N_0=0,\quad N_1=n_1,\quad N_2=n_1+n_2,\quad\cdots\quad N_p=n_1+\cdots+n_p=N.
\]
The quantum mechanics on the vortex worldline is a 1D ${\cal N}=2$ theory with a global symmetry $G=SU(N)$ which is gauged by the 3D vectormultiplet fields. Also, its Witten index should reproduce the corrected version of (\ref{VbetaQM}):
\begin{equation}
 V_\beta(\widehat\sigma) = \sum_{w\in{\cal W}/{\cal W}_K} V_\beta^\text{v}(w(\widehat\sigma)),\qquad
V^\text{v}_\beta(\widehat\sigma)~=~\frac{e^{2\pi\sum_i(\lambda+\tilde\rho)_i\widehat\sigma_i}}{\prod_{\beta_i>\beta_j}2\sinh\pi(\widehat\sigma_i-\widehat\sigma_j)},
\label{VbetaSUN}
\end{equation}
where ${\cal W},{\cal W}_K,\tilde\rho,\Pi^+$ are defined around (\ref{rhotilde}) and $\lambda,\tilde\rho$ are $N$-component vectors
\begin{alignat}{2}
\lambda= (\lambda_1,\cdots,\lambda_N) &= \big(\,\underbrace{\lambda_{(1)},\cdots,\lambda_{(1)}}_{n_1}\,,\,\cdots\,,\,\underbrace{\lambda_{(p)},\cdots,\lambda_{(p)}}_{n_p}\,\big)\,,\quad&\lambda_{(a)}&=k\beta_{(a)}\,,\nonumber\\[-2mm] 
\tilde\rho=(\tilde\rho_1,\cdots,\tilde\rho_N)&= \big(\,\underbrace{\tilde\rho_{(1)},\cdots,\tilde\rho_{(1)}}_{n_1}\,,\,\cdots\,\,,\,\underbrace{\tilde\rho_{(p)},\cdots,\tilde\rho_{(p)}}_{n_p}\,)\,,\quad&\tilde\rho_{(a)}&=\frac12(N-N_a-N_{a-1})\,. 
\end{alignat}
By noticing that each $w\in {\cal W}/{\cal W}_K$ is in one-to-one correspondence with a division of $\{1,\cdots,N\}$ into subsets $d_1,\cdots,d_p$ of order $|d_a|=n_a$, (\ref{VbetaSUN}) can also be written as the sum over divisions
\begin{equation}
 V_\beta(\widehat\sigma) = \sum_{\{d_1,\cdots,d_p\}}\frac{e^{2\pi\sum_i(\lambda+\tilde\rho)_i\widehat\sigma_i}}{\prod_{a<b}\prod_{i\in d_a}\prod_{j\in d_b}2\sinh\pi(\widehat\sigma_i-\widehat\sigma_j)}.
\label{VbetaSUN2}
\end{equation}

The $V_\beta(\widehat\sigma)$ in (\ref{VbetaSUN}) or (\ref{VbetaSUN2}) equals the character for the representation of $SU(N)$ with the highest weight $\lambda$. The same character formulae work also for $G=U(N)$ by relaxing the tracelessness condition for $\widehat\sigma_i$ and modifying the quantization condition for $\lambda_i$. To be more explicit, recall that we have described the highest weights of $SU(N)$ representations as $N$-component vectors $\lambda=(\lambda_1,\cdots,\lambda_N)$ satisfying
\[
 \lambda_i-\lambda_j\in\mathbb Z_{\ge0}~~(i>j),\qquad
 \sum_{i=1}^N\lambda_i=0.
\]
So, $\lambda_i$ are all equal modulo $\mathbb Z$ to $m/N$ for some integer $m$ which gives the charge of the representation under the central subgroup $\mathbb Z_N\subset SU(N)$. The highest weight of a $U(N)$ representation is obtained from that of an $SU(N)$ representation $\lambda$ by a uniform shift of $\lambda_i$ to make them all integer.

\paragraph{A GLSM and its quiver representation.}

\begin{figure}[ht]
\begin{center}
\begin{overpic}[scale=1.5]{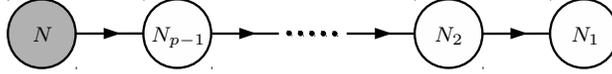}
\put(5.7,6){\scriptsize $N$}
\put(24.5,6){\scriptsize $N_{p-1}$}
\put(69.2,6){\scriptsize $N_2$}
\put(90.8,6){\scriptsize $N_1$}
\end{overpic}
\end{center}
\caption{The quiver diagram for a GLSM  on a vortex loop. It has a 1D ${\cal N}=2$ vectormultiplet for each node and a bifundamental chiral multiplet for each solid line connecting the neighboring nodes. The shaded node represents the 3D gauge symmetry.}
\label{fig:quiver1}
\end{figure}

The GLSM for flag manifolds has been discussed in many places; see \cite{Donagi:2007hi,Gadde:2013dda} for example. Here we study the 1D ${\cal N}=2$ version of it. The models can be conveniently described by the quiver diagram of Figure \ref{fig:quiver1}.
 It is a $U(N_{p-1})\times \cdots\times U(N_1)$ gauge theory with $N$ chiral multiplets in the anti-fundamental of $U(N_{p-1})$ and one bi-fundamental chiral multiplet for each neighboring pair of unitary groups, namely ${\bf N}_{a+1}\times\overline{\bf N}_a$ of $U(N_{a+1})\times U(N_a)$ for each $a\in\{1,\cdots,p-2\}$. The FI couplings for the diagonal $U(1)^{p-1}$ are chosen to be all negative. In addition, we turn on the following Wilson line for the $U(1)^{p-1}$:
\begin{equation}
 L_\text{WL} = -\sum_{a=1}^{p-1}q_a\mathrm{Tr}\big(iA_t^{(a)}+\sigma^{(a)}\big),\qquad
q_a = k(\beta_{(a)}-\beta_{(a+1)})+\frac12(N_{a+1}-N_{a-1}).
\end{equation}
The first term in the formula for $q_a$ is needed so that the model agrees with the adjoint orbit quantization with $\lambda = k\beta/2$. The second term is needed to cancel the global anomaly.

Let us denote the constant value of the $U(N_a)$ vectormultiplet fields at saddle points as
\[
 \sigma^{(a)}+iA_t^{(a)} = \mathrm{diag}(u_1^{(a)},\cdots,u_{N_a}^{(a)}).
\]
The index is then given by the JK residue integral of the holomorphic function
\begin{equation}
 e^{-S_\text{WL}(u)}\Delta(u,\widehat\sigma) ~=~
\frac{\displaystyle\exp\bigg(\sum_{a=1}^{p-1}\sum_{i=1}^{N_a}2\pi q_au_i^{(a)}\bigg)\cdot\prod_{a=1}^{p-1}\prod_{i\ne j}^{N_a}2\sinh\pi(u^{(a)}_i-u^{(a)}_j)}{\displaystyle\prod_{i=1}^N\prod_{j=1}^{N_{p-1}}2\sinh\pi(\widehat\sigma_i-u_j^{(p-1)})\prod_{a=1}^{p-2}\prod_{i=1}^{N_{a+1}}\prod_{j=1}^{N_a}2\sinh\pi(u^{(a+1)}_i-u^{(a)}_j)}\,.
\label{DeltaGLSM}
\end{equation}

At each pole of $\Delta$, the value of the variables $u_i^{(a)}$ are determined one by one through an iterated residue integral. At some of the poles, they are determined according to the following steps. First, each of $u_j^{(p-1)}\;(j=1,\cdots,N_{p-1})$ is set equal to one of $\{\widehat\sigma_1,\cdots,\widehat\sigma_{N}\}$. Their values must be all different so that the numerator of $\Delta$ is nonzero. Once $\{u^{(a)}_j\}_{j=1,\cdots,N_a}$ are determined, then the values of $\{u_j^{(a-1)}\}_{j=1,\cdots,N_{a-1}}$ are chosen in the same way as in the previous step, until all the $u_i^{(a)}$ are determined and a pole is thus specified. Each such pole corresponds to a division of $\{1,\cdots,N\}$ into subsets $d_1,\cdots,d_p$ of order $|d_a|=n_a$. There are $\prod_{a=1}^{p-1}N_a!$ different poles corresponding to the same division, and they all have the same residue. As we will explain shortly, for negative FI couplings these are the only poles which contribute to the JK-residue integral.

The index $I(\widehat\sigma)$ of the GLSM thus obtained is related to $V_\beta(\widehat\sigma)$ (\ref{VbetaSUN}) as follows:
\begin{equation}
 V_\beta(\widehat\sigma) ~=~ I(\widehat\sigma)\cdot e^{2\pi q\sum_{i=1}^N\widehat\sigma_i}~=~ I(\widehat\sigma)\cdot W_q(\widehat\sigma),
\label{IvsV}
\end{equation}
where $q=k\beta_{(p)}-\frac12N_{p-1}$. The index reproduces $V_\beta(\widehat\sigma)$ precisely for $G=SU(N)$. If the 3D gauge group is $G=U(N)$, the GLSM has to be accompanied by a Wilson line of charge $q$ for the diagonal $U(1)$ subgroup of $U(N)$.

\paragraph{Detail of JK-residue integral (1).} Here we explain some detail of the JK residue integral for our present problem. Let us denote by $\{{\bf e}_i^{(a)}\}^{a=1,\cdots,p-1}_{i=1,\cdots,N_a}$ the basis vectors for the space of charges. The singular hyperplanes of $\Delta$ (\ref{DeltaGLSM}) are then labeled by the charge vectors of the form
\begin{equation}
 {\bf q}_j\equiv -{\bf e}_j^{(p-1)}\qquad\text{or}\qquad
 {\bf q}_{ij}^{(a)}\equiv {\bf e}_i^{(a+1)}-{\bf e}_j^{(a)}.
\label{charges1}
\end{equation}
The dimension of the space of charges is $r=\sum_{a=1}^{p-1}N_a$.

At each pole, the values of $u_i^{(a)}$ are determined one by one through an iterated residue integral. The process can be regarded as if the $u$-variables are connected together into some trees each starting at one of the $\widehat\sigma_i$'s. At the same time, a set $\Pi$ of $r$ charge vectors are chosen from (\ref{charges1}), and all the basis vectors ${\bf e}_i^{(a)}$ are expressed as their linear combinations. As an example, take $N=4, (N_3,N_2,N_1)=(3,2,1)$ and consider a pole
\begin{alignat}{4}
\widehat\sigma_1 \;=\;& u_1^{(3)} &&,\nonumber\\
\widehat\sigma_2 \;=\;& u_2^{(3)} &&\;=\;u_1^{(2)}&&\;=\; u_1^{(1)},\nonumber\\
\widehat\sigma_4 \;=\;& u_3^{(3)} &&\;=\;u_2^{(2)}&&.
\label{trees}
\end{alignat}
Then all the basis vectors ${\bf e}_i^{(a)}$ are expressed as linear combinations of the 6 charge vectors in $\Pi=\{{\bf q}_1,{\bf q}_2,{\bf q}_3,{\bf q}_{21}^{(2)},{\bf q}_{32}^{(2)},{\bf q}_{11}^{(1)}\}$:
\begin{alignat}{3}
{\bf e}_1^{(3)} &= -{\bf q}_1,\qquad&
\nonumber \\
{\bf e}_2^{(3)} &= -{\bf q}_2,\qquad&
{\bf e}_1^{(2)} &= -{\bf q}_2-{\bf q}^{(2)}_{21},\qquad&
{\bf e}_1^{(1)} &= -{\bf q}_2-{\bf q}^{(2)}_{21}-{\bf q}^{(1)}_{11},
\nonumber \\
{\bf e}_3^{(3)} &= -{\bf q}_3,\qquad&
{\bf e}_2^{(2)} &= -{\bf q}_3-{\bf q}^{(2)}_{32}.
\label{paths}
\end{alignat}
The form of the trees can be read from (\ref{trees}), and the relations (\ref{paths}) indicate how each $u_i^{(a)}$ is connected to one of the $\widehat\sigma_i$'s by a unique path along the trees. In the above example, the elements of $\Pi$ always appear in the RHS of the relations (\ref{paths}) with {\it negative} coefficients because the trees have grown only in the decreasing direction of $a$. Since the reference charge vector $\delta$ is given by
\begin{equation}
 \delta=\sum_{a=1}^{p-1}\zeta_a\sum_{i=1}^{N_a}{\bf e}_i^{(a)},\qquad(\zeta_a<0)
\end{equation}
it is a {\it positive} linear combination of the elements of $\Pi$, and therefore the pole (\ref{trees}) contributes to the JK-residue integral. The same argument applies to all the poles described in the paragraph after (\ref{DeltaGLSM}): all of them contribute to the index since the corresponding trees extend only in the decreasing direction of $a$.

In fact, $\Delta$ (\ref{DeltaGLSM}) has other poles corresponding to (i) trees with branchings or (ii) trees part of which grow in the wrong direction. Both types of the poles have vanishing residues, but those of type (ii) are also excluded by the rule of JK-residue. If parts of the trees grow in the wrong direction, some elements of $\Pi$ appear in the expression for $\delta$ with wrong sign.

\paragraph{Another GLSM.}

\begin{figure}[ht]
\begin{center}
\begin{overpic}[scale=1.5]{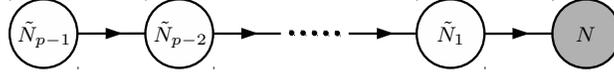}
\put(3,6){\scriptsize $\tilde N_{p-1}$}
\put(24.5,6){\scriptsize $\tilde N_{p-2}$}
\put(69.2,6){\scriptsize $\tilde N_1$}
\put(91.2,6){\scriptsize $N$}
\end{overpic}
\end{center}
\caption{The quiver diagram of another GLSM for the same flag manifold.}
\label{fig:quiver2}
\end{figure}

There is another GLSM whose Witten index reproduces $V_\beta(\widehat\sigma)$ (\ref{VbetaSUN}) up to sign. It has the gauge group $U(\tilde N_{p-1})\times\cdots\times U(\tilde N_1)$, where $\tilde N_a=N-N_a$, with one bifundamental chiral for each neighboring pair of unitary groups and $N$ chirals in the fundamental of $U(\tilde N_1)$. The model is described by the quiver diagram of Figure \ref{fig:quiver2}. The $p-1$ FI couplings are all chosen to be positive. In addition, we need Wilson line with the $U(1)^{p-1}$ charge
\begin{equation}
 \tilde q_a = k(\beta_{(a+1)}-\beta_{(a)})+\frac12(\tilde N_{a+1}-\tilde N_{a-1}).
\end{equation}
The index for this GLSM can be computed in the same way as in the previous model. It satisfies (\ref{IvsV}) with $q=k\beta_{(1)}+\frac12\tilde N_1$.

\paragraph{More alternatives.}

\begin{figure}[ht]
\begin{center}
\begin{overpic}[scale=1.5]{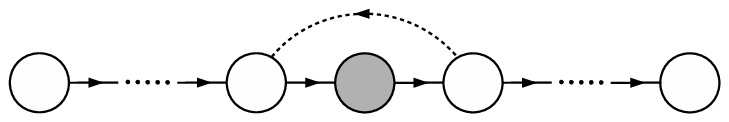}
\put(2.3,4.5){\scriptsize $\tilde N_{p-1}$}
\put(33.5,4.5){\scriptsize $\tilde N_s$}
\put(49,4.5){\scriptsize $N$}
\put(61.8,4.5){\scriptsize $N_{s-1}$}
\put(93,4.5){\scriptsize $N_1$}
\end{overpic}
\end{center}
\caption{The quiver diagram describing a series of GLSMs for the same flag manifold. The dashed line represents a bifundamental Fermi multiplet.}
\label{fig:quiver3}
\end{figure}

In addition to the two quiver theories presented above, there are two series of alternative quiver theory realizations for the sigma model whose target is the same flag manifold. The first is defined by the quiver diagram of Figure \ref{fig:quiver3}. In addition to the bifundamental chiral multiplets for neighboring pairs of nodes, the theory has a Fermi multiplet in ${\bf N}_{s-1}\times\overline{\tilde{\bf N}}_s$ of $U(N_{s-1})\times U(\tilde N_s)$. The FI couplings for the $U(N_a)$ are all negative while those for $U(\tilde N_a)$ are all positive. In addition, we need the Wilson line with $U(1)^{p-1}$ charge
\begin{alignat}{2}
q_a &= k(\beta_{(a)}-\beta_{(a+1)})+\frac12(N_{a+1}-N_{a-1}),\qquad&&(a=1,\cdots,s-1)\nonumber\\
\tilde q_a &= k(\beta_{(a+1)}-\beta_{(a)})+\frac12(\tilde N_{a+1}-\tilde N_{a-1}).\qquad&&(a=s,\cdots,p-1)
\label{qs}
\end{alignat}
The index of the model satisfies (\ref{IvsV}) with $q=k\beta_{(s)}+\frac12(\tilde N_s-N_{s-1})$.
This series interpolates the previous two GLSM descriptions.

\begin{figure}[ht]
\begin{center}
\begin{overpic}[scale=1.5]{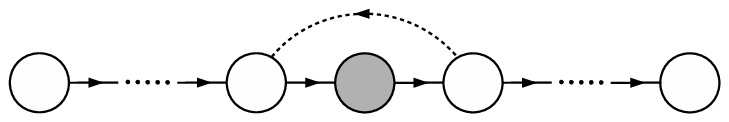}
\put(2.3,4.5){\scriptsize $\tilde N_{p-1}$}
\put(33.5,4.5){\scriptsize $\tilde N_s$}
\put(49,4.5){\scriptsize $N$}
\put(63.2,4.5){\scriptsize $N_s$}
\put(93,4.5){\scriptsize $N_1$}
\end{overpic}
\end{center}
\caption{Quivers for another series of GLSMs for the same flag manifold.}
\label{fig:quiver4}
\end{figure}

The second series of GLSMs is defined by the quiver diagrams of Figure \ref{fig:quiver4} which have one more node than the previous ones. The FI couplings for the $U(N_a)$ are all negative while those for $U(\tilde N_a)$ are all positive. In addition, we need the Wilson line with the following $U(1)^p$ charges:
\begin{alignat}{2}
q_a &= k(\beta_{(a)}-\beta_{(a+1)})+\frac12(N_{a+1}-N_{a-1})~(a<s),\quad&
q_s &= k(\beta_{(s)}-\beta_\ast)+\frac12(N-N_{s-1}-\tilde N_s),\nonumber \\
\tilde q_a &= k(\beta_{(a+1)}-\beta_{(a)})+\frac12(\tilde N_{a+1}-\tilde N_{a-1})~(a>s),\quad&
\tilde q_s &= k(\beta_{(s+1)}-\beta_\ast)+\frac12(\tilde N_{s+1}+N_s-N),
\end{alignat}
where $\beta_\ast$ is a parameter which is constrained only by the anomaly cancellation condition. The index of this model satisfies (\ref{IvsV}) with $q=k\beta_\ast+\frac12(\tilde N_s-N_s)$. Note that the second series for $\beta$ (\ref{betaSUN}) can be thought of as the first series for
\[
 \beta = \mathrm{diag}\big(\,\underbrace{\beta_{(1)},\cdots,\beta_{(1)}}_{n_1}\,,\,\cdots\,,\,\underbrace{\beta_{(s)},\cdots,\beta_{(s)}}_{n_s}\,,\,\underbrace{\beta_\ast,\cdots,\beta_\ast}_{0}\,,\,\underbrace{\beta_{(s+1)},\cdots,\beta_{(s+1)}}_{n_{s+1}}\,,\,\cdots\,,\,\underbrace{\beta_{(p)},\cdots,\beta_{(p)}}_{n_p}\,\big)\,.
\]

It is tempting to identify $\beta_{(a)}$'s as some kind of position coordinates. The formulae for $q_a,\tilde q_a$ suggest that the $a$-th gauge node (white node) corresponds to {\it branes} stretching between $\beta=\beta_{(a)}$ and $\beta=\beta_{(a+1)}$. One might also think that the position of the 3D gauge node (shaded node) should be determined according to the value of $\beta_{(a)}$'s, but this is not the case. The bulk 3D $U(N)$ gauge theory has fields in the adjoint representation only, so there are no fields charged under the diagonal $U(1)$ subgroup of $U(N)$. The expectation value of a vortex loop should therefore be invariant under the uniform shift $\beta_{(a)}\to\beta_{(a)}+c$. As we will see in the next section, the situation changes when matters in (anti-)fundamental representation of $U(N)$ are introduced.

\begin{figure}[ht]
\begin{center}
\begin{overpic}[scale=1.5]{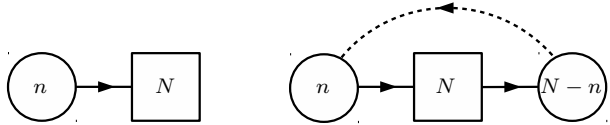}
\put(6,6.5){\scriptsize $n$}
\put(25.6,6.5){\scriptsize $N$}
\put(51.5,6.5){\scriptsize $n$}
\put(70.7,6.5){\scriptsize $N$}
\put(87.6,6.5){\scriptsize $N-n$}
\end{overpic}
\end{center}
\caption{The usual and alternative quivers of the GLSM for the Grassmannian $\text{Gr}(n,N)$.}
\label{fig:quiverGr}
\end{figure}

Let us briefly explain how these alternatives give the same flag manifolds as the space of classical vacua, by taking the example for the Grassmannian $\mathrm{Gr}(n,N)=U(N)/\big(U(n)\times U(N-n)\big)$. The usual model is the $U(n)$ gauge theory with $N$ fundamental chiral multiplets $A_{iI}~(i=1,\cdots,n,I=1,\cdots,N)$, as described by the quiver diagram of Figure \ref{fig:quiverGr} left. The classical vacuum equation is
\[
 \sum_IA_{iI}\bar A_{Ij} = \zeta\delta_{ij},
\]
where $\zeta$ is the FI coupling for the diagonal $U(1)$ subgroup of $U(n)$. For $\zeta>0$, each solution gives a set of $n$ orthonormal $N$-component complex vectors. The equivalence classes of solutions with respect to $U(n)$ define $n$-dimensional hyperplanes in $\mathbb{C}^N$, and the space of such hyperplanes is $\mathrm{Gr}(n,N)$.

The alternative model is the $U(n)\times U(N-n)$ gauge theory with $N$ chiral multiplets in the fundamental of $U(n)$, $N$ chiral multiplets in the anti-fundamental of $U(N-n)$ and a Fermi multiplet in the bifundamental of $U(N-n)\times U(n)$ as described by the quiver of Figure \ref{fig:quiverGr} right. Let us denote the bottom components of these multiplets as
\[
 A_{iI},~~B_{I\tilde\jmath},~~ \eta_{\tilde\jmath i}\qquad(i=1,\cdots,n\,;\,\tilde\jmath=1,\cdots,N-n\,;\,I=1,\cdots,N).
\]
In the presence of the superpotential $W = \sum_{i,\tilde\jmath,I}A_{iI}B_{I\tilde\jmath}\eta_{\tilde\jmath i}$, the classical vacuum equations are
\[
 \sum_IA_{iI}\bar A_{Ij} = \zeta\delta_{ij},\quad
 \sum_I \bar B_{\tilde\imath I}B_{I\tilde\jmath}=-\tilde\zeta\delta_{\tilde\imath\tilde\jmath},\quad
 A_{iI}B_{I\tilde\jmath}=0,
\]
where $\zeta,\tilde\zeta$ are the FI couplings for $U(n)$ and $U(N-n)$. If $\zeta>0$ and $\tilde\zeta<0$, each solution of these equations defines an $n$-plane and a $(N-n)$-plane in $\mathbb C^N$ that are orthogonal to each other. The space of such pairs is again given by $\mathrm{Gr}(n,N)$.

In view of the fact that many alternative GLSMs presented in this section give the same Witten index and vacuum manifold, we suspect they are all dual to one another.

\section{Theories with matters}\label{sec:add-matters}

Here we study vortex loops in 3D $U(N)$ gauge theories with various matter chiral multiplets. The path integral with respect to the added chiral multiplets on the vortex background modifies $V_\beta(\widehat\sigma)$ (\ref{VbetaSUN}) according to the formula in Section \ref{sec:exactPI}. We would like to find the corresponding modification of the quiver GLSMs introduced in the last section.

\paragraph{Global symmetry of the 1D theory.}

For a vortex loop in a theory with chiral multiplets of real mass $m$ and R-charge $r$, the function $V_\beta(\widehat\sigma)$ will also depend on $m,r$ and the squashing parameter $b$. Since $m$ is in a 3D vectormultiplet, $\widehat m\equiv\ell m$ appears in the 1D theory on the vortex worldline according to the same rule as that for $\widehat\sigma$. In fact, the other parameters $r,b$ also appear in the 1D theory through the background gauging of a specific global $U(1)$ symmetry.

The 3D ${\cal N}=2$ theory on an ellipsoid has the translation symmetry $U(1)_\tau\times U(1)_\varphi$ and the R-symmetry $U(1)_\text{R(3D)}$. The $U(1)_\tau$ descends to the translation symmetry along the vortex loop, whereas $U(1)_\varphi$ appears in the 1D theory as a global symmetry. The R-symmetry of the 1D ${\cal N}=2$ SUSY theory should be a linear combination of $U(1)_\varphi$ and $U(1)_\text{R(3D)}$ (and other abelian global symmetries if there are any). However, the Witten index is independent of the assignment of this R-charge on matters because the square of the 1D SUSY (\ref{1Dvec}), (\ref{SUSYGLQM}) does not contain the R-symmetry. But the index does depend on the charge assignments of the other non-R linear combination of $U(1)_\varphi$ and $U(1)_\text{R(3D)}$, as we now explain.

The SUSY of the 3D theory on an ellipsoid squares to
\[
 \SUSY_\text{(3D)}^2 = \frac1\ell\mathrm H+\frac1{\tilde\ell}\mathrm M-\frac12\left(\frac1\ell+\frac1{\tilde\ell}\right)\mathrm R_\text{(3D)}+i\Big(\sigma+\frac i\ell A_\tau+\frac i{\tilde\ell}A_\varphi\Big)+im,
\]
where $\mathrm H$ and $\mathrm M$ are operators that act on dynamical fields as $-i{\cal L}_{\partial_\tau}$ and $-i{\cal L}_{\partial_\varphi}$, respectively. In section \ref{sec:QM-Adj-Orb} we have made contact of this $\SUSY_\text{(3D)}$ with the 1D SUSY on the vortex worldline using the fact that the cohomological variables transform under $\SUSY_\text{(3D)}$ like 1D ${\cal N}=2$ multiplets. So, let us study the action of $\SUSY^2_\text{(3D)}$ on cohomological variables on top of the vortex worldline. As an example take $\Psi=\xi\psi$ (\ref{defcoh}) which is the superpartner of a chiral scalar $\phi$. With the understanding that $\mathrm H,\mathrm M,\mathrm R_\text{(3D)}$ act only on a dynamical field $\psi$ and not $\xi$, one finds
\begin{align}
 \SUSY_\text{(3D)}^2\Psi\Big|_{\theta=0} &~=~ \xi\cdot\left\{\frac1\ell\mathrm H+\frac1{\tilde\ell}\mathrm M-\frac12\left(\frac1\ell+\frac1{\tilde\ell}\right)\mathrm R_\text{(3D)}+i\Big(\sigma+\frac i\ell A_\tau\Big)+im\right\}\psi
\nonumber \\ &~=~ \frac i\ell\left\{-\partial_\tau+\frac{ibQ}2\big(\mathrm R_\text{(3D)}-2\mathrm M\big)+(\widehat\sigma+iA_\tau)+\widehat m\right\}\Psi.
\label{QPsi}
\end{align}
Here we used ${\cal L}_{\partial_\tau}\xi={\cal L}_{\partial_\varphi}\xi=\frac i2\xi$ and also that ${\cal L}_{\partial_\varphi}\Psi=0$ along the vortex worldline because $\Psi$ is a Lorentz scalar. The above computation works for all the cohomological variables. Thus the SUSY squared of the vortex worldline theory should take the form (here $t$ is the worldline coordinate of period $2\pi$):
\begin{align}
 \SUSY^2_\text{(1D)} &~\sim~ -\partial_t + \frac{ibQ}{2}\mathrm G +(\widehat\sigma+iA_\tau)+\widehat m + (\text{1D vectormultiplet fields})\,,\nonumber\\
\mathrm G &~\equiv~\mathrm R_\text{(3D)}-2\mathrm M\,.
\label{defG}
\end{align}
Note that $\mathrm G$ is a non-R global symmetry in the sense of both 3D and 1D. The interpretation of the second term in the RHS is that the global $U(1)_\text{G}$ symmetry of the vortex worldline theory is gauged by the background field:
\begin{equation}
 \sigma^\mini{\text{G}}+iA_t^\mini{\text{G}} = \frac{ibQ}2.
\end{equation}

\subsection{Adjoint representation}\label{sec:adj-matter}

Let us first consider the case with an adjoint chiral multiplet with mass $m$ and R-charge $r$. According to the result of Section \ref{sec:exactPI}, the function $V_\beta$ now consists of the contribution from vector and chiral multiplets:
\begin{equation}
 V_\beta(\widehat\sigma) = \sum_{w\in{\cal W}/{\cal W}_K}V^\text{v}_\beta(w(\widehat\sigma))V^\text{c}_\beta(w(\widehat\sigma)).
\label{VbetawMat1}
\end{equation}
Here $V_\beta^\text{v}(\widehat\sigma)$ is given in (\ref{VbetaSUN}) and
\begin{align}
 V_\beta^{\text{c}{\bf 1}}(\widehat\sigma) &= \prod_{\beta_i>\beta_j}2\sinh\pi\Big(\widehat\sigma_i-\widehat\sigma_j-\widehat m-\frac{irbQ}2\Big), \nonumber \\
 V_\beta^{\text{c}{\bf 2}}(\widehat\sigma) &= \prod_{\beta_i>\beta_j}\Big(2\sinh\pi\Big(\widehat\sigma_i-\widehat\sigma_j+\widehat m-\frac{i(2-r)bQ}2\Big)\Big)^{-1},
\end{align}
depending on the choice of boundary condition \textbf{BC1} or \textbf{BC2}. Suitable 1D ${\cal N}=2$ SUSY theories should reproduce these as the Witten index up to a freedom of additional Wilson lines. It is natural to expect that such theories can be obtained by modifying the GLSMs introduced in the previous section. We take the theory of Figure \ref{fig:quiver3} as the starting point.

\paragraph{BC1.}

\begin{figure}[ht]
\begin{center}
\begin{overpic}[scale=1.5]{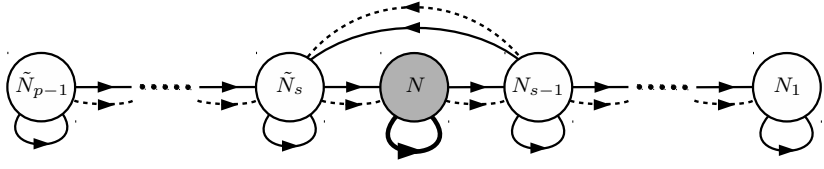}
\put(2.3,9){\scriptsize $\tilde N_{p-1}$}
\put(33.5,9){\scriptsize $\tilde N_s$}
\put(49,9){\scriptsize $N$}
\put(61.7,9){\scriptsize $N_{s-1}$}
\put(93,9){\scriptsize $N_1$}
\end{overpic}
\end{center}
\caption{The worldline theory for a vortex loop in 3D $U(N)$ gauge theory with an adjoint chiral multiplet (represented by a thick line) satisfying {\bf BC1}.}\label{fig:quiverAd1}
\end{figure}

Let us consider a GLSM corresponding to the quiver diagram of Figure \ref{fig:quiverAd1} which is obtained by adding links to the quiver of Figure \ref{fig:quiver3}. The matters corresponding to the added links are charged under $U(1)_\mathrm{G}$ as well as $U(1)_\mathrm{m}$ corresponding to the 3D real mass. We denote their generators by $\text G$ and $\text m$.

The matter multiplets and their charges are as follows. Each gauge group has an adjoint chiral multiplet with $\text{m}=1$ and $\text{G}=r$. Each pair of neighboring nodes has a bifundamental chiral multiplet with $\text{m}=\text{G}=0$ and a bifundamental Fermi multiplet with $\text{m}=1,\text{G}=r$. In addition, there is a Fermi multiplet with $\text{m}=\text{G}=0$ and a chiral multiplet with $\text{m}=-1,\text{G}=-r$ in the bifundamental of $U(N_{s-1})\times U(\tilde N_s)$. The FI couplings are negative for $U(N_a)$ and positive for $U(\tilde N_a)$ gauge groups.

The theory is free of global anomaly, so the charge of Wilson line is determined by the CS coupling and $\beta$ only.
\begin{alignat}{2}
q_a&=k(\beta_{(a)}-\beta_{(a+1)}),\qquad&&(a=1,\cdots,s-1) \nonumber \\
\tilde q_a&=k(\beta_{(a+1)}-\beta_{(a)}).\qquad&&(a=s,\cdots,p-1)
\label{qsAdj}
\end{alignat}
However, when $V_\beta^\text{v}(\widehat\sigma)$ was re-defined in (\ref{VbetaSUN}), we also included the Wilson line factor $e^{2\pi\sum_i\tilde\rho_i\widehat\sigma_i}$ which cancels the global anomaly of the quiver theory for Figure \ref{fig:quiver3}. The added massive 1D matters bring about another global anomaly, but it can be canceled by a Wilson line factor $e^{-2\pi\sum_i\tilde\rho_i\widehat\sigma_i}$. Thus $V_\beta^{\text{c}{\bf 1}}(\widehat\sigma)$ needs to be corrected by this Wilson line factor. 

The Witten index is the JK-residue integral of the following one-loop determinant $\Delta$ multiplied by the Wilson line with charges (\ref{qsAdj}):
\begin{align}
\Delta ~=~& (2\sinh\pi\widetilde m)^{-\sum_aN_a-\sum_a\tilde N_a}\cdot\prod_{a=1}^{s-1}\prod_{i\ne j}^{N_a}\frac{2\sinh\pi(u_i^{(a)}-u_j^{(a)})}{2\sinh\pi(u_i^{(a)}-u_j^{(a)}+\widetilde m)}\prod_{a=s}^{p-1}\prod_{i\ne j}^{\tilde N_a}\frac{2\sinh\pi(\tilde u_i^{(a)}-\tilde u_j^{(a)})}{2\sinh\pi(\tilde u_i^{(a)}-\tilde u_j^{(a)}+\widetilde m)} \nonumber \\
~\times~& 
 \prod_{a=s}^{p-2}\prod_{i=1}^{\tilde N_{a+1}}\prod_{j=1}^{\tilde N_a}\frac{2\sinh\pi(\tilde u_i^{(a+1)}-\tilde u_j^{(a)}+\widetilde m)}{2\sinh\pi(\tilde u_i^{(a+1)}-\tilde u_j^{(a)})}\cdot
\prod_{i=1}^{\tilde N_s}\prod_{j=1}^{N}\frac{2\sinh\pi(\tilde u_i^{(s)}-\widehat\sigma_j+\widetilde m)}{2\sinh\pi(\tilde u_i^{(s)}-\widehat\sigma_j)}\nonumber \\
~\times~&
 \prod_{i=1}^N\prod_{j=1}^{N_{s-1}}\frac{2\sinh\pi(\widehat\sigma_i-u_j^{(s-1)}+\widetilde m)}{2\sinh\pi(\widehat\sigma_i-u_j^{(s-1)})}\cdot
 \prod_{a=1}^{s-2}\prod_{i=1}^{N_{a+1}}\prod_{j=1}^{N_a}\frac{2\sinh\pi(u_i^{(a+1)}-u_j^{(a)}+\widetilde m)}{2\sinh\pi(u_i^{(a+1)}-u_j^{(a)})}
\nonumber \\
 ~\times~&\prod_{i=1}^{N_{s-1}}\prod_{j=1}^{\tilde N_s}\frac{2\sinh(u^{(s-1)}_i-\tilde u_j^{(s)})}{2\sinh(u^{(s-1)}_i-\tilde u_j^{(s)}-\widetilde m)}\,,
\label{DeltaGLSMA}
\end{align}
where we used $\widetilde m\equiv\widehat m+\frac{ibQr}2$. In the limit $\widehat m\to-\infty$ the one-loop determinants for the massive multiplets turn into Wilson lines. The above $\Delta$ then reduces to
\[
\exp\Big({-}\pi\widetilde m\sum_{a<b}n_an_b\Big)\cdot W_{\frac12(\tilde N_s-N_{s-1})} (\widehat\sigma)
\]
times the $\Delta$ for the quiver GLSM of Figure \ref{fig:quiver3} and a Wilson line factor that shift the charges (\ref{qsAdj}) back to (\ref{qs}). On the other hand, we will see in Section \ref{sec:N=4} that the 1D theory has an enhanced SUSY when $\widehat m=0$.

The above one-loop determinant $\Delta$ has more poles than the one corresponding to Figure \ref{fig:quiver3} due to the added chiral multiplets. However, as we explain in the next paragraph, none of those new poles contribute to the index according to the rule of JK-residue. This is in accordance with the fact that $V_\beta(\widehat\sigma)$ (\ref{VbetawMat1}) is given by a sum over elements of ${\cal W}/{\cal W}_K$ as in pure CS theory. Once one accepts this fact, it is straightforward to check that the index reproduces (\ref{VbetawMat1}) for \textbf{BC1}.

\paragraph{Detail of JK-residue integral (2).}

Here we discuss some detail of the JK-residue integral with the above $\Delta$ in the integrand. The space of charges is of dimension $r=\sum_{a=1}^{s-1}N_a+\sum_{a=s}^{p-1}\tilde N_a$ and we denote its basis vectors by $\{{\bf e}_i^{(a)}\}_{i=1,\cdots,N_a}^{a=1,\cdots,s-1},~\{\tilde{\bf e}_i^{(a)}\}_{i=1,\cdots,\tilde N_a}^{a=s,\cdots,p-1}$. Let us first list the charge vectors labeling the singular hyperplanes of $\Delta$. The hyperplanes which are present before introducing the 3D adjoint chiral multiplet are labeled by the charges
\[
\tilde{\bf q}^{(a)}_{ij}\equiv \tilde{\bf e}_i^{(a)}-\tilde{\bf e}_j^{(a-1)},\qquad \tilde{\bf q}_i\equiv\tilde{\bf e}_i^{(s)},\qquad
{\bf q}_i\equiv-{\bf e}_i^{(s-1)},\qquad
{\bf q}^{(a)}_{ij}\equiv {\bf e}_i^{(a+1)}-{\bf e}_j^{(a)}.
\]
The hyperplanes corresponding to the added chiral multiplets are labeled by
\[
 {\bf p}_{ij}^{(a)}\equiv {\bf e}_i^{(a)}-{\bf e}_j^{(a)}+{\bf m},\qquad
 \tilde{\bf p}_{ij}^{(a)}\equiv \tilde{\bf e}_i^{(a)}-\tilde{\bf e}_j^{(a)}+{\bf m},\qquad
 {\bf r}_{ij}\equiv {\bf e}_j^{(s-1)}-\tilde{\bf e}_i^{(s)}-{\bf m},
\]
where we included the generator $\bf m$ of the $U(1)_\mini{\text{m}}$ for convenience.

As in the previous example, the iterative residue integral at each pole determines the values of the variables $u_j^{(a)},\tilde u_j^{(a)}$ one by one. The process can be viewed as if those variables are linked together to form trees each starting from one of the $\widehat\sigma_i$. At the same time, the process also picks up from the above list a set $\Pi$ of $r$ charge vectors that play the role of the links. All the basis vectors ${\bf e}_i^{(a)},\tilde{\bf e}_i^{(a)}$ are then expressed as linear combinations of the elements of $\Pi$. Now, to decide whether the pole contributes to the JK-residue integral, one expresses the reference charge vector
\[
 \delta ~=~ \sum_{a=1}^{s-1}\sum_{i=1}^{N_a}\zeta_a{\bf e}_i^{(a)} +\sum_{a=s}^{p-1}\sum_{i=1}^{\tilde N_a}\tilde\zeta_a\tilde{\bf e}_i^{(a)}\qquad(\zeta_a<0,~~\tilde\zeta_a>0)
\]
as a linear combination of the elements of $\Pi$, and checks if the coefficients are all positive. As we observed in the previous simpler example, the sign of the coefficient of a given element of $\Pi$ is to a large extent related to the direction in which the trees grow at the corresponding link.

There are a few conditions that a pole must satisfy in order to contribute to the integral. One can prove them step by step. First, ${\bf r}_{ij}$ cannot participate in $\Pi$. Then, all the basis vectors $-{\bf e}_i^{(a)}$ must be expressed as non-negative linear combinations of $\{{\bf q}_j,{\bf q}_{jk}^{(b)}, {\bf p}_{jk}^{(b)}\}$, and similarly all $\tilde{\bf e}_i^{(a)}$ must be non-negative linear combinations of $\{\tilde{\bf q}_j,\tilde{\bf q}_{jk}^{(b)}, \tilde{\bf p}_{jk}^{(b)}\}$. In terms of the formation of trees these conditions can be phrased as follows: each tree consists of $u$-variables only or $\tilde u$-variables only. A tree of $u$-variables can only be extended by attaching a new variable $u_j^{(b)}$ according to
\[
\widehat\sigma_j=  u_j^{(b)}~~(\text{for }b=s-1)\qquad\text{or}\qquad
 u_k^{(b+1)}=u_j^{(b)}\qquad\text{or}\qquad
 u_k^{(b)}+\tilde m=u_j^{(b)}.
\]
Likewise, a tree of $\tilde u$-variables can only be extended by attaching $\tilde u_j^{(b)}$ according to
\[
 \tilde u_j^{(b)} = \widehat\sigma_j~~(\text{for }b=s)\qquad\text{or}\qquad
 \tilde u_j^{(b)} = \tilde u_k^{(b-1)}\qquad\text{or}\qquad
 \tilde u_j^{(b)} = \tilde u_k^{(b)}-\tilde m.
\]

For each pole satisfying the above conditions we study whether the residue is nonvanishing. In fact, due to the determinants of vector and Fermi multiplets in the numerator of $\Delta$, the residue vanishes if $\Pi$ contains ${\bf p}_{ij}^{(a)}$ or $\tilde{\bf p}_{ij}^{(a)}$. The residue also vanishes when two or more trees start from a single $\widehat\sigma_i$, or when there are trees with branchings. Thus the trees must consist only of (i) linear chains of $u$-variables extending in the decreasing direction of $a$ and (ii) linear chains of $\tilde u$-variables extending in the increasing direction of $a$. Moreover, each $\widehat\sigma_i$ can have at most one chain starting from it. The set of poles contributing to the JK-residue integral is therefore the same as before introducing the adjoint chiral multiplet in 3D, and it is precisely what is needed for the integral to reproduce (\ref{VbetawMat1}).

\paragraph{BC2.}

\begin{figure}[ht]
\begin{center}
\begin{overpic}[scale=1.5]{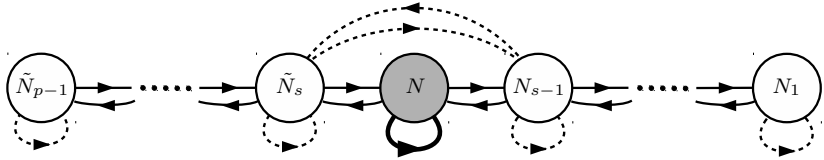}
\put(2.3,9){\scriptsize $\tilde N_{p-1}$}
\put(33.5,9){\scriptsize $\tilde N_s$}
\put(49,9){\scriptsize $N$}
\put(61.8,9){\scriptsize $N_{s-1}$}
\put(92.9,9){\scriptsize $N_1$}
\end{overpic}
\end{center}
\caption{The quiver GLSM on the vortex worldline for 3D $U(N)$ gauge theory with an adjoint chiral multiplet satisfying {\bf BC2}.}
\label{fig:quiverAdj2}
\end{figure}

For this boundary condition, the GLSM on the vortex worldline is described by the quiver diagram of Figure \ref{fig:quiverAdj2} which has extra links compared to the quiver of Figure \ref{fig:quiver3}.

The matter content and the charge assignment are as follows. For each $U(N_a)$ or $U(\tilde N_a)$ gauge node, it has a vectormultiplet as well as an adjoint Fermi multiplet with $\mathrm m=-1$ and $\mathrm G=2-r$. Each pair of neighboring nodes has a bifundamental and an anti-bifundamental chiral multiplets, and the latter has $\text{m}=+1, \mathrm G=r-2$. In addition, the pair $U(N_{s-1})\times U(\tilde N_s)$ has one bifundamental and one anti-bifundamental Fermi multiplets, the latter carrying $\text{m}=-1$ and $\mathrm G=2-r$. The FI couplings for $U(N_a)$ are all negative while those for $U(\tilde N_a)$ are all positive. As in the previous case of {\bf BC1}, the model is free of global anomaly. The charge of the Wilson line can be chosen the same way as (\ref{qsAdj}), and the function $V_\beta^{\text{c}{\bf 2}}(\widehat\sigma)$ needs to be corrected by a Wilson line factor $e^{-2\pi\sum_i\tilde\rho_i\widehat\sigma_i}$. We will not go into the detail of the JK-residue evaluation as it is somewhat simpler than the previous case.

In the limit $\widehat m\to-\infty$ the massive matters turn into a Wilson line of appropriate $U(1)^{p-1}$ charge and the model reduces to that for the quiver of Figure \ref{fig:quiver3}. On the other hand, the 1D field content is such that the supersymmetry enhances to ${\cal N}=4$ if $\widehat m$ is turned off and an appropriate superpotential interaction is turned on. The $\text{m},\text{G}$-charges of the adjoint Fermi multiplets were chosen so that the superpotential terms are invariant. However, the enhanced ${\cal N}=4$ SUSY here is qualitatively different from the one for {\bf BC1}: they have different kind of multiplets and R-symmetries. Also, as we will see in the next section, the SUSY enhancement here does not seem to be related to the enhancement of bulk 3D SUSY.

\subsection{Fundamental representation}\label{sec:fund-matter}

Next we consider vortex loops in 3D $U(N)$ gauge theory with a fundamental chiral multiplet of mass $m$ and R-charge $r$. We regard that the matter is in a bifundamental of $U(N)\times U(1)_\mathrm{m}$. According to the result of section \ref{sec:exactPI}, the function $V_\beta(\widehat\sigma)$ is given by (\ref{VbetawMat1}) with
\begin{align}
 V_\beta^\text{c{\bf 1}}(\widehat\sigma)&~=~\prod_{\beta_i<0}2\sinh\pi\Big(\widehat\sigma_i-\widehat m+\frac{irbQ}2\Big),\nonumber \\
 V_\beta^\text{c{\bf 2}}(\widehat\sigma)&~=~\prod_{\beta_i>0}\Big(2\sinh\pi\Big(\widehat\sigma_i-\widehat m-\frac{i(2-r)bQ}2\Big)\Big)^{-1},
\label{VbetawMat2}
\end{align}
depending on the choice of boundary condition. But a simple multiplication of these products of sinh functions will lead to a global anomaly, so we also need a suitable Wilson line. It is also known that the introduction of (anti-)fundamental chiral multiplets shifts the effective CS and FI couplings \cite{Intriligator:2013lca}.

\begin{figure}[ht]
\begin{center}
\begin{tabular}{cc}
~~~
\begin{overpic}[scale=1.5]{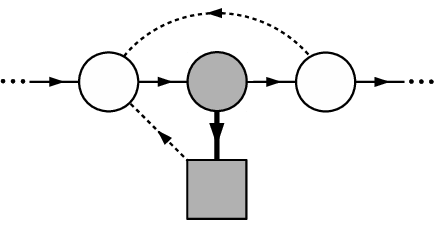}
\put(48,32.5){\scriptsize $N$}
\put(22.2,32.5){\scriptsize $\tilde N_s$}
\put(70.3,32.5){\scriptsize $N_{s-1}$}
\put(49,8){\scriptsize $1$}
\end{overpic}
~~~&~~~
\begin{overpic}[scale=1.5]{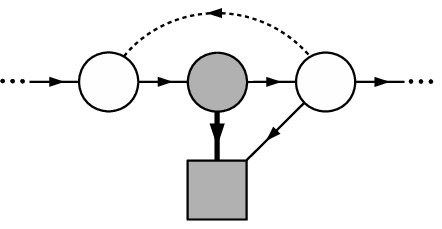}
\put(48,32.5){\scriptsize $N$}
\put(22.2,32.5){\scriptsize $\tilde N_s$}
\put(70.3,32.5){\scriptsize $N_{s-1}$}
\put(49,8){\scriptsize $1$}
\end{overpic}
~~~
\\ {\bf BC1} & {\bf BC2}
\end{tabular}
\end{center}
\caption{Quivers for the worldline theory of a vortex loop in 3D $U(N)$ gauge theory with a fundamental chiral multiplet satisfying \textbf{BC1} or  \textbf{BC2}. In both diagrams, the 3D gauge and global symmetries $U(N)\times U(1)_\mathrm{m}$ are represented by shaded nodes and the 3D fundamental chiral multiplet is represented by a thick link.}
\label{fig:quiverFn1}
\end{figure}

As in the previous subsection, we construct the vortex worldline theory as a modification of the quiver GLSM of Figure \ref{fig:quiver3}. Let us also assume
\begin{equation}
 \beta_{(s+1)}<\beta_{(s)}=0<\beta_{(s-1)}.
\label{betaseq0} 
\end{equation}
Then it turns out that the necessary modification of the quiver is to add just one link connecting a 1D gauge node and the flavor $U(1)_\text{m}$ node as shown in Figure \ref{fig:quiverFn1}. Depending on the choice of boundary condition, we introduce\\[1mm]
\hskip5mm({\bf BC1}) a Fermi multiplet in the bifundamental of $U(1)_\text{m}\times U(\tilde N_s)$ with $\mathrm G=-r$, \\[1mm]
\hskip5mm({\bf BC2}) a chiral multiplet in the bifundamental of $U(N_{s-1})\times U(1)_\text{m}$ with $\mathrm G=r-2$. \\[1mm]
The added links reproduce precisely the contribution of the 3D fundamental chiral multiplet to $V_\beta(\widehat\sigma)$ (\ref{VbetawMat2}), but the 1D theory now has global anomaly. It can be canceled by shifting the charge of the Wilson line $\tilde q_s$ or $q_{s-1}$ by $\pm1/2$.

Under the interpretation of $\beta_{(a)}$ as position coordinates, the assumption (\ref{betaseq0}) means that the 3D gauge node is at $\beta=0$. Note that this assumption is not mandatory. One may start with a quiver realization in which the 3D gauge node is not at $\beta=0$ and find necessary modifications, though the answer will not be as simple as the one given above.

\begin{figure}[ht]
\begin{center}
\begin{tabular}{cc}
~~~
\begin{overpic}[scale=1.5]{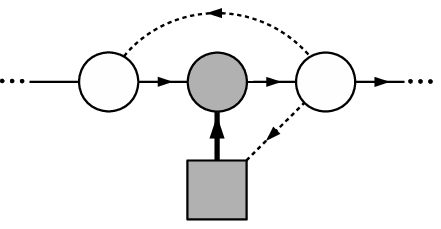}
\put(48,32.5){\scriptsize $N$}
\put(22.2,32.5){\scriptsize $\tilde N_s$}
\put(70.3,32.5){\scriptsize $N_{s-1}$}
\put(49,8){\scriptsize $1$}
\end{overpic}
~~~&~~~
\begin{overpic}[scale=1.5]{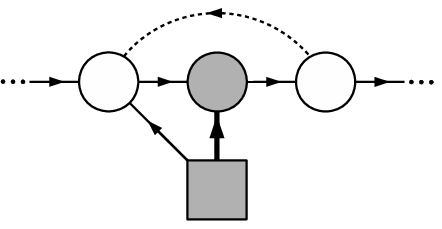}
\put(48,32.5){\scriptsize $N$}
\put(22.2,32.5){\scriptsize $\tilde N_s$}
\put(70.3,32.5){\scriptsize $N_{s-1}$}
\put(49,8){\scriptsize $1$}
\end{overpic}
~~~
\\ {\bf BC1} & {\bf BC2}
\end{tabular}
\end{center}
\caption{Addition of an anti-fundamental chiral multiplet to 3D $U(N)$ theory and the corresponding modification of the GLSM on vortex worldline.}
\label{fig:quiver8}
\end{figure}

The vortex loops in 3D $U(N)$ gauge theory with an anti-fundamental chiral multiplet can be studied in the same way. Depending on the boundary condition, the function $V_\beta(\widehat\sigma)$ is given by (\ref{VbetawMat1}) with
\begin{align}
 V_\beta^\text{c{\bf 1}}(\widehat\sigma)&~=~\prod_{\beta_i>0}2\sinh\pi\Big(\widehat\sigma_i-\widehat m-\frac{irbQ}2\Big),\nonumber \\
 V_\beta^\text{c{\bf 2}}(\widehat\sigma)&~=~\prod_{\beta_i<0}\Big(2\sinh\pi\Big(\widehat\sigma_i-\widehat m+\frac{i(2-r)bQ}2\Big)\Big)^{-1}.
\label{VbetawMat3}
\end{align}
The corresponding vortex worldline theories are given by the two quivers of Figure \ref{fig:quiver8}. They are modifications of the quiver theory of Figure \ref{fig:quiver3} by adding\\[1mm]
\hskip5mm ({\bf BC1}) a Fermi multiplet in the bifundamental of $U(N_{s-1})\times U(1)_\text{m}$ with $\mathrm G=-r$,\\[1mm]
\hskip5mm ({\bf BC2}) a chiral multiplet in the bifundamental of $U(1)_\text{m}\times U(\tilde N_{s})$ with $\mathrm G=r-2$.\\[1mm]
Also, the charge of the Wilson line needs to be modified to take care of global anomaly.

Let us explain how we determined the orientation of the arrows for the 1D matter multiplets just added. For the cases with {\bf BC2}, the added 1D chiral multiplets contribute to the denominator of $\Delta$ (\ref{DeltaGLSM}) and give rise to more poles. But those new poles must not contribute to the index. This determines the orientation of the arrow for the added chiral multiplets. For the case with {\bf BC1}, the orientation of the arrow for Fermi multiplets has been determined from the consistency with SUSY enhancement. As we will discuss in the next section, when 3D bulk theory has ${\cal N}=4$, the vortex worldline theory also has an enhanced ${\cal N}=4$ SUSY.

\paragraph{Large mass limit.}

Integration of massive chiral multiplets in 3D sometimes yields an effective CS coupling \cite{Redlich:1983kn,Redlich:1983dv}. In the presence of vortex loop, it also gives rise to an effective Wilson line for the worldline theory. Let us study this effect in a simple example.

Consider a 3D $U(N)_k$ CS theory with one fundamental and one anti-fundamental chiral multiplets with the masses $m_\text{f}, m_\text{a}$ and R-charges $r_\text{f},r_\text{a}$. They contribute the following one-loop determinant to the ellipsoid partition function (\ref{Zell}):
\begin{equation}
 \Delta_\text{1-loop}^\text{c} = \prod_{i=1}^Ns_b\Big(\frac{i(1-r_\text{f})Q}2-\hat\sigma_i+\hat m_\text{f}\Big)s_b\Big(\frac{i(1-r_\text{a})Q}2+\hat\sigma_i-\hat{m}_\text{a}\Big)\,.
\end{equation}
By using the asymptotics of the double sine function
\[
 s_b(x) \sim \exp\frac{\pm i\pi}2\Big(x^2+\frac{b^2+b^{-2}}{12}\Big)\qquad(\text{Re}(x)\to\pm\infty)
\]
and comparing with (\ref{Scl}), one finds that the integration of the heavy chiral multiplets in the limit $m_\text{f}\to \pm\infty,\, m_\text{a}\to \mp\infty$ shifts the CS and FI couplings by
\begin{equation}
\delta k= \pm1,\qquad
\delta\zeta = \pm\frac{\widehat m_\text{f}+\widehat m_\text{a}}2\pm\frac{i(r_\text{f}-r_\text{a})Q}4\,.
\end{equation}

Let us introduce a vortex loop with vorticity $\beta$ and put the boundary condition {\bf BC1} for the fundamental, {\bf BC2} for the anti-fundamental chirals. As explained above, the 1D theory has an additional pair of chiral and Fermi multiplets in the antifundamental of $U(\tilde N_s)$. The added matters do not produce anomaly, so the $U(1)^{p-1}$ charge of the Wilson line may be chosen as (\ref{qs}). The one-loop determinant $\Delta$ of the worldline theory is modified by the factor
\begin{equation}
 \prod_{\beta_i<0}\frac{2\sinh\pi(\widehat\sigma_i-\widehat m_\text{f}+\frac{ir_\text{f}bQ}2)}{2\sinh\pi(\widehat\sigma_i-\widehat m_\text{a}+\frac{i(2-r_\text{a})bQ}2)}~\longrightarrow~ \exp\bigg({\mp}2\pi\sum_{\beta_i<0}\widehat\sigma_i\bigg).
\end{equation}
This corresponds to the shift of the charge of the Wilson line $\tilde q_s$ by $\mp1$.

Here we recall that the charges (\ref{qs}) of the Wilson line was determined from the consistency with the relation $\lambda_i=k\beta_i$ in pure CS theory. However, after the massive matters are introduced and integrated out, the parameters $k,q_a,\tilde q_a$ will get corrected and (\ref{qs}) will no longer be satisfied. Taking account of this effect, perhaps one should regard $\lambda_i$ or $(q_a,\tilde q_a)$ as more important label than $\beta$ since they determine the value of BPS vortex loop observables more directly. But $\beta$ still has an important role to set the pattern of gauge symmetry breaking and the orderings of unbroken gauge group factors.

\section{${\cal N}=4$ theories}\label{sec:N=4}

In this section we extend our description of vortex loops to those in 3D ${\cal N}=4$ theories. We will first find out the condition on the singular behavior of fields near 1/2 BPS vortex loops, and then identify the corresponding worldline quantum mechanics with 1D ${\cal N}=4$ supersymmetry.

We begin by reviewing the basic properties of 3D ${\cal N}=4$ gauge theories. For the theories on flat $\mathbb R^3$, the four sets of supercharges transform as a bispinor under the R-symmetry $SU(2)_\text{C}\times SU(2)_\text{H}$. We denote its Cartan generators as $\mathrm{J}^3_\text{C}$ and $\mathrm{J}^3_\text{H}$. A 3D ${\cal N}=4$ vectormultiplet is made from an ${\cal N}=2$ vectormultiplet $(A_m, \sigma, \lambda, \bar\lambda, D)$ and an adjoint chiral multiplet $(\phi,\psi,F)$. The three scalars $(\sigma,\phi,\bar\phi)$, three auxiliary scalars $(D,F,\bar F)$ and four spinors $(\lambda,\bar\lambda,\psi,\bar\psi)$ form the representations $({\bf 3},{\bf 1}),\,({\bf 1,{\bf 3}})$ and $({\bf 2},{\bf 2})$ of $SU(2)_\text{C}\times SU(2)_\text{H}$, respectively. In our convention $\phi$ has $\mathrm{J}^3_\text{C}=1$ whereas $F$ has $\mathrm{J}^3_\text{H}=1$. The charges of the fields are summarized in Table \ref{tab:N4qn1}.
\begin{table}[ht]
\begin{center}
\begin{tabular}{c||cccc|cccc|ccc}
\hline
field & $A_m$ & $\sigma$ & $\phi$ & $\bar\phi$ & $\lambda$ & $\bar\lambda$ & $\psi$ & $\bar\psi$ & $D$ & $F$ & $\bar F$\\\hline
$\mathrm J^3_\text{C}$ & $0$ & $0$ & $+1$ & $-1$ & $+\frac12$ & $-\frac12$ & $+\frac12$ & $-\frac12$ & $0$ & $0$ & $0$\\\hline
$\mathrm J^3_\text{H}$ & $0$ & $0$ & $0$ & $0$ & $-\frac12$ & $+\frac12$ & $+\frac12$ & $-\frac12$ & $0$ & $+1$ & $-1$\\\hline
\end{tabular}
\end{center}
\caption{R-charges of ${\cal N}=4$ vectormultiplet fields.}
\label{tab:N4qn1}
\end{table}

Let us turn to the theory on $S^3$. The SYM Lagrangian for an ${\cal N}=4$ vectormultiplet is given by the sum of ${\cal L}_\text{YM}$ for the vectormultiplet and $g^{-2}{\cal L}_\text{mat}$ for the adjoint chiral multiplet in (\ref{Lagrangians}). It is not $SU(2)_\text{C}\times SU(2)_\text{H}$ R-symmetric due to the coupling with the background auxiliary field. But when $\ell=\tilde\ell=f$ and the adjoint chiral multiplet has $r=1$ the Lagrangian has a $\mathbb Z_2$ invariance:
\begin{align}
&~ {\cal L}(A_m\,;\,\sigma,\phi,\bar\phi\,;\,\lambda,\bar\lambda,\psi,\bar\psi\,;\,D,F,\bar F\,;\,H) \nonumber \\ =&~
 {\cal L}(A_m\,;\,-\sigma,\bar\phi,\phi\,;\,\bar\psi,-\psi,-\bar\lambda,\lambda\,;\,D,-F,-\bar F\,;\,-H).
\label{Z2N4S3}
\end{align}
This implies that ${\cal L}$ with $r=1$ on a round $S^3$ has an enhanced supersymmetry: in addition to the original ${\cal N}=2$ SUSY corresponding to the four independent solutions of (\ref{KSsugra}), it has the second set of ${\cal N}=2$ SUSY corresponding to four independent solutions of the same equation (\ref{KSsugra}) with $H$ sign-flipped. The $U(1)$ R-charge of the original ${\cal N}=2$ SUSY is identified with $\mathrm J^3_\text{C}-\mathrm J^3_\text{H}$ because the fields $\phi,\psi,F$ have the charges $\mathrm R_{U(1)}=1,0,-1$. Similarly, the $U(1)$ R-charge of the second ${\cal N}=2$ SUSY is identified as $\mathrm R'_{U(1)}=-\mathrm J^3_\text{C}-\mathrm J^3_\text{H}$.

A hypermultiplet in a representation $\Lambda$ of the gauge group consists of ${\cal N}=2$ chiral multiplets in the representations $\Lambda$ and $\bar\Lambda$. We will denote the chiral scalars as $q,\tilde q$, and their spinor superpartners as $\chi,\tilde\chi$. It is known that $(q,\bar{\tilde q})$ form a doublet of $SU(2)_\text{H}$ and $(\chi,\bar{\tilde\chi})$ form a doublet of $SU(2)_\text{C}$. On $S^3$, these two chiral multiplets both need to have $r=1/2$. Then the $\mathbb Z_2$ symmetry (\ref{Z2N4S3}) of the theory on $S^3$ can be easily extended to hypermultiplet sector by identifying it with an element of $SU(2)_\text{C}\times SU(2)_\text{H}$. The charges of the hypermultiplet fields are summarized in Table \ref{tab:N4qn2}.

\begin{table}[ht]
\begin{center}
\begin{tabular}{c||cccc|cccc}
\hline
field & $q$ & $\bar q$ & $\tilde q$ & $\bar{\tilde q}$ & $\chi$ & $\bar\chi$ & $\tilde\chi$ & $\bar{\tilde\chi}$ \\\hline
$\mathrm R_{U(1)}$ & $+\frac12$ & $-\frac12$ & $+\frac12$ & $-\frac12$ & $-\frac12$ & $+\frac12$ & $-\frac12$ & $+\frac12$ \\\hline
$\mathrm R'_{U(1)}$ & $+\frac12$ & $-\frac12$ & $+\frac12$ & $-\frac12$ & $+\frac12$ & $-\frac12$ & $+\frac12$ & $-\frac12$ \\\hline
$\mathrm J^3_\text{C}$ & $0$ & $0$ & $0$ & $0$ & $-\frac12$ & $+\frac12$ & $-\frac12$ & $+\frac12$ \\\hline
$\mathrm J^3_\text{H}$ & $-\frac12$ & $+\frac12$ & $-\frac12$ & $+\frac12$ & $0$ & $0$ & $0$ & $0$ \\\hline
\end{tabular}
\end{center}
\caption{R-charges of hypermultiplet fields.}
\label{tab:N4qn2}
\end{table}

\subsection{BPS boundary condition}\label{sec:N=4bc}

Let us now turn to the definition of vortex loops. Consider first a vortex line stretching along the $x^3$-axis of flat $\mathbb R^3$. As in the cases with ${\cal N}=2$ SUSY, the gauge field behaves as
\[
A\sim\beta \mathrm d\varphi,\qquad F_{12}=2\pi\beta\delta^2(x^1,x^2)+\cdots.
\]
The vortex configuration can be made half-BPS by turning on the $SU(2)_\text{H}$-triplet auxiliary scalars $D^a= (F,\bar F,D)$ appropriately. The unbroken SUSY then corresponds to solutions of the BPS equation of the form
\begin{equation}
 0 ~=~ \SUSY\lambda_\mini{A\bar B} ~=~ F_{12}\gamma^{12}\xi_\mini{A\bar B}-D^a\xi_\mini{A\bar C}(\sigma^a)^\mini{\bar C}_\mini{~~\bar B},
\end{equation}
where $A,B,\cdots$ and $\bar A,\bar B,\cdots$ are doublet indices for $SU(2)_\text{C}$ and $SU(2)_\text{H}$ respectively, and $\sigma^a$ is Pauli's matrix. It has nontrivial solutions if one sets, for example,
\[
D^3 = D = iF_{12}\,.
\]
The Lorentz symmetry $SU(2)_\text{Lorentz}$ and the R-symmetry $SU(2)_\text{C}\times SU(2)_\text{H}$ are then broken to $U(1)_{\mathrm{M}}\times SU(2)_\text{C}\times U(1)_{\mathrm{J}^3_\text{H}}$, where $\mathrm M$ generates the rotation about the $x^3$-axis. Four of the eight supercharges corresponding to the SUSY parameter $\xi_\mini{A\bar B}$ with $\gamma^3=\pm1~(\text M=\pm\frac12)$ and $\mathrm{J}^3_\text{H}=\mp\frac12$ remain unbroken.

Let us next consider the theory on a round $S^3$ with a half-BPS vortex loop along $S^1_{(\tau)}$ at $\theta=0$. Four of the eight supercharges are broken as in flat space. Two of the four unbroken supercharges correspond to the Killing spinors $\xi,\bar\xi$ of (\ref{KS-ell}), and the other two correspond to new Killing spinors
\begin{equation}
 \xi' = e^{\frac i2(\varphi-\tau)}\left(\begin{array}{r}\cos\frac\theta2 \\ -i\sin\frac\theta2\end{array}\right),\qquad
 \bar\xi' = e^{-\frac i2(\varphi-\tau)}\left(\begin{array}{r}-i\sin\frac\theta2 \\ \cos\frac\theta2\end{array}\right).
\label{xi2nd}
\end{equation}
These four Killing spinors satisfy
\[
\nabla_m\xi=\frac{i}{2\ell}\gamma_m\xi,\quad
\nabla_m\bar\xi=\frac{i}{2\ell}\gamma_m\bar\xi,\quad
\nabla_m\xi'=-\frac{i}{2\ell}\gamma_m\xi',\quad
\nabla_m\bar\xi'=-\frac{i}{2\ell}\gamma_m\bar\xi'.
\]
One can check that the new Killing spinors $\xi',\bar\xi'$ have $\mathrm{M}=\pm\frac12$, so the flat space analysis implies they have $\mathrm{J}_\text{H}^3=\mp\frac12$. The quantum numbers of the four Killing spinors are thus determined as in Table \ref{tab:KSqn}. The $\mathbb Z_2$ transformation (\ref{Z2N4S3}) acts as
\begin{equation}
 \xi\leftrightarrow\xi',\qquad\bar\xi\leftrightarrow\bar\xi'.
\label{Z2xi}
\end{equation}
\begin{table}[ht]
\begin{center}
\begin{tabular}{c||cccc|cc}
\hline
{\small Killing spinor} & $\mathrm R_{U(1)}$ & $\mathrm R'_{U(1)}$ & $~~\mathrm J^3_\text{C}~~$ & $~~\mathrm J^3_\text{H}~~$ & $-i{\cal L}_{\partial_\tau}$ & $-i{\cal L}_{\partial_\varphi}$ \\\hline
$\xi$ & $+1$ & $0$ & $+\frac12$ & $-\frac12$ & $+\frac12$ & $+\frac12$ \\
$\bar\xi$ & $-1$ & $0$ & $-\frac12$ & $+\frac12$ & $-\frac12$ & $-\frac12$ \\
$\xi'$ & $0$ & $+1$ & $-\frac12$ & $-\frac12$ & $-\frac12$ & $+\frac12$ \\
$\bar\xi'$ & $0$ & $-1$ & $+\frac12$ & $+\frac12$ & $+\frac12$ & $-\frac12$ \\\hline
\end{tabular} 
\end{center}
\caption{Quantum numbers of Killing spinors on $S^3$.}\label{tab:KSqn}
\end{table}

\paragraph{Boundary condition on fluctuations.}

It remains to check if there is a set of boundary conditions on the fluctuation of fields preserving $1/2$ of the ${\cal N}=4$ SUSY. We continue to work with a vortex loop in $S^3$ winding along the $S^1_{(\tau)}$ at $\theta=0$.

Let us first study the fluctuation of ${\cal N}=4$ vectormultiplet using the decomposition into ${\cal N}=2$ multiplets. According to what we found in Section \ref{sec:exactPI} for the fluctuation of ${\cal N}=2$ vectormultiplet fields, $\xi\lambda,\bar\xi\bar\lambda$ may diverge mildly but $\xi\bar\lambda,\bar\xi\lambda$ must be finite near the vortex loop. The boundary condition also preserves the SUSY corresponding to $\xi',\bar\xi'$ if it respects the $\mathbb Z_2$ invariance (\ref{Z2N4S3}) and (\ref{Z2xi}). So $\xi'\bar\psi,\bar\xi'\psi$ may diverge but $\xi'\psi,\bar\xi'\bar\psi$ must be finite. Here one can replace $\xi'$ by $\xi$ (and similarly $\bar\xi'$ by $\bar\xi$) because they are proportional to each other along the vortex loop. The resulting boundary conditions on $\psi$ and $\bar\psi$ imply that the ${\cal N}=2$ adjoint chiral multiplet must obey {\bf BC1}.

To be fully explicit, let us list the boundary condition for all the fields in an ${\cal N}=4$ vectormultiplet near a BPS vortex loop:
\begin{alignat}{4}
&\xi\gamma^m\xi\delta A_m,~\bar\xi\gamma^m\bar\xi\delta A_m,\qquad
&& \xi\lambda,~\xi\bar\psi,~\bar\xi\bar\lambda,~\bar\xi\psi,\qquad
&& F,~\bar F\qquad
&& \text{may diverge,}\nonumber\\
&\bar\xi\gamma^m\xi\delta A_m,~\delta\sigma,~\phi,~\bar\phi,\qquad
&& \xi\bar\lambda,~\xi\psi,~\bar\xi\lambda,~\bar\xi\bar\psi,\qquad
&& \delta D\qquad
&& \text{must be finite.}
\end{alignat}
This preserves the SUSY corresponding to $\xi,\bar\xi$ as well as $\xi',\bar\xi'$.

The above form of boundary condition can also be used for a vortex line lying along, say, the $x^3$-axis of flat $\mathbb R^3$. In that case $\xi,\xi'$ are eigenspinors of $\gamma^3=1$ and $\bar\xi,\bar\xi'$ are eigenspinors of $\gamma^3=-1$. The above set of boundary conditions is clearly consistent with the unbroken $SU(2)_\text{C}$ symmetry.

A hypermultiplet in a representation $\Lambda$ consists of an ${\cal N}=2$ chiral multiplet $q,\chi$ in $\Lambda$ and another chiral multiplet $\tilde q,\tilde\chi$ in $\bar\Lambda$. To preserve the SUSY corresponding to $\xi$ and $\bar\xi$, each of the two chiral multiplets must obey the boundary condition {\bf BC1} or {\bf BC2}. Then, as in the previous paragraph, one can argue that the unbroken SUSY enhances if the boundary condition respects the $SU(2)_\text{C}$ symmetry. Recall that, whichever boundary conditions we choose, the fields in the representation $\Lambda$ are divided into four groups of cohomological variables as follows:
\begin{alignat}{3}
 q,\xi\chi~&\in~{\cal H}~&\begin{array}{c}\xrightarrow{~~~\;{\cal J}\;~~~}\\[-4mm] \xleftarrow[~~~\;\bar{\cal J}\;~~~]{}\end{array}&&~{\cal H}'~&\ni~ \bar\xi\chi,
\nonumber \\
 \bar{\tilde q},\bar\xi\bar{\tilde\chi}~&\in~\tilde{\cal H}^\ast~&\begin{array}{c}\xrightarrow{~~(\tilde{\cal J})^\dagger~~}\\[-4mm] \xleftarrow[~~(\bar{\tilde{\cal J}})^\dagger~~]{}\end{array}&&~\tilde{\cal H}'^\ast~&\ni~ \xi\bar{\tilde\chi}.
\end{alignat}
Here the differential operators $\tilde{\cal J},\bar{\tilde{\cal J}}$ are defined in the same way as ${\cal J}\equiv i\bar\xi\gamma^m\bar\xi\nabla_m$ and $\bar{\cal J}\equiv -i\xi\gamma^m\xi\nabla_m$ using the covariant derivative for fields in $\bar\Lambda$. So in fact $(\tilde{\cal J})^\dagger=\bar{\cal J}$ and $(\bar{\tilde{\cal J}})^\dagger={\cal J}$. Hence one can preserve $SU(2)_\text{C}$ by imposing the same boundary condition on $\xi\chi$ and $\xi\bar{\tilde\chi}$, and similarly on $\bar\xi\chi$ and $\bar\xi\bar{\tilde\chi}$, which form doublets. This leads us to conclude that there are the following two BPS boundary conditions on a hypermultiplet:

\begin{itemize}
\item $\xi\chi,\xi\bar{\tilde\chi}$ are finite but $\bar\xi\chi,\bar\xi\bar{\tilde\chi}$ may diverge near the vortex loop. Namely, the chiral multiplet $(q,\chi)$ obeys {\bf BC1} and $(\tilde q,\tilde\chi)$ obeys {\bf BC2}.
\item The opposite of the above. Namely, $(q,\chi)$ obeys {\bf BC2} and $(\tilde q,\tilde\chi)$ obeys {\bf BC1}.
\end{itemize}
Note that our result is similar to the one obtained in \cite{Dimofte:2019zzj}. There the fluctuation of fields with more general (i.e. not necessarily mild) singular behavior near vortex lines is considered.

\subsection{${\cal N}=4$ SUSY quantum mechanics}\label{sec:N=4QM}

Let us next turn to the study of the vortex worldline theories. For a straight vortex line in a flat $\mathbb R^3$, the worldline theory has a global symmetry $SU(2)_\text{C}\times U(1)_{\mathrm J^3_\text{H}}\times U(1)_\mathrm{M}$. The four unbroken supercharges transform under its $SU(2)\times U(1)$ subgroup as two $SU(2)$-doublets of $U(1)$ charge $\pm1$. The 1D ${\cal N}=4$ SUSY with this R-symmetry is a dimensional reduction of the 4D ${\cal N}=1$ SUSY.

A 1D ${\cal N}=4$ vectormultiplet is made from an ${\cal N}=2$ vectormultiplet $(A_t,\sigma,\lambda,\bar\lambda,D)$ and an adjoint chiral multiplet $(\phi,\psi)$. The quantum numbers of the fields are determined as in Table \ref{tab:qn1Dv} from the fact that $\epsilon,\bar\epsilon$ in the transformation rules (\ref{1Dvec}) and (\ref{SUSYGLQM}) carry the same quantum numbers as $\xi,\bar\xi$. The $U(1)$ R-charge of 1D ${\cal N}=4$ SUSY is identified with a linear combination
\[
 c_1\mathrm J^3_\text{H}+c_2\mathrm M.\quad(c_2-c_1=2)
\]
\begin{table}[ht]
\begin{center}
\begin{tabular}{c||cccc|cccc|c||cc}
\hline
 field & $A_t$ & $\sigma$ & $\phi$ & $\bar\phi$ & $\lambda$ & $\bar\lambda$ & $\psi$ & $\bar\psi$ & $D$ & $\epsilon$ & $\bar\epsilon$\\\hline
$\mathrm J^3_\text{C}$ & $0$ & $0$ & $+1$ & $-1$ & $+\frac12$ & $-\frac12$ & $+\frac12$ & $-\frac12$ & $0$ & $+\frac12$ & $-\frac12$\\\hline
$\mathrm J^3_\text{H}$ & $0$ & $0$ & $0$ & $0$ & $-\frac12$ & $+\frac12$ & $+\frac12$ & $-\frac12$ & $0$ & $-\frac12$ & $+\frac12$\\\hline
$\mathrm M$ & $0$ & $0$ & $0$ & $0$ & $+\frac12$ & $-\frac12$ & $-\frac12$ & $+\frac12$ & $0$ & $+\frac12$ & $-\frac12$\\\hline
\end{tabular}
\end{center}
\caption{Quantum numbers of ${\cal N}=4$ vectormultiplet fields on the vortex worldline.}\label{tab:qn1Dv}
\end{table}

For the computation of Witten index, one chooses a pair of supercharges (such as the pair we have been using in the previous sections) that generate an ${\cal N}=2$ subalgebra. The index can be generalized by twisting the periodic boundary condition of fields by global symmetries that commute with the chosen supercharges. Of particular importance is the symmetry generated by $\mathrm G\equiv \mathrm J^3_\text{C}-\mathrm J^3_\text{H}-2\mathrm M$, as it shows up in the Witten index for vortex loops inside $S^3$. This $\text G$ was already introduced in the previous section at (\ref{defG}) as a non-R global symmetry of ${\cal N}=2$ SUSY theories. One can easily find from Table \ref{tab:qn1Dv} that ${\cal N}=2$ vectormultiplet has $\mathrm G=0$ while the adjoint chiral multiplet has $\mathrm G=+1$.

A 1D ${\cal N}=4$ chiral multiplet is made from an ${\cal N}=2$ chiral multiplet $(q,\chi)$ and a Fermi multiplet $(\eta,F)$ in the same representation of the gauge group. The quantum numbers of fields under $\mathrm{J}^3_\text{H}$ and $\mathrm M$ are constrained only by the requirement that the fermions $(\chi, \eta)$ form an $SU(2)$-doublet, so generally they take values as summarized in Table \ref{tab:qn1Dc}. This implies that, if $q$ and $\chi$ have $\mathrm G=g$, then $\eta$ and $F$ should have $\mathrm G=g+1$. We call such a set of fields an ${\cal N}=4$ chiral multiplet of $\mathrm G=g$.

\begin{table}[ht]
\begin{center}
\begin{tabular}{c||c|cc|c}
\hline
field & $q$ & $\chi$ & $\eta$ & $F$ \\\hline
$\mathrm J^3_\text{C}$ & $0$ & $-\frac12$ & $+\frac12$ & $0$ \\\hline
$\mathrm J^3_\text{H}$ & $a$ & $a+\frac12$ & $a+\frac12$ & $a+1$ \\\hline
$\mathrm M$ & $b$ & $b-\frac12$ & $b-\frac12$ & $b-1$ \\\hline
\end{tabular}
\end{center}
\caption{Quantum numbers of ${\cal N}=4$ chiral multiplet fields on the vortex worldline.}\label{tab:qn1Dc}
\end{table}

\begin{figure}[ht]
\begin{center}
\begin{overpic}[scale=1.5]{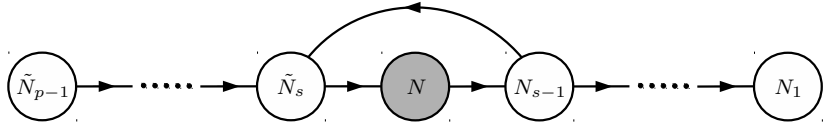}
\put(2.3,4.5){\scriptsize $\tilde N_{p-1}$}
\put(33.5,4.5){\scriptsize $\tilde N_s$}
\put(49,4.5){\scriptsize $N$}
\put(61.8,4.5){\scriptsize $N_{s-1}$}
\put(93,4.5){\scriptsize $N_1$}
\end{overpic}
\end{center}
\caption{1D ${\cal N}=4$ quiver diagram describing the GLSM for a vortex loop in 3D ${\cal N}=4$ $U(N)$ pure SYM. The white nodes and solid lines represent 1D ${\cal N}=4$ vector and chiral multiplets.}
\label{fig:quiverN4YM}
\end{figure}

\paragraph{Examples.} As the most basic example of vortex loops in 3D ${\cal N}=4$ theories, let us consider those in ${\cal N}=4$ $U(N)$ pure SYM. The worldline theory is a special case ($m=0, r=1$ and {\bf BC1}) of the quiver GLSM studied in Section \ref{sec:adj-matter}. In 1D ${\cal N}=2$ terminology, it is a gauge theory with the 1D and 3D gauge groups
\[
 U(\tilde N_{p-1})\times\cdots U(\tilde N_s)\times U(N)_\mini{\text{(3D)}}\times U(N_{s-1})\times\cdots U(N_1).
\]
The ${\cal N}=2$ vectormultiplet for each 1D gauge group factor is paired with an adjoint chiral multiplet with $\mathrm G=1$ to form an ${\cal N}=4$ vectormultiplet. For each pair of neighboring gauge group factors one has a pair of bifundamental ${\cal N}=2$ chiral and Fermi multiplets of $\mathrm{G}=0$ and $1$, which form an ${\cal N}=4$ chiral multiplet of $\mathrm{G}=0$. The theory also has a pair of a chiral and Fermi multiplets of $G=-1,0$ in the bifundamental of $U(N_{s-1})\times U(\tilde N_{s})$, which form an ${\cal N}=4$ chiral multiplet with $\mathrm G=-1$. The field content is described by the ${\cal N}=4$ quiver diagram of Figure \ref{fig:quiverN4YM}. The FI couplings for $U(N_a)$ are all negative while those for $U(\tilde N_a)$ are all positive. The theory has no Wilson line since it is free of global anomaly and one cannot turn on 3D CS coupling without breaking SUSY to ${\cal N}\le3$.

\begin{figure}[ht]
\begin{center}
\begin{tabular}{cc}
~~~
\begin{overpic}[scale=1.5]{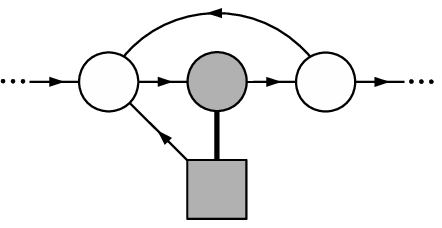}
\put(48,32.5){\scriptsize $N$}
\put(22.2,32.5){\scriptsize $\tilde N_s$}
\put(70.3,32.5){\scriptsize $N_{s-1}$}
\put(49,8){\scriptsize $1$}
\end{overpic}
~~~&~~~
\begin{overpic}[scale=1.5]{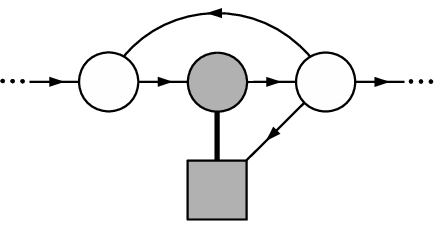}
\put(48,32.5){\scriptsize $N$}
\put(22.2,32.5){\scriptsize $\tilde N_s$}
\put(70.3,32.5){\scriptsize $N_{s-1}$}
\put(49,8){\scriptsize $1$}
\end{overpic}
~~~\\ \hskip10mm
\parbox[t]{70mm}{{\bf BC1} for fundamental chiral\\ {\bf BC2} for anti-fundamental chiral} & \hskip10mm
\parbox[t]{70mm}{{\bf BC2} for fundamental chiral\\ {\bf BC1} for anti-fundamental chiral}
\end{tabular}
\end{center}
\caption{Addition of a fundamental hypermultiplet to 3D ${\cal N}=4$ $U(N)$ theory and the corresponding modification of the vortex worldline GLSM.}
\label{fig:quiverN4QCD}
\end{figure}

The next simplest are the vortex loops in 3D ${\cal N}=4$ $U(N)$ gauge theory with a fundamental hypermultiplet. The worldline theory is obtained by adding some more fields to the theory described previously according to the discussion of Section \ref{sec:fund-matter}. The corresponding quiver diagram is presented in Figure \ref{fig:quiverN4QCD}. As was explained in Section \ref{sec:N=4bc}, there are two consistent boundary conditions on the hypermultiplet, which result in two different modification of the quiver diagram of Figure \ref{fig:quiverN4YM}. The added ${\cal N}=4$ chiral multiplet is either in the anti-fundamental of $U(\tilde N_s)$ or in the fundamental of $U(N_{s-1})$, and it has $\mathrm G=-3/2$ in both cases. Note that the model agrees with the one discussed in \cite{Gadde:2013dda, Assel:2015oxa} if the 3D $U(N)$ gauge node is at either end of the linear quiver.

\paragraph{Background fields for vortex loops in $S^3$.}

Let us explain what kind of background fields appear on the worldline of vortex loops in 3D ${\cal N}=4$ gauge theories on $S^3$.

We recall that the 1D ${\cal N}=2$ SUSY of the vortex worldline theory was defined in accordance with the 3D SUSY acting on cohomological variables. For vortex loops of ${\cal N}=4$ theory on $S^3$, the square of the supercharge is given by (here $t$ is the worldline coordinate of period $2\pi$)
\begin{equation}
\SUSY_\text{(1D)}^2  \sim -\partial_t+\sigma+iA_t + i(\mathrm J^3_\text{C}-\mathrm J^3_\text{H}-2\mathrm M) + (\text{3D vectormultiplet fields}).
\end{equation}
${\cal N}=4$ theories on $S^3$ have the second set of ${\cal N}=2$ SUSY corresponding to the Killing spinors $\xi',\bar\xi'$ (\ref{xi2nd}). It can be used to define the second 1D supercharge $\SUSY'_\text{(1D)}$ which squares to
\begin{equation}
\SUSY'^2_\text{(1D)}  \sim -\partial_t-\sigma+iA_t - i(-\mathrm J^3_\text{C}-\mathrm J^3_\text{H}-2\mathrm M) + (\text{3D vectormultiplet fields})',
\label{Q1D'}
\end{equation}
where the prime on the 3D vectormultiplet fields stands for the $\mathbb Z_2$ action defined in (\ref{Z2N4S3}). Here one needs to be careful for the fact that the two supercharges are defined by identifying different sets of cohomological variables as 1D multiplets. The set of 1D variables on which $\SUSY_\text{(1D)}$ acts as (\ref{1Dvec}) or (\ref{SUSYGLQM}) is therefore different from the set on which $\SUSY_\text{(1D)}'$ acts the same way. But the two sets of variables are related by a simple ``gauge transformation'' as we now explain.

Let $\Phi$ be a cohomological variable made of 3D fields and $\xi,\bar\xi$ such as $\Psi$ that we considered in (\ref{QPsi}), and $\Phi'$ the same cohomological variable with $(\xi,\bar\xi)$ replaced by $(\xi',\bar\xi')$. Using the quantum number of Killing spinors listed in Table \ref{tab:KSqn} and the fact that cohomological variables are all 3D Lorentz scalar, one generally finds
\begin{equation}
 \partial_\tau\Phi = i(\mathrm H-\mathrm M)\Phi,\qquad
 \partial_\tau\Phi' = i(\mathrm H+\mathrm M)\Phi'.
\end{equation}
So the two cohomological variables are related by
\begin{equation}
\Phi' = e^{2i\mathrm M\tau}\Phi.
\end{equation}
The gauge transformation relating the two sets of 1D variables explained above is given by the same formula. Therefore, when considering the action of $\SUSY'^2_\text{(1D)}$ on $\Phi$ instead of $\Phi'$, the RHS of (\ref{Q1D'}) has to be shifted by $-2i\mathrm M$ due to the above gauge transformation. The value of the background 1D vectormultiplet field is thus determined as follows.
\begin{equation}
\begin{array}{rcl}
\sigma^\mini{\text{bg}}+iA_t^\mini{\text{bg}} &=& i(\mathrm J^3_\text{C}-\mathrm J^3_\text{H}-2\mathrm M), \nonumber \\
-\sigma^\mini{\text{bg}}+iA_t^\mini{\text{bg}} &=& -i(-\mathrm J^3_\text{C}-\mathrm J^3_\text{H}),
\end{array}
\qquad\therefore\qquad
\begin{array}{rcl}
 A_t^\mini{\text{bg}}&=& \mathrm J^3_\text{C}-\mathrm M, \\
\sigma^\mini{\text{bg}}&=& -i\mathrm J^3_\text{H}-i\mathrm M.
\end{array}
\end{equation}
Thus we recovered the result in Section 5.2 of \cite{Assel:2015oxa} using a slightly different argument.

\section{Concluding remarks}\label{sec:conclusion}

In this paper we studied different descriptions of BPS vortex loops in 3D ${\cal N}=2$ SUSY gauge theories and worked out exact formulae for their expectation values on an ellipsoid. Hopefully it will be useful in studying aspects of 3D SUSY gauge theories and their dualities \cite{Aharony:1997bx, Aharony:1997gp, Giveon:2008zn, Intriligator:2013lca}. See \cite{Dimofte:2019zzj,Dey:2020hfe,Dey:2021jbf,Nawata:2021nse} for recent work on line defects in relation to 3D mirror symmetry. The relations among different descriptions that we found in this paper may be useful for the study of codimension-two defects in gauge theories in other dimensions.

\paragraph{Open problems and future directions.}

As we have seen, vortex loops can be defined either via path integral over fields with prescribed singular behaviors or via 1D-3D coupled systems. We first studied the relation between the two definitions in pure CS theories where the problem turned out to be related to the equivalence of Wilson and vortex loops. We found that in ${\cal N}=2$ theories the equivalence holds up to a parameter shift ($\lambda\to\lambda-\tilde\rho$). We resolved this mismatch by relating it to the global anomaly of 1D theory on vortex worldline and canceling it by Wilson lines. However, it remains unclear what this Wilson line corresponds to in the other description. In retrospect, we were able to perform exact path integration on singular vortex backgrounds but the result turned out anomalous. It might be interesting to understand the source and resolutions of this anomaly without moving to the description in terms of 1D-3D coupled systems.

We studied the equivalence of the two descriptions of vortex loops by translating them to the boundary actions with or without averaging, i.e. (\ref{N2-SCSQM}) and (\ref{N2-SCSQMg}). It would be nice to check the equivalence further by working out expectation values of multiple coexisting loops, such as Hopf link invariants, using the same argument.

It would also be interesting to study in more detail the series of GLSMs we found for flag manifolds and their duality relation. For vortex loops in 3D $U(N)$ theory with fundamental matters, we gave the worldline theory only when the modification due to the 3D matters is the simplest. In principle one can work out all other dual theories and study their relation. We also expect that the quiver GLSMs on vortex worldline and their duality relations have a simple explanation in terms of type IIB brane construction.

The 3D ${\cal N}=4$ theories considered in this paper are those made of vector and hypermultiplets only. By analyzing them using decomposition into ${\cal N}=2$ multiplets we found that the adjoint chiral multiplet (which is a part of ${\cal N}=4$ vectormultiplet) must always obey {\bf BC1}, though the SUSY on the vortex worldline seems to enhance for both choices of boundary conditions. It may be the case that the mechanism of SUSY enhancement is different for vortex loops in the ABJM model (for a recent work, see \cite{Griguolo:2021rke}) or other CS-matter theories with ${\cal N}\ge4$ SUSY that were classified in \cite{Hosomichi:2008jd,Hosomichi:2008jb}.

Finally, let us point out that we managed to reproduce the worldline theory for only a part of the vortex loops that were identified in \cite{Assel:2015oxa} as the mirror of Wilson loops. The main limitation for our analysis arises from that the function $V_\beta(\widehat\sigma)$ has to be expressed as a sum over elements of ${\cal W}/{\cal W}_K$ as in (\ref{VbetaSUN}) or (\ref{VbetawMat1}). This turns into a constraint on the set of poles contributing to the JK-residue integral for the index $I(\widehat\sigma)$. On the other hand, \cite{Assel:2015oxa} has examples of vortex loops for which the index receives contributions from more poles. Perhaps this means there are more vortex loops defined by worldline quantum mechanics than those described by singular behavior of fields. Or it might be the case that we could reproduce more vortex loops in \cite{Assel:2015oxa} by relaxing the assumption of small $\beta$ (\ref{betasmall}). Either way, more thorough study of the correspondence is needed for a full understanding.

\appendix

\section{Vortex loops for abelian symmetry}\label{sec:U(1)}

Here we focus on vortex loops for $U(1)$ symmetry and make some detailed comparison of our result with those in the earlier work \cite{Kapustin:2012iw,Drukker:2012sr}. For this special case, it is helpful to consider first the vortex loops for global symmetry rather than gauge symmetry. As was argued in \cite{Kapustin:2012iw} and will be reviewed in the following, such global vortex loops are exchanged with Wilson loops under an $SL(2,\mathbb Z)$ \cite{Witten:2003ya} which acts on the space of 3D CFTs with a global $U(1)$ symmetry.

\paragraph{Action of $SL(2,\mathbb Z)$ on 3D CFTs and loops.}

For a 3D CFT with a global $U(1)$ symmetry, one can define its $S$-transform by the same CFT plus a gauge field $A$ which gauges the $U(1)$. The resulting theory also has a global (often called topological) $U(1)$ symmetry generated by the current $\ast F/2\pi$, where $F=\mathrm dA$. Let $Z[A]$ be the generating function of correlation functions of conserved currents for the original theory, defined by coupling an external gauge field $A$ to the current in a gauge-invariant way. The generating function for the S-transformed theory is then given by
\begin{equation}
 \tilde Z[\tilde A] = \int {\cal D}A  Z[A]\exp\left(\frac{i}{2\pi}\int A\mathrm d\tilde A\right)\,.
\end{equation}
The $T$-transform shifts the two-point function of the current by a contact term, which amounts to
\begin{equation}
  \tilde Z[A]= Z[A]\exp\left(\frac{i}{4\pi}\int A\mathrm dA\right)
\end{equation}
in terms of generating function. It was found in \cite{Witten:2003ya} that the above $S$ and $T$ generate an $SL(2,\mathbb Z)$. This $SL(2,\mathbb Z)$ action has a natural SUSY lift in which the external gauge field $A$ is promoted to a whole vectormultiplet. The action of $S$ was considered originally in \cite{Kapustin:1999ha} in explaining mirror symmetry of 3D SUSY abelian gauge theories.

In terms of generating functions, the insertion of a global vortex loop of vorticity $\beta$ along a curve $\gamma$ can formally be described by
\begin{equation}
\left\langle V_\beta(\gamma)\right\rangle[A] ~=~ Z[A+A_{\beta,\gamma}],
\label{Vbshift}
\end{equation}
where $A_{\beta,\gamma}$ is the singular gauge field whose field strength has a delta-functional support on $\gamma$. In fact, using the same gauge field one can express the Wilson loop of charge $q$ along the curve $\gamma$ as
\begin{equation}
 W_q(\gamma)[A] ~=~ \exp\left(iq\int_\gamma A\right) ~=~ \exp\left(\frac{i}{2\pi}\int A_{q,\gamma}\;\mathrm dA\right).
\end{equation}
This relation holds even for non-integer $q$ as long as the homology class $q\cdot[\gamma]$ is integral. Using this one can show \cite{Kapustin:2012iw} that $S$ maps a CFT with a global Wilson loop $W_q(\gamma)$ to another CFT with a global vortex loop $V_q(\gamma)$. Similarly, $S$ maps a global vortex loop $V_\beta(\gamma)$ to a Wilson loop $W_{-\beta}(\gamma)$.

However, recall that vortex loops must be periodic in $\beta$ while Wilson loops are not periodic in $q$. It is therefore the most reasonable to expect that the above simple map between Wilson and vortex loops is valid for $\beta$ within a certain range, and that (\ref{Vbshift}) behaves discontinuously when $\beta$ is at the ends of the range. We have indeed found in Section \ref{sec:exactPI} that the path integral over a charged chiral multiplet on a vortex loop background is a periodic but discontinuous function of $\beta$.

\paragraph{SCFTs on ellipsoids.}

For Lagrangian ${\cal N}=2$ SCFTs on an ellipsoid (\ref{ellipsoid}), one can compute the generating function using localization if the external vectormultiplet fields preserve SUSY, i.e. if their values are as given in (\ref{saddlept}). Let us turn on a constant value $m=(\ell\tilde\ell)^{1/2}\,\hat m$ for its scalar component, and denote by $Z(\hat m)$ the generating function for an ${\cal N}=2$ SCFT on that background. The generating function for the $S$-transformed theory is then given by
\begin{equation}
 \tilde Z(\hat m) = \int\mathrm d\hat\sigma Z(\hat\sigma) e^{-2\pi i\hat\sigma\hat m}.
\end{equation}
The insertion of a global Wilson loop $W_q$ along the circle $S^1_\tau$ at $\theta=0$ is described by the multiplication of $e^{2\pi ibq\hat m}$ to $Z(\hat m)$. The argument in the previous paragraph then implies that the expectation value of a global vortex loop $V_\beta$ should be given by
\begin{equation}
 \langle V_\beta\rangle(\hat m) ~=~ Z(\hat m+ib\beta),
\label{simpleshift}
\end{equation}
for a certain range of $\beta$.

As the simplest example, consider the free SCFT of a single chiral multiplet with $r=1/2$. According to the result in Section \ref{sec:exactPI}, the generating function for this case equals $Z(\hat\sigma)=s_b(\frac{iQ}4-\hat\sigma)$, and the insertion of a global vortex loop $V_\beta$ modifies it as
\begin{equation}
\langle V_\beta\rangle(\hat\sigma) ~=~\left\{
\begin{array}{ll}
\displaystyle s_b\left(\frac{iQ}4-\hat\sigma-ib\beta+ib\lfloor\beta\rfloor\right)& (\textbf{BC1})\\[3mm]
\displaystyle s_b\left(\frac{iQ}4-\hat\sigma-ib\beta+ib\lceil\beta\rceil\right)& (\textbf{BC2})
\end{array}
\right.\,.
\label{VbBC}
\end{equation}
Note that both expressions are periodic in $\beta$. The vortex loops with \textbf{BC1} satisfy (\ref{simpleshift}) for $\beta\in[0,1)$, so they are the S-transform of the Wilson loops $W_q$ with charge $q\in[0,1)$. Likewise, the vortex loops with \textbf{BC2} are the S-transform of the Wilson loops with $q\in(-1,0]$.

\paragraph{Smearing.}

Global vortex loops were studied in \cite{Kapustin:2012iw} as a limit of smooth gauge field configuration. Let us explain how this definition leads to a different answer for the vortex loop observables. As a concrete example, consider an ellipsoid with a background $U(1)$ vortex loop configuration (\ref{saddlept}). The singularity at $\theta=0$ can be smoothed by replacing $A_\varphi=\beta$ with
\begin{equation}
 A_\varphi = \beta -g(\theta),
\label{smear1}
\end{equation}
where $g(\theta)$ is a smooth monotonically decreasing function such that $g(0)=\beta$ and $g(\epsilon)=0$ for some small $\epsilon$. The magnetic flux is then smeared over a tube of diameter $\sim 2\tilde\ell\epsilon$. The full saddle point condition is then solved by the following {\it complex} field configuration
\begin{equation}
 \sigma = \sigma_0+\frac i{\tilde\ell}g(\theta),\qquad
 D = \frac1f\left(\sigma_0+\frac i{\tilde\ell}\left\{g(\theta)-g'(\theta)\cot\theta\right\}\right).
\label{smear2}
\end{equation}
Note that the periodicity in $\beta$ is lost after the smearing since the smeared configurations with vorticity $\beta$ and $\beta+1$ are not gauge-equivalent.

Now consider the theory of a free chiral multiplet coupled to the above background vectormultiplet. The path integral can be performed most easily by using the machinery introduced in Section \ref{sec:exactPI}. Since the smeared background is regular everywhere, the zeromodes of ${\cal J},\bar{\cal J}$ should be regular near $\theta=0$ (that is, $\theta\ll\epsilon$) as well as $\theta=\pi/2$. The computation of the determinant goes the same way as in the absence of the vortex loop, except that $\SUSY^2$ on the background (\ref{smear1}), (\ref{smear2}) is given by
\begin{equation}
\SUSY^2 ~=~ -\frac i{\tilde\ell}{\cal L}_{\partial_\varphi}-\frac i\ell{\cal L}_{\partial_\tau}+i\left(\sigma_0+\frac{i\beta}{\tilde\ell}\right)-\frac12\left(\frac1{\tilde\ell}+\frac1\ell\right)R_{U(1)}.
\end{equation}
We thus recover the result in \cite{Kapustin:2012iw}:
\begin{equation}
 \langle V_\beta\rangle(\hat\sigma_0)~=~s_b\left(\frac{iQ}4-\hat\sigma_0-ib\beta\right)\quad(\textbf{smearing})\,.
\end{equation}
The same result was also obtained in \cite{Drukker:2012sr} under some assumption on the mode expansion of fields, which amounts to that $\mathrm{ker}({\cal J})$ and $\mathrm{ker}(\bar{\cal J})$ depend continuously on $\beta$.

The above result may look very plausible from the viewpoint of the $SL(2,\mathbb Z)$ action. However, the smeared background actually leads to a rather peculiar rule of mode expansion for charged chiral fields. In order to explain this point, let us move to the system of cohomological variables introduced in Section \ref{sec:exactPI} and focus on the fields valued in ${\cal H}$, because the fields valued in ${\cal H}'$ can be studied in the same way. The most natural choice for the basis wavefunctions of ${\cal H}$ is the simultaneous eigenfunctions of the operators $\bar{\cal J}{\cal J}, -i{\cal L}_{\partial_\varphi}$ and $-i{\cal L}_{\partial_\tau}$.

For nonzero eigenvalues of $\bar{\cal J}{\cal J}$, the general solution to the eigenmode equation takes the following approximate form at $\theta\ll\epsilon$:
\begin{equation}
 \Phi~\sim~ e^{im\varphi+in\tau}\left\{A\,\theta^{|m|}+B\,\theta^{-|m|}\right\}.
\end{equation}
Regularity at $\theta=0$ requires the coefficient $B$ to be zero. The regular solution, when continued to the region $\epsilon\le\theta\ll 1$, should behave as
\begin{equation}
\Phi~\sim~e^{im\varphi+in\tau}\left\{C\theta^{|m-\beta|}+D\theta^{-|m-\beta|}\right\}
\end{equation}
with some coefficients $C$ and $D$. On general grounds, the two terms in the bracket should be of the same order at $\theta\sim\epsilon$, which implies $D/C\sim \epsilon^{2|m-\beta|}$. So the first term becomes more and more dominant as $\epsilon\to 0$. This means that all the nonzero eigenmodes of $\bar{\cal J}{\cal J}$ are required to be finite around the smeared vortex loop.

On the other hand, zero eigenmodes of $\bar{\cal J}{\cal J}$ (zeromodes of ${\cal J}$) behave as
\begin{align}
 \Phi~\sim~& \left\{\begin{array}{ll}
 e^{im\varphi+in\tau}\cdot\theta^{-m} & (\theta\ll\epsilon) \\[1mm]
 e^{im\varphi+in\tau}\cdot\theta^{\beta-m} & (\epsilon\le\theta\ll1)
 \end{array}\right.\,.
\end{align}
Regularity requires $m\le0$. In the limit $\epsilon\to0$, the ${\cal J}$-zeromodes which behave as the $\beta$-th or higher power of $\theta$ are included in ${\cal H}$. The condition for a zeromode to be in ${\cal H}$ is therefore quite different from the one for non-zeromodes: many singular zeromodes are to be included in ${\cal H}$ for large negative $\beta$, whereas many finite zeromodes are excluded for large positive $\beta$.

\paragraph{Triviality of gauge vortex loops.}

Finally let us consider vortex loops for $U(1)$ gauge symmetry. Suppose that the partition function of a $U(1)$ gauge theory is given by a path integral with respect to the gauge field $A$:
\begin{equation}
 Z = \int {\cal D}A\,Z[A].
\label{ZU1g}
\end{equation}
Then the expectation value of a vortex loop of vorticity $\beta$ along $\gamma$ is formally given by
\begin{equation}
 \langle V_\beta(\gamma)\rangle = \int{\cal D}A\,Z[A+A_{\beta,\gamma}]\,.
\label{VbU1g}
\end{equation}
If $A_{\beta,\gamma}$ were a smooth $U(1)$ gauge field, then $\langle V_\beta(\gamma)\rangle$ would equal $Z$ by redefining the path-integration variable. Therefore, if a gauge vortex loop is defined by expressing $A_{\beta,\gamma}$ as (a limit of) a smooth configuration, it will always lead to a trivial operator.

What about the gauge vortex loop defined by \textbf{BC1} or \textbf{BC2} on matter fields? As a concrete example, consider some chiral multiplets and assign the R-charge $r_i$ and the $U(1)$ gauge charge $q_i$ to the $i$-th chiral multiplet. The partition function for the resulting $U(1)$ gauge theory on an ellipsoid is $\int\mathrm d\hat\sigma Z(\hat\sigma)$, where
\begin{equation}
 Z(\hat\sigma) = \prod_{i}s_b\left(\frac{i(1-r_i)Q}2-q_i\hat\sigma\right).
\end{equation}
The introduction of a vortex loop $V_\beta$ along $S^1_\tau$ at $\theta=0$ modifies the integrand as follows:
\begin{align}
 Z(\hat\sigma) ~\rightarrow~&
 \prod_{i(\text{BC1})} s_b\left(\frac{i(1-r_i)Q}2-q_i\hat\sigma-ibq_i\beta+ib\lfloor q_i\beta\rfloor\right)
\nonumber \\ \cdot&
 \prod_{\tilde\imath(\text{BC2})} s_b\left(\frac{i(1-r_{\tilde\imath})Q}2-q_{\tilde\imath}\hat\sigma-ibq_{\tilde\imath}\beta+ib\lceil q_{\tilde\imath}\beta\rceil\right)\,.
\end{align}
In general the result cannot be expressed as $Z(\hat\sigma)$ with a shifted argument. But something special happens if all the matters have $U(1)$ charge $\pm q$. If one imposes \textbf{BC1} on all chirals with charge $+q$ and \textbf{BC2} on all those with charge $-q$, then the modification amounts to $\hat\sigma\to\hat\sigma+ib\beta-ibq^{-1}\lfloor q\beta\rfloor$. The simplest such example is the theory of a single chiral multiplet (\ref{VbBC}). In such cases the effect of inserting a vortex loop can be eliminated by shifting the integration contour of $\hat\sigma$.

Similar results were obtained for vortex defects in 2D ${\cal N}=(2,2)$ SUSY abelian gauge theories \cite{Hosomichi:2017dbc}. Indeed, some nontrivial BPS defect operators with intriguing properties were found there only in theories with multiple chiral matters.

For vortex loops in theories with CS coupling, the argument of Section \ref{sec:WVequiv} applies. In particular, for $U(1)$ pure CS theories the vortex loop with vorticity $\beta$ is equivalent to the Wilson loop with charge $q=k\beta$. This can be formally reproduced by modifying (\ref{ZU1g}) and (\ref{VbU1g}) as
\[
 Z = \int{\cal D}AZ[A]e^{-S_\text{CS}[A]},\qquad
\langle V_\beta(\gamma)\rangle = \int{\cal D}AZ[A+A_{\beta,\gamma}]e^{-S_\text{CS}[A]}
\]
and shifting the integration variable.

\newpage

\providecommand{\href}[2]{#2}

\end{document}